\documentclass[a4paper,11pt]{article}
\pdfoutput=1
\usepackage{jcappub}
\usepackage[T1]{fontenc}
\usepackage{amssymb}
\usepackage{graphicx}
\usepackage{amsmath}
\usepackage{hyperref}
\usepackage{tensor}

\title{\boldmath Energy budget of cosmological first-order phase transition in FLRW background}

\author[1,3]{Rong-Gen Cai}
\author[1,2,3]{Shao-Jiang Wang}

\affiliation[1]{CAS Key Laboratory of Theoretical Physics, Institute of Theoretical Physics, Chinese Academy of Sciences, No.55 Zhong Guan Cun East Road, Beijing 100190, China}
\affiliation[2]{Department of Physics and Helsinki Institute of Physics, P.O. Box 64, FI-00014 University of Helsinki, Finland}
\affiliation[3]{School of Physical Sciences, University of Chinese Academy of Sciences, No.19A Yuquan Road, Beijing 100049, China}

\emailAdd{schwang@itp.ac.cn}

\abstract{We study the hydrodynamics of bubble expansion in cosmological first-order phase transition in the Friedmann-Lema\^{\i}tre-Robertson-Walker (FLRW) background with probe limit. Different from previous studies for fast first-order phase transition in flat background, we find that, for slow first-order phase transition in FLRW background with a given peculiar velocity of the bubble wall, the efficiency factor of energy transfer into bulk motion of thermal fluid is significantly reduced, thus decreasing the previously-thought dominated contribution from sound wave to the stochastic gravitational-wave background.
\begin{flushleft}
\end{flushleft}
\begin{flushleft}
\end{flushleft}}

\begin{document}
\maketitle
\flushbottom

\section{Introduction}

Our world is symmetry-broken. For some symmetry breaking in the early Universe, the induced phase transition is of first-order. The first-order phase transition has drawn much attention over the past three decades, because it could be relevant to the electroweak baryogenesis\cite{Cohen:1990py,Cohen:1990it,Nelson:1991ab,Cohen:1994ss,Cohen:1993nk}, stochastic gravitational-waves (GWs) background\cite{Witten:1984rs,Hogan:1986qda,Kosowsky:1991ua,Kosowsky:1992rz,Kosowsky:1992vn,Kamionkowski:1993fg}, primordial magnetic fields \cite{Hogan:1983zz,Quashnock:1988vs,Vachaspati:1991nm,Cheng:1994yr,Baym:1995fk} and primordial black holes\cite{Hawking:1982ga,Kodama:1982sf,Moss:1994iq}, to name a few. The first-order phase transition proceeds with the nucleation, expansion and percolation of true vacuum bubbles within the false vacuum environment. See \cite{Binetruy:2012ze,Caprini:2015zlo,Cai:2017cbj,Weir:2017wfa} for brief reviews of GWs from first-order phase transitions.

During the stage of bubble nucleation, the phase transition can be either of fast or slow types \cite{Kobakhidze:2017mru,Cai:2017tmh} according to the nucleation rate, which describes the number of nucleated bubbles per unit volume and per unit time. For the nucleation rate of exponential growth with decreasing temperature, the phase transition ends shortly after the nucleation rate catches up the Hubble expansion rate. This is what we usually study as fast first-order phase transition. However, the nucleation rate can also be of quadratic growth with decreasing temperature as studied in \cite{Megevand:2016lpr,Jinno:2017ixd}. When the nucleation rate measured by Hubble rate becomes non-monotonic with decreasing temperature, there could be a situation \cite{Cai:2017tmh} when the nucleation rate is at most slightly smaller than the order of unity. In this case, the expanding bubbles have to wait for more than one Hubble time then to percolate at last. This kind of slow first-order phase transition \cite{Cai:2017tmh} is thus identified as a small window before the regime when the phase transition is too slow to be ended. It is worth noting that, the fast and slow first-order phase transitions here are classified according to their nucleation rate of either monotonic or non-monotonic types. The first-order phase transition can also be regarded as slow if the bubble wall velocity is small enough on its own as discussed in \cite{Megevand:2017vtb}.

During the stage of bubble expansion, the bubbles continually expand under the driving force from the pressure difference inside and outside the bubbles. If the bubbles are nucleated within vacuum background \cite{Coleman:1977py,Callan:1977pt}, they will shortly approach the speed-of-light after their nucleations. However, the bubbles are actually nucleated within thermal background \cite{Linde:1980tt,Linde:1981zj}, thus the interaction with other particle species will exert friction force against bubble expansion. In the early study, the bubble wall velocity was fixed in \cite{Steinhardt:1981ct} by the Chapman-Jouguet condition observed in chemical combustion, which was later recognized in \cite{Laine:1993ey} as unrealistic condition for cosmological phase transitions. The terminal velocity of the bubble wall can only be rigorously settled down with input from microscopic physics \cite{Moore:1995ua,Moore:1995si,John:2000zq}, namely solving the combined equations of equation-of-motion (EOM) and Boltzman equation, which only succeeded for the standard model (SM) \cite{Moore:1995ua,Moore:1995si}, minimal supersymmetric standard model (MSSM) \cite{John:2000zq,Cline:2000nw,Carena:2000id,Carena:2002ss,Konstandin:2005cd,Cirigliano:2006dg} and SM extended with real singlet scalar \cite{Kozaczuk:2015owa} so far. As a result, it is usually difficult to solve Botlzman equation \cite{Huber:2013kj,Konstandin:2014zta}, hence some phenomenological approaches \cite{KurkiSuonio:1984ba,Ignatius:1993qn,Moore:1995si} are adopted to parameterize the friction term in a model-independent manner \cite{Megevand:2009ut,Megevand:2009gh,Espinosa:2010hh,Megevand:2013hwa}. The outcome for the bubble wall velocity can be either steady or runaway \cite{Bodeker:2009qy} according to the competition between driving force and friction force. However, it is found recently in \cite{Bodeker:2017cim} that, contrary to the leading order result \cite{Bodeker:2009qy}, at next-to-leading order an extra friction arises due to transition splitting at the bubble wall, which may prevent the bubble wall from runaway. We will revisit this issue in detail in future works.

With input value of the bubble wall velocity from microscopic physics, the macroscopic hydrodynamics could help us to understand the energy distribution of total released vacuum energy among kinetic energy of the bubble wall, kinetic energy of bulk fluid motion, and thermal energy of plasma without going to the details of microphysics. There is an efficiency factor to measure how much energy has been converted into bulk motion of thermal fluid, which can be analytically determined from macroscopic hydrodynamics in terms of the bubble wall velocity as free parameter. This efficiency factor was only studied and summarized in \cite{Espinosa:2010hh} with respect to bag equation-of-state (EOS) in Minkowski spacetime without backreactions (probe limit). Several generalizations of macroscopic hydrodynamics of bubble expansion have been made for a non-spherical bubble wall \cite{Leitao:2010yw} and non-standard constant sound velocity \cite{Leitao:2014pda} as well as some stability analysis \cite{Megevand:2013yua,Megevand:2014yua,Megevand:2014dua}. See \cite{Jackson:2018maa} for the appreciation of the fluctuation-dissipation theorem on the studies of non-equilibrium dynamics of cosmological phase transitions and references therein. We will study in this paper the macroscopic hydrodynamics of bubble expansion in the Friedmann-Lema\^{\i}tre-Robertson-Walker (FLRW) background, and leave the work beyond bag EOS and with backreaction for future.

During the stage of bubble percolation, the kinetic energy stored in the bubble wall would be transformed into the GW energy through bubble collisions. In a series of numerical simulations of bubble collisions \cite{Kosowsky:1991ua,Kosowsky:1992rz,Kosowsky:1992vn,Kamionkowski:1993fg,Huber:2008hg}, it was found that the GW energy spectrum can be characterized with a few parameters, including the bubble wall velocity and efficiency factors. These numerical simulations were carried out under thin-wall and envelope approximations \cite{Weir:2016tov}, where the GWs are mainly from the uncollided envelopes of bubble thin walls. Remarkably some analytic results \cite{Caprini:2007xq,Caprini:2009fx} of bubble collisions can be obtained, either by reserving both \cite{Jinno:2016vai} the thin-wall and envelope approximations or relaxing only \cite{Jinno:2017fby} the envelope approximation. Recently, new simulations of thermal first-order phase transition \cite{Hindmarsh:2013xza,Hindmarsh:2015qta,Hindmarsh:2017gnf} without envelope approximation \footnote{Recently, new simulation \cite{Cutting:2018tjt} of vacuum first-order phase transition without envelope approximation found that, the GWs energy spectrum falls off at high wavenumber as $k^{-1.5}$ instead of $k^{-1}$ from envelope approximation. It is also found that, there is a linear growth as an additional bump in the tail of GWs energy spectrum when scalar field settles down in the true vacuum during oscillation phase after bubble collisions.} found that, the dominated contribution of GWs energy spectrum comes from the sound waves \cite{Hogan:1986qda} of bulk fluid motion in addition to another negligible contribution from magneto-hydrodynamics (MHD) turbulence \cite{Kamionkowski:1993fg,Kosowsky:2001xp,Dolgov:2002ra,Nicolis:2003tg,Caprini:2006jb,Gogoberidze:2007an,Caprini:2009yp} (see \cite{Niksa:2018ofa} for recent progresses on GWs from MHD turbulence and references therein). However, almost all analytic estimations \footnote{Early analytic estimations \cite{Caprini:2007xq,Caprini:2009fx} of GW spectrum have used FLRW background, but not for the latest analytic estimations \cite{Jinno:2016vai,Jinno:2017fby}.} and numerical simulations \footnote{We only found one relevant paper \cite{KurkiSuonio:1996rk} on numerical simulation for phase transition with account for Hubble expansion. All later numerical simulations worked in a flat background.} to date of bubble collisions are implemented in a flat background without account for the background Hubble expansion.

The primary motivation behind the ignorance of background expansion is that, the fast first-order phase transition is usually completed in a short period compared to the Hubble time, therefore it seems reasonable to assume that the background spacetime is not expanding at all. However, the background expansion cannot be simply neglected for slow first-order phase transition, especially at its late-time stage. There is an extra term in the EOM of bulk fluid as we will see in this paper, which gives rise to a thinner profile for bulk fluid peculiar velocity. Therefore, for a slow first-order phase transition in FLRW background, there is less energy transfer into bulk fluid motion compared to the total released vacuum energy than that for a fast first-order phase transition in flat background, thus reducing the contributions to GWs from sound waves and MHD turbulence. The detection of stochastic GWs background from slow first-order phase transitions might be even harder than we previously thought.

The outline of the paper is as follows: In section \ref{sec:thermo}, the thermodynamical description for the scalar-fluid system \ref{subsec:scalar-fluid} is given along with its equation-of-state \ref{subsec:EOS}; In section \ref{sec:hydro}, the hydrodynamical description for the scalar-fluid system is given according to its junction equation \ref{subsec:junction} and EOM \ref{subsec:EOM}; In section \ref{sec:modes}, three modes of bubble expansion are solved for the given EOM, including detonation wave \ref{subsec:detona}, deflagration wave \ref{subsec:deflag} and hybrid wave \ref{subsec:hybrid}, which are the velocity profiles \ref{subsec:profile} for the bulk fluid motions; In section \ref{sec:efficiency}, the analytic estimations and numerical fittings for the efficiency factor are given in \ref{subsec:analytic} and \ref{subsec:fitting}, respectively; The section \ref{sec:conclusion} is devoted to conclusions. Some discussions on the bubble wall velocity are also presented in appendix.

\section{Thermodynamics}\label{sec:thermo}

In this section, we will give a brief review of the thermodynamical basis for the scalar-fluid system with bag EOS.

\subsection{Scalar-fluid system}\label{subsec:scalar-fluid}

The physical picture behind the bubble expansion in thermal plasma is described by the scalar-fluid system, of which the scalar field part is described by
\begin{align}
T_{\mu\nu}^\phi=\nabla_\mu\phi\nabla_\nu\phi-g_{\mu\nu}\left(\frac12(\nabla\phi)^2+V_{T=0}(\phi)\right),
\end{align}
and the thermal fluid part is described by
\begin{align}
T_{\mu\nu}^f=\sum_i\int\frac{\mathrm{d}^3k}{(2\pi)^32E_i}2k_\mu k_\nu f_i(k,x),
\end{align}
where $f_i$ is the distribution function of particle species $i$.
If the thermal fluid is in local equilibrium, then the scalar-fluid system can be parameterized as perfect fluid with energy-mentum tensor of form
\begin{align}
T_{\mu\nu}=(e+p)u_\mu u_\nu+pg_{\mu\nu},
\end{align}
here $e$ is the internal energy density, $p$ is the pressure, $g_{\mu\nu}$ is the background metric, and $u^\mu$ is the usual four-velocity.

For the flat background with metric of form
\begin{align}
\mathrm{d}s^2=-\mathrm{d}t^2+\mathrm{d}r^2+r^2\mathrm{d}\Omega_2^2,\qquad \mathrm{d}\Omega_2^2=\mathrm{d}\theta^2+\sin^2\theta\mathrm{d}\varphi^2,
\end{align}
the four-velocity is
\begin{align}
u^\mu=\frac{\mathrm{d}x^\mu}{\sqrt{-\mathrm{d}s^2}}=\gamma(v)(1,v,0,0),\qquad v\equiv\frac{\mathrm{d}r}{\mathrm{d}t},
\end{align}
where $\gamma(v)=1/\sqrt{1-v^2}$ is the Lorentz factor of three-velocity $v$, and we have assumed a spherically expanding bubble so that $\mathrm{d}\theta/\mathrm{d}t=\mathrm{d}\varphi/\mathrm{d}t=0$. From now on, we will not write down explicitly the velocity components along $\theta$ and $\varphi$ directions. For FLRW spacetime, introducing $\bar{t}_n$ as the conformal time of bubble nucleation since the beginning of radiation era, and $\bar{t}$ as the elapsed conformal time of bubble expansion, then the FLRW metric with comoving coordinates $\bar{t}$ and $\bar{r}$ reads
\begin{align}
\mathrm{d}s^2
=a(\bar{t}+\bar{t}_n)^2(-\mathrm{d}(\bar{t}+\bar{t}_n)^2+\mathrm{d}\bar{r}^2+\bar{r}^2\mathrm{d}\Omega_2^2)
=a(\bar{t}+\bar{t}_n)^2(-\mathrm{d}\bar{t}^2+\mathrm{d}\bar{r}^2+\bar{r}^2\mathrm{d}\Omega_2^2),
\end{align}
The four-velocity field is then
\begin{align}
u^\mu=\frac{\bar{\gamma}(\bar{v})}{a(\bar{t}+\bar{t}_n)}(1,\bar{v}), \qquad \bar{v}\equiv\frac{\mathrm{d}\bar{r}}{\mathrm{d}\bar{t}},
\end{align}
where $\bar{\gamma}(\bar{v})\equiv1/\sqrt{1-\bar{v}^2}$ is the Lorentz factor for the peculiar velocity $\bar{v}$. It can be checked that $u_\mu u^\mu=-1$ for both cases. We have also assumed a spherically expanding bubble so that $\mathrm{d}\theta/\mathrm{d}\bar{t}=\mathrm{d}\varphi/\mathrm{d}\bar{t}=0$. From now on, we will also not write down explicitly the velocity components along $\theta$ and $\varphi$ directions.

The thermodynamical properties of the scalar-fluid system are characterized by its free energy density at finite temperature $\mathcal{F}(\phi,T)$. To see this, note that, the pressure is just the minus free energy density
\begin{align}
p=-\mathcal{F}(\phi,T),
\end{align}
and the entropy density is by definition
\begin{align}
s=\frac{\partial p}{\partial T}=-\frac{\partial\mathcal{F}}{\partial T}.
\end{align}
The energy density is given by
\begin{align}
e=\mathcal{F}+Ts=\mathcal{F}-T\frac{\partial\mathcal{F}}{\partial T},
\end{align}
and the enthalpy density is simply the sum of energy density and pressure, namely
\begin{align}
w=e+p=Ts=T\frac{\partial p}{\partial T}=-T\frac{\partial\mathcal{F}}{\partial T}.
\end{align}
As long as the free energy density is provided, the thermodynamics of scalar-fluid system is determined.

For the scalar-fluid system, the free energy density is defined by the effective potential at finite temperature, namely
\begin{align}
\mathcal{F}(\phi,T)\equiv V_\mathrm{eff}(\phi,T)=V_0(\phi)+V_T(\phi,T).
\end{align}
At 1-loop order, the zero-temperature part $V_0(\phi)=V_\mathrm{tree}(\phi)+V_\mathrm{CW}(\phi)$ is just the sum of tree potential and Coleman-Weinberg potential, and the finite-temperature part is
\begin{align}
V_T(\phi,T)=\sum_{i=\mathrm{B,F}}\pm g_iT\int\frac{\mathrm{d}^3k}{(2\pi)^3}\log\left(1\mp e^{-\sqrt{k^2+m_i^2}/T}\right)
=\frac{T^4}{2\pi^2}\sum_ig_iY_{\mathrm{B/F}}\left(\frac{m_i}{T}\right),
\end{align}
where for bosons/fermions the form of $Y_{\mathrm{B/F}}(x)$ is
\begin{align}
Y_{\mathrm{B/F}}(x)=\pm\int_0^\infty\mathrm{d}y y^2\log\left(1\mp\exp(-\sqrt{x^2+y^2})\right).
\end{align}
Therefore, the finite-temperature part can be computed as
\begin{align}
V_T(\phi,T)=\sum_{i=\mathrm{Boson}}g_i^Bf_i^B+\sum_{i=\mathrm{Fermion}}g_i^Ff_i^F,
\end{align}
here the free energy densities for bosons and fermions are
\begin{align}
f_i^B=&-\frac{\pi^2}{90}T^4+\frac{m_i^2}{24}T^2-\frac{m_i^3}{12\pi}T-\frac{m_i^4}{64\pi^2}\log\frac{m_i^2}{b_BT^2}
     -\frac{m_i^4}{16\pi^{5/2}}\sum_lc_l^B\left(\frac{m_i^2}{4\pi^2T^2}\right)^l;\\
f_i^F=&-\frac{7}{8}\frac{\pi^2}{90}T^4+\frac{m_i^2}{48}T^2+\frac{m_i^4}{64\pi^2}\log\frac{m_i^2}{b_FT^2}
     +\frac{m_i^4}{16\pi^{5/2}}\sum_lc_l^F\left(\frac{m_i^2}{4\pi^2T^2}\right)^l,
\end{align}
respectively, where
\begin{align}
b_B&=16\pi^2\ln\left(\frac32-2\gamma_E\right);\\
b_F&=\pi^2\ln\left(\frac32-2\gamma_E\right);\\
c_l^B&=(-1)^l\frac{\zeta(2l+1)}{(l+1)!};\\
c_l^F&=(-1)^l\frac{\zeta(2l+1)}{(l+1)!}(1-2^{-2l-1})\Gamma(l+\frac12),
\end{align}
with Euler constant $\gamma_E$ and Riemann zeta function $\zeta(s)$.
For massless particle species, note that
\begin{align}
Y_{\mathrm{B}}(x=0)=-\frac{\pi^4}{45},\quad Y_{\mathrm{F}}(x=0)=-\frac{7}{8}\frac{\pi^4}{45},
\end{align}
one recovers the usual expression for the relativistic particles,
\begin{align}
V_T(\phi,T)=-\frac13aT^4,\quad a=\frac{\pi^2}{30}\sum_i\left(g_i^B+\frac78g_i^F\right),
\end{align}
where the Stefan parameter $a$ should not be confused with scale factor.

\subsection{Equation-of-state}\label{subsec:EOS}

In the literatures, the dubbed bag EOS \cite{Chodos:1974je} is often used to approximate the free energy density of scalar-fluid system as a simple combination of the constant vacuum energy and ideal thermal gas, namely
\begin{align}
\mathcal{F}(\phi_\pm(T),T)=V_0(\phi_\pm(T))-\frac13a_\pm T^4, \quad a_\pm=\frac{\pi^2}{30}\sum_i\left(g_i^B+\frac78g_i^F\right),
\end{align}
where the plus and minus signs stand for the symmetric and broken phases, respectively. The vacuum-expectation-value (VEV) $\phi_\pm(T)$ has mild dependence on temperature, which can be ignored in bag EOS. The essential idea of bag EOS is that, the symmetric phase consists of light particles with $m_i/T\ll1$, which contribute to the free energy density as the radiation energy density $\frac13a_+T^4$. When the bubble wall sweeps over the symmetric phase, some particles will acquire very large masses in the broken phase, which contribute to the free energy density in an exponentially suppressed manner. In this case, bag EOS simply ignores all orders of contributions to the free energy density from these would-be heavy particles. Therefore, the broken phase consists of remaining light particles with radiation contribution $\frac13a_-T^4$ to the free energy density. Assuming bag EOS, in the symmetric/broken phases, the pressure and energy density are of form
\begin{align}
\begin{split}\label{eq:bagEOS}
p_+=\frac13a_+T^4-\epsilon_+,\quad e_+&=a_+T^4+\epsilon_+, \quad \epsilon_+\equiv V_0(\phi_+);\\
p_-=\frac13a_-T^4-\epsilon_-,\quad e_-&=a_-T^4+\epsilon_-, \quad \epsilon_-\equiv V_0(\phi_-).
\end{split}
\end{align}
It is conventional to characterize the strength of phase transition dubbed strength factor with the released vacuum energy density normalized by the background radiation energy density,
\begin{align}
\alpha_+=\frac{\Delta\epsilon}{a_+T_+^4}=\frac{4\Delta\epsilon}{3w_+}, \quad \Delta\epsilon=\epsilon_+-\epsilon_-.
\end{align}

Deviations from bag EOS \cite{Leitao:2014pda} are caused by the phase transition from the symmetric phase to the broken phase when some of light particles acquire masses $m_i\approx T$ comparable to the temperature. In this case, the general form of EOS is given by
\begin{align}
p_\pm(T)&=-V_\mathrm{eff}(\phi_\pm(T),T),\\
e_\pm(T)&=V_\mathrm{eff}(\phi_\pm(T),T)-T\frac{\partial}{\partial T}V_\mathrm{eff}(\phi_\pm(T),T).
\end{align}
As an example, for particle species acquiring masses $m_i\lesssim T$, the free energy density in the broken phase could preserve only the quadratic contribution in temperature,
\begin{align}
\mathcal{F}_+(T)&=V_0(\phi_+(T))-\frac13a_+T^4;\\
\mathcal{F}_-(T)&=V_0(\phi_-(T))-\frac13a_+T^4+bT^2.
\end{align}
It is worth noting that, the leading term in $T^4$ is the same for both phases since massive particles are not ignored at all orders. Formally we can still make bag-like decomposition
\begin{align}
p_\pm(T)=\frac13a_\pm(T)T^4-\epsilon_\pm(T), \quad e_\pm(T)=a_\pm(T)T^4+\epsilon_\pm(T).
\end{align}
However, such bag-like decomposition is meaningless unless one specifies
\begin{align}
a_\pm(T)&=\frac{3}{4T^3}\frac{\partial p_\pm(T)}{\partial T}=\frac{3w_\pm(T)}{4T^4}=\frac{3}{4T^3}\frac{\partial}{\partial T}V_\mathrm{eff}(\phi_\pm(T),T);\\
\epsilon_\pm(T)&=\frac14(e_\pm(T)-3p_\pm(T))=V_\mathrm{eff}(\phi_\pm(T),T)-\frac{T}{4}\frac{\partial}{\partial T}V_\mathrm{eff}(\phi_\pm(T),T).
\end{align}
Therefore, the deviations from bag EOS can be characterised by the sound speed,
\begin{align}
c_\pm^2(T)=\frac{\partial p_\pm(T)}{\partial e_\pm(T)}=\frac13(1+\delta c_\pm^2(T)),
\end{align}
which is in general temperature dependent over spacetime. See \cite{Leitao:2014pda} for constant sound speed $c_s^2\neq1/3$ in the case of a planner bubble wall. Generalization to spherical bubble should be straightforward. Further generalization for temperature-dependent sound speed would be important, which is reserved for future work.

\section{Hydrodynamics}\label{sec:hydro}

In this section, we will derive the EOM for fluid peculiar velocity of bulk motion in FLRW background, and then we solve the EOM with the matching conditions at some discontinuity interfaces like the bubble wall and shockwave front.

\subsection{Junction equations}\label{subsec:junction}

\begin{figure}
  \centering
  \includegraphics[width=0.7\textwidth]{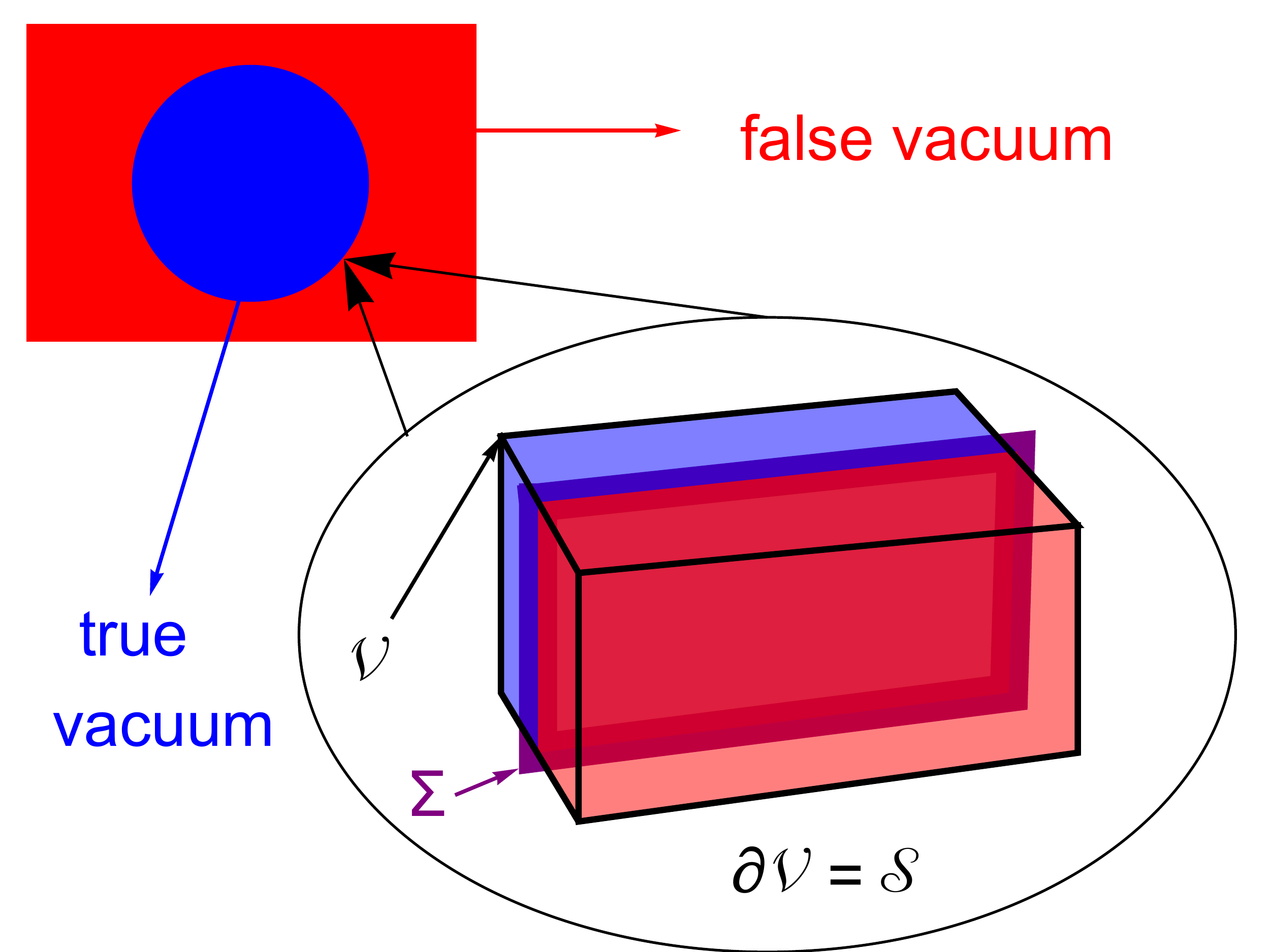}\\
  \caption{Illustration for the junction condition at the bubble wall. The blue shaded region is in true vacuum and the red shade region is in false vacuum. For a local region at the bubble wall, the junction condition can be derived at an interface $\Sigma$ inside some volume $\mathcal{V}$ enclosed by surface $\mathcal{S}$. }\label{fig:junction}
\end{figure}

Before deriving the EOM for fluid peculiar velocity of bulk motion in FLRW background, it is often useful to first specify the matching conditions at the bubble wall. Such matching conditions at the interface of discontinuity in hydrodynamics are dubbed as Taub (Rankine-Hugoniot) junction conditions with relativistic (Newtonian) treatments. The junction conditions at the local interface $\Sigma$ of some spherical discontinuity can be derived from the conservation of energy-momentum tensor for arbitrary $\lambda_\mu$,
\begin{align}
\nabla_\mu(T^{\mu\nu}\lambda_\nu)=T^{\mu\nu}\nabla_\mu\lambda_\nu,
\end{align}
which will be integrated over a volume $\mathcal{V}$ enclosed by surface $\mathcal{S}$ and containing interface $\Sigma$. See Fig.\ref{fig:junction}
for demonstration. Using Stokes's theorem, one has
\begin{align}
\int_\mathcal{V}\nabla_\mu(T^{\mu\nu}\lambda_\nu)\mathrm{d}^4x=\int_\mathcal{V}T^{\mu\nu}\nabla_\mu\lambda_\nu\mathrm{d}^4x=\oint_\mathcal{S}T^{\mu\nu}\lambda_\nu n_\mu\mathrm{d}V,
\end{align}
where $n_\mu$ is an unit vector $n_\mu=(0,1,0,0)$ along radial direction. Shrinking $\mathcal{V}$ down to the interface $\Sigma$, the left-hand-side (LHS) is simply zero, thus the right-hand-side (RHS) becomes
\begin{align}
\int_\Sigma\lambda_\mu(T_+^{\mu\nu}-T_-^{\mu\nu})n_\nu\mathrm{d}V=0.
\end{align}
Since $\lambda_\mu$ is arbitrary, the junction condition reads as
\begin{align}
(T_+^{\mu\nu}-T_-^{\mu\nu})n_\nu=0,
\end{align}
namely
\begin{align}
\begin{split}\label{eq:junction}
T_+^{rt}&=T_-^{rt};\\
T_+^{rr}&=T_-^{rr}.
\end{split}
\end{align}

The junction conditions \eqref{eq:junction} are usually written in the bubble wall frame. For flat background, the junction conditions \eqref{eq:junction} simply imply
\begin{align}
\begin{split}\label{eq:flatjunction}
w_+v_+\gamma_+^2&=w_-v_-\gamma_-^2;\\
w_+v_+^2\gamma_+^2+p_+&=w_-v_-^2\gamma_-^2+p_-,
\end{split}
\end{align}
where $v_\pm$ is the fluid velocity with respect to the bubble wall and $\gamma_\pm$ is the Lorentz factor for the corresponding fluid velocity.
For FLRW background in comoving coordinate system, the junction conditions \eqref{eq:junction} also give rise to
\begin{align}
\begin{split}
\frac{1}{a_+^2}(w_+\bar{v}_+\bar{\gamma}_+^2)&=\frac{1}{a_-^2}(w_-\bar{v}_-\bar{\gamma}_-^2),\\
\frac{1}{a_+^2}(w_+\bar{v}_+^2\bar{\gamma}_+^2+p_+)&=\frac{1}{a_-^2}(w_-\bar{v}_-^2\bar{\gamma}_-^2+p_-),
\end{split}
\end{align}
where $\bar{v}_\pm$ is the fluid peculiar velocity with respect to the bubble wall in comoving coordinate system, and $\bar{\gamma}_\pm$ is the Lorentz factor for the corresponding fluid peculiar velocity. In the probe limit when the backreactions on the background expansion from the bubbles can be neglected, one has the same scale factors $a_+=a_-$ inside and outside of bubbles, therefore the junction conditions for FLRW background are very similar to the case of flat background,
\begin{align}
\begin{split}\label{eq:FLRWjunction}
w_+\bar{v}_+\bar{\gamma}_+^2&=w_-\bar{v}_-\bar{\gamma}_-^2,\\
w_+\bar{v}_+^2\bar{\gamma}_+^2+p_+&=w_-\bar{v}_-^2\bar{\gamma}_-^2+p_-.
\end{split}
\end{align}
Beyond the probe limit, the metric ansatz can be of form
\begin{align}
\mathrm{d}s^2=-f(t,r)\mathrm{d}t^2+\frac{\mathrm{d}r^2}{g(t,r)}+h(t,r)r^2(\mathrm{d}\theta^2+\sin^2\theta\mathrm{d}\varphi^2),
\end{align}
which will be reserved for future work.
\begin{figure}
  \centering
  \includegraphics[width=0.7\textwidth]{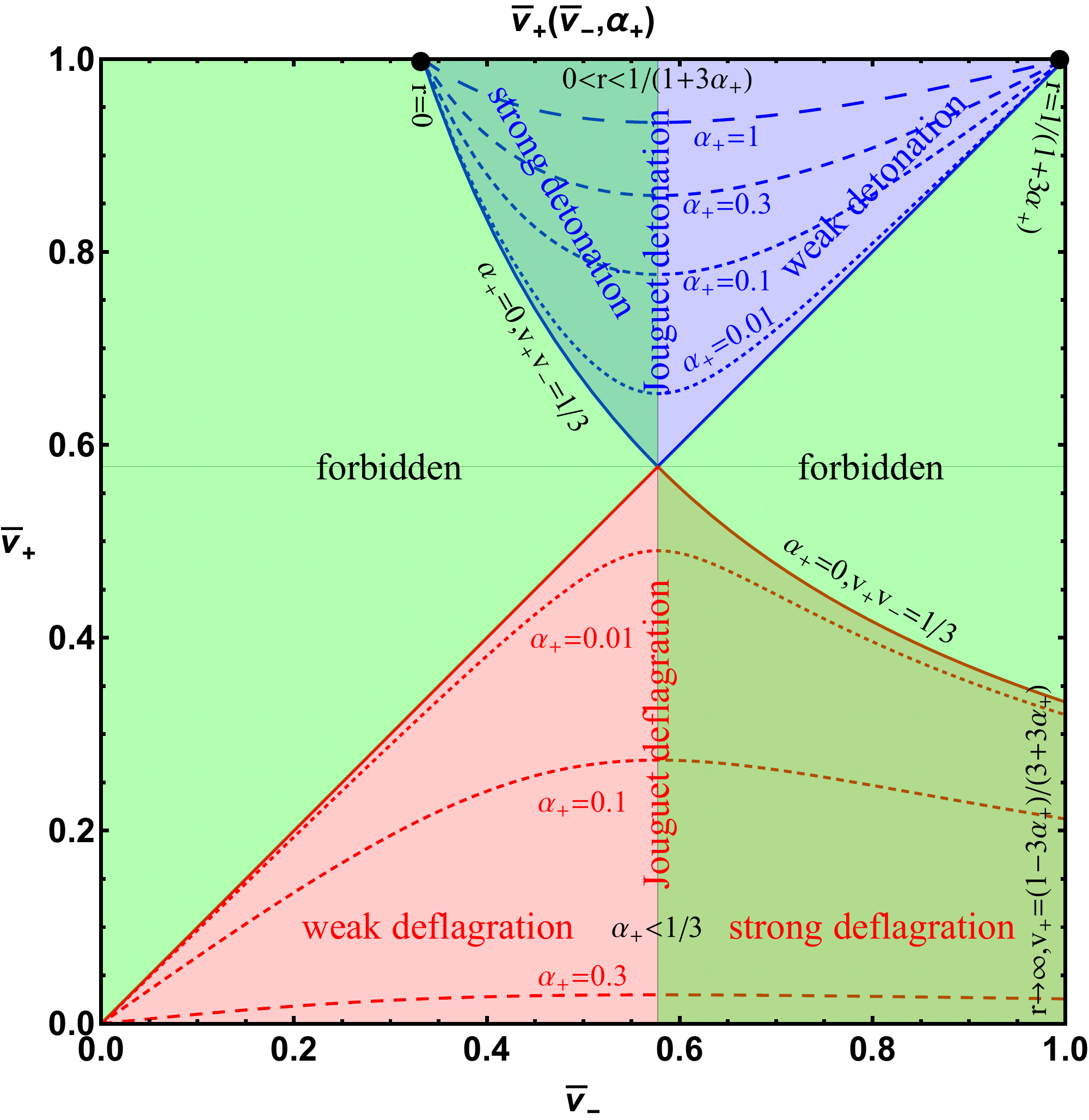}
  \caption{The wall frame peculiar velocity $\bar{v}_-$ and $\bar{v}_+$ of bulk fluid in the back (x-axis) and front (y-axis) of the bubble wall for given strength factor $\alpha_+$. The blue shaded region is detonation mode, and the red dashed region is deflagration mode. The green shaded region is forbidden so that only the weak detonation and weak deflagration are allowed. The strong deflagration will be decay into the Jouguet deflagration, which is the hybrid mode with $\bar{v}_-=c_s^-$. }\label{fig:WallVelocity}
\end{figure}
From matching conditions \eqref{eq:FLRWjunction}, one can obtain following relations,
\begin{align}
\bar{v}_+\bar{v}_-=\frac{p_+-p_-}{e_+-e_-},\quad \frac{\bar{v}_+}{\bar{v}_-}=\frac{e_-+p_+}{e_++p_-}.
\end{align}
Adopting the bag EOS \eqref{eq:bagEOS}, one has
\begin{align}\label{eq:vpandvm}
\bar{v}_+\bar{v}_-=\frac{1-(1-3\alpha_+)r}{3-3(1+\alpha_+)r},\quad \frac{\bar{v}_+}{\bar{v}_-}=\frac{3+(1-3\alpha_+)r}{1+3(1+\alpha_+)r},
\end{align}
where
\begin{align}
\alpha_+=\frac{\Delta\epsilon}{a_+T_+^4}=\frac{4\Delta\epsilon}{3w_+},\quad r=\frac{a_+T_+^4}{a_-T_-^4}=\frac{w_+}{w_-}.
\end{align}
Hence for given $\alpha_+$ and $r$ one recovers $\bar{v}_\pm$ by
\begin{align}
\begin{split}\label{eq:vpvmalphar}
\bar{v}_+(\alpha_+,r)&=\sqrt{\frac{1-(1-3\alpha_+)r}{3-3(1+\alpha_+)r}\cdot\frac{3+(1-3\alpha_+)r}{1+3(1+\alpha_+)r}};\\
\bar{v}_-(\alpha_+,r)&=\sqrt{\left.\frac{1-(1-3\alpha_+)r}{3-3(1+\alpha_+)r}\right/\frac{3+(1-3\alpha_+)r}{1+3(1+\alpha_+)r}}.
\end{split}
\end{align}
One can also kill $r$ from above relations and immediately derive following solutions with two branches,
\begin{align}\label{eq:vpvmalpha}
\bar{v}_+=\frac{1}{1+\alpha_+}\left[\left(\frac{\bar{v}_-}{2}+\frac{1}{6\bar{v}_-}\right)
\pm\sqrt{\left(\frac{\bar{v}_-}{2}+\frac{1}{6\bar{v}_-}\right)^2+\alpha_+^2+\frac23\alpha_+-\frac13}\right],
\end{align}
which can be presented in Fig.\ref{fig:WallVelocity} as a function $\bar{v}_+(\bar{v}_-,\alpha_+)$ for given strength factor $\alpha_+$. The expanding modes for bubbles can be classified according to the wall frame fluid peculiar velocity $\bar{v}_\pm$ just in the front/back of a bubble wall with respect to the sound speed $c_s$. The expanding mode is of detonation wave if $\bar{v}_+>\bar{v}_-$, which can be further classified into strong, Jouguet and weak types if $\bar{v}_+>c_s>\bar{v}_-$, $\bar{v}_-=c_s$ and $\bar{v}_->c_s$, respectively. Similarly, the expanding mode is of deflagration wave if $\bar{v}_->\bar{v}_+$, which can be further classified into strong, Jouguet and weak types if $\bar{v}_->c_s>\bar{v}_+$, $\bar{v}_-=c_s$ and $\bar{v}_-<c_s$, respectively. In \cite{Laine:1993ey}, it has been convinced that both strong detonation and strong deflagration are forbidden. As we will see in Section \ref{sec:modes}, the detonation wave proceeds with rarefaction wave in the back of the bubble wall, and the deflagration wave proceeds with compression shockwave in the front of the bubble wall. In \cite{Espinosa:2010hh}, it has been shown clearly that the Jouguet deflagration is actually a hybrid wave with shock wave and rarefaction wave in the front and back of the bubble wall. The same characteristic is also manifested in FLRW background as we will see later.

\subsection{Equation-of-motion}\label{subsec:EOM}

Similar to the case in flat background, the EOM of peculiar velocity of bulk fluid in FLRW background can be derived from the conservation of energy-momentum tensor,
\begin{align}
\nabla^\mu T_{\mu\nu}=\nabla^\mu(wu_\mu)u_\nu+wu_\mu\nabla^\mu u_\nu+\nabla_\nu p=0.
\end{align}
When projected along the direction of fluid flow, namely the four-velocity $u^\mu=\frac{\bar{\gamma}(\bar{v})}{a(\bar{t}+\bar{t}_n)}(1,\bar{v})$, the conservation of energy-momentum tensor becomes
\begin{align}
u_\nu\nabla_\mu T^{\mu\nu}=u_\nu u^\nu\nabla_\mu(wu^\mu)+wu^\mu u_\nu\nabla_\mu u^\nu+u_\nu\nabla^\nu p=0,
\end{align}
which, after using $u_\nu u^\nu=-1$ and $u_\nu\nabla_\mu u^\nu=0$, turns into
\begin{align}
-\nabla_\mu(wu^\mu)+u_\mu\nabla^\mu p=0,
\end{align}
namely
\begin{align}\label{eq:FLRWEOM1}
w\nabla_\mu u^\mu+u^\mu\nabla_\mu e=0.
\end{align}
Then we construct $\tilde{u}^\mu$ so that $\tilde{u}_\mu u^\mu=0$ and $\tilde{u}_\mu\tilde{u}^\mu=1$, namely
\begin{align}
\tilde{u}^\mu=\frac{\bar{\gamma}(\bar{v})}{a(\bar{t}+\bar{t}_n)}(\bar{v},1),
\end{align}
which is perpendicular to the bulk fluid flow. Hence one can also projecting the conservation of energy-momentum tensor along the perpendicular direction of fluid flow,
\begin{align}
\tilde{u}^\nu\nabla^\mu T_{\mu\nu}=\tilde{u}^\nu u_\nu\nabla^\mu(wu_\mu)+w\tilde{u}^\nu u_\mu\nabla^\mu u_\nu+\tilde{u}^\nu\nabla_\nu p=0,
\end{align}
which, after using $\tilde{u}^\nu u_\nu=0$, becomes
\begin{align}\label{eq:FLRWEOM2}
w\tilde{u}^\nu u^\mu\nabla_\mu u_\nu+\tilde{u}^\nu\nabla_\nu p=0.
\end{align}
Assuming a spherically expanding bubble wall in the bubble center frame (comoving frame), there is no characteristic distance scale when comoving coordinates $\bar{t}$ and $\bar{r}$ are adopted. The peculiar velocity of bulk fluid
\begin{align}
\bar{\mathbf{v}}=\bar{v}(t,r)\hat{\bar{r}}=\bar{v}(\bar{\xi}\equiv\frac{\bar{r}}{\bar{t}})\hat{\bar{r}}
\end{align}
thus depends only on the so-called comoving similarity coordinate $\bar{\xi}\equiv\bar{r}/\bar{t}$, where $\bar{r}$ is the comoving distance from the bubble center and $\bar{t}$ is the conformal time since bubble nucleation. Therefore, for steady configuration of velocity profile of bulk fluid motion, $\bar{\xi}$ is the peculiar velocity of a given point in the wave profile and the fluid element at a position traced by $\bar{\xi}$ in the wave profile move with peculiar velocity $\bar{v}(\bar{\xi})$, which is the fluid peculiar velocity in the bubble center frame.

Equipped with comoving similarity coordinate, the energy-momentum conservation \eqref{eq:FLRWEOM1} and \eqref{eq:FLRWEOM2} can be greatly simplified. Since the energy $e$ and pressure $p$ are all scalar functions in \eqref{eq:FLRWEOM1} and \eqref{eq:FLRWEOM2}, the covariant derivative is equivalent to the normal derivative, which can be rewritten with respect to the comoving similarity coordinate $\bar{\xi}$,
\begin{align}
u^\mu\nabla_\mu e&=\frac{\bar{\gamma}}{a}(1,\bar{v})(-\frac{\bar{\xi}}{\bar{t}},\frac{1}{\bar{t}})^T\partial_{\bar{\xi}}e=\frac{\bar{\gamma}}{a\bar{t}}(\bar{v}-\bar{\xi})\partial_{\bar{\xi}}e;\\
\tilde{u}^\nu\nabla_\nu p&=\frac{\bar{\gamma}}{a}(\bar{v},1)(-\frac{\bar{\xi}}{\bar{t}},\frac{1}{\bar{t}})^T\partial_{\bar{\xi}}p=\frac{\bar{\gamma}}{a\bar{t}}(1-\bar{\xi}\bar{v})\partial_{\bar{\xi}}p.
\end{align}
Plugging the FLRW metric into \eqref{eq:FLRWEOM1} and \eqref{eq:FLRWEOM2}, one arrives at the following equations,
\begin{align}
\frac{\bar{\gamma}}{a\bar{t}}(\bar{\xi}-\bar{v})\frac{\partial_{\bar{\xi}}e}{w}&=\nabla_\mu u^\mu=\frac{2\bar{v}}{\bar{\xi}}\frac{\bar{\gamma}}{a\bar{t}}
+3\bar{t}\frac{\partial_{\bar{t}}a}{a}\frac{\bar{\gamma}}{a\bar{t}}
+\frac{\bar{\gamma}}{a\bar{t}}\bar{\gamma}^2(1-\bar{\xi}\bar{v})\partial_{\bar{\xi}}\bar{v};\\
\frac{\bar{\gamma}}{a\bar{t}}(1-\bar{\xi}\bar{v})\frac{\partial_{\bar{\xi}}p}{w}&=-\tilde{u}^\nu u^\mu\nabla_\mu u_\nu=-\bar{v}\bar{t}\frac{\partial_{\bar{t}}a}{a}\frac{\bar{\gamma}}{a\bar{t}}
+\frac{\bar{\gamma}}{a\bar{t}}\bar{\gamma}^2(\bar{\xi}-\bar{v})\partial_{\bar{\xi}}\bar{v},
\end{align}
which, after abbreviate the $\bar{t}\partial_{\bar{t}}a(\bar{t}+\bar{t}_n)/a(\bar{t}+\bar{t}_n)\equiv n$, turns into
\begin{align}
(\bar{\xi}-\bar{v})\frac{\partial_{\bar{\xi}}e}{w}&=2\frac{\bar{v}}{\bar{\xi}}+3n+\bar{\gamma}^2(1-\bar{\xi}\bar{v})\partial_{\bar{\xi}}\bar{v};\label{eq:FLRWEOM3}\\
(1-\bar{\xi}\bar{v})\frac{\partial_{\bar{\xi}}p}{w}&=-\bar{v}n+\bar{\gamma}^2(\bar{\xi}-\bar{v})\partial_{\bar{\xi}}\bar{v}.\label{eq:FLRWEOM4}
\end{align}
Combining the above equations into a single equation, we obtain the final form of the EOM,
\begin{align}\label{eq:FLRWEOM}
2\frac{\bar{v}}{\bar{\xi}}+n\left(3+\frac{\bar{\mu}\bar{v}}{c_s^2}\right)
=\bar{\gamma}^2(1-\bar{\xi}\bar{v})\left(\frac{\bar{\mu}^2}{c_s^2}-1\right)\partial_{\bar{\xi}}\bar{v},
\end{align}
where the definition of speed-of-sound $c_s^2=\partial_{\bar{\xi}}p/\partial_{\bar{\xi}}e$ is used. The local Lorentz-transformed fluid peculiar  velocity
\begin{align}
\bar{\mu}(\bar{\xi},\bar{v}(\bar{\xi}))\equiv\frac{\bar{\xi}-\bar{v}(\bar{\xi})}{1-\bar{\xi}\bar{v}(\bar{\xi})}
\end{align}
is used to transform between the bubble center frame (with prime symbol) and the bubble wall frame (without prime symbol) in comoving coordinate system,
\begin{align}
\mathbf{\bar{v}}'_\pm&=\frac{\mathbf{\bar{v}}_\pm+\mathbf{\bar{\xi}}_w}{1+\mathbf{\bar{v}}_\pm\cdot\mathbf{\bar{\xi}}_w}\Rightarrow \bar{v}'_\pm=\frac{\bar{\xi}_w-\bar{v}_\pm}{1-\bar{\xi}_w\bar{v}_\pm}\equiv\bar{\mu}(\bar{\xi}_w,\bar{v}_\pm);\\
\mathbf{\bar{v}}_\pm&=\frac{\mathbf{\bar{v}}'_\pm-\mathbf{\bar{\xi}}_w}{1-\mathbf{\bar{v}}'_\pm\cdot\mathbf{\bar{\xi}}_w}\Rightarrow \bar{v}_\pm=\frac{\bar{\xi}_w-\bar{v}'_\pm}{1-\bar{\xi}_w\bar{v}'_\pm}\equiv\bar{\mu}(\bar{\xi}_w,\bar{v}'_\pm).
\end{align}
It should not be confused that $\bar{v}(\bar{\xi})$, although without prime symbol, always denotes fluid peculiar velocity in bubble center frame.

\begin{figure}
  \centering
  \includegraphics[width=0.49\textwidth]{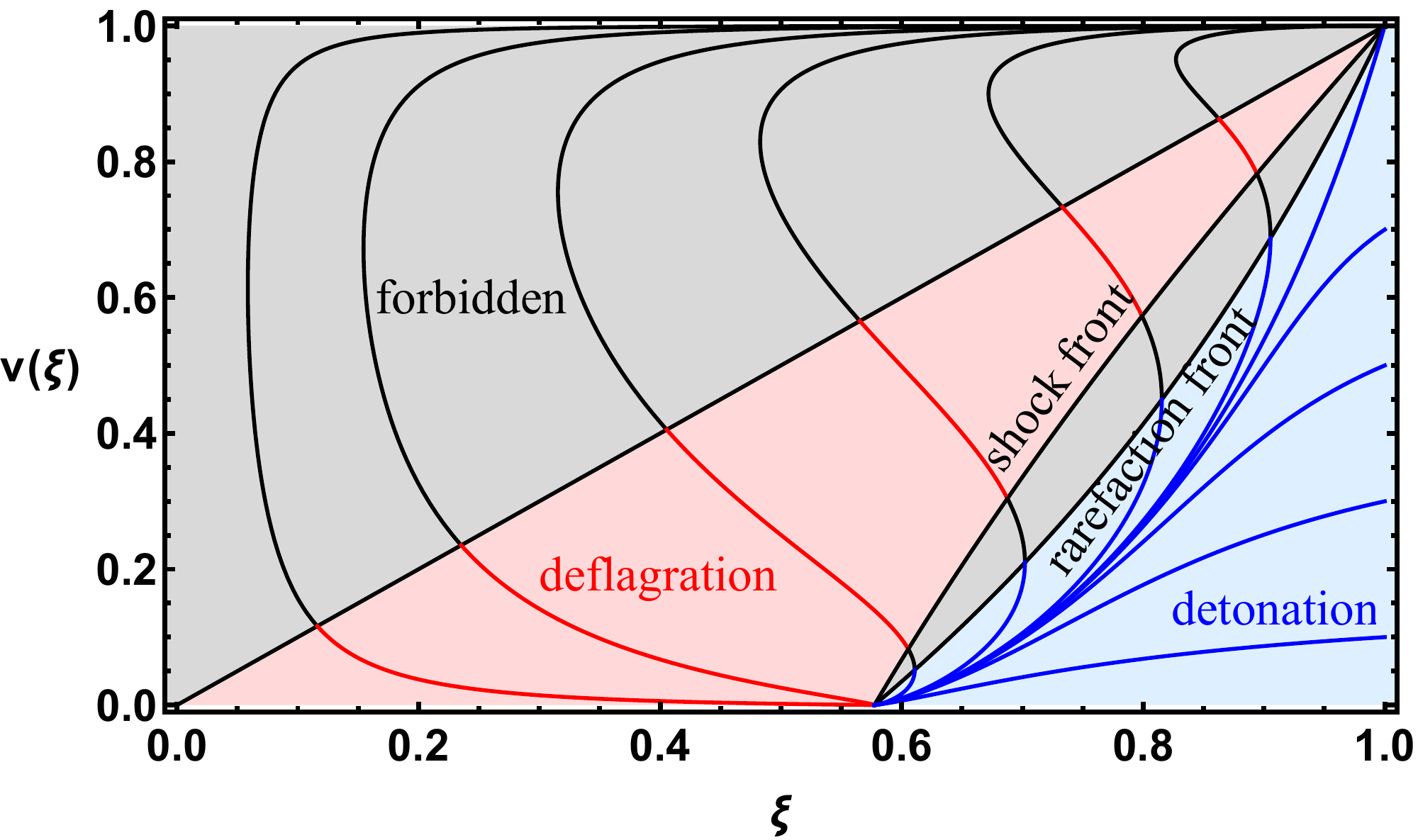}
  \includegraphics[width=0.49\textwidth]{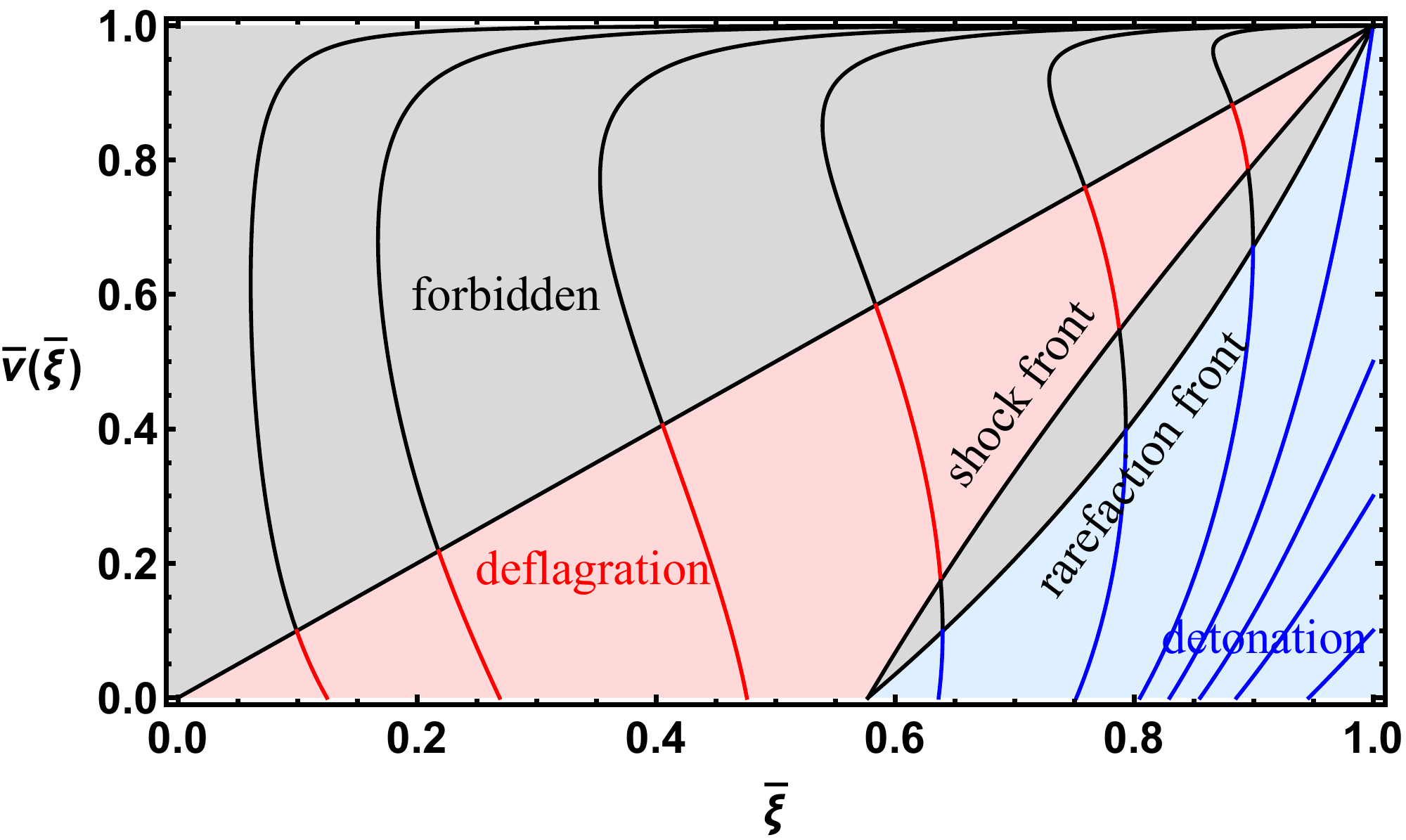}\\
  \caption{The similarity solutions $v(\xi)$ and $\bar{v}(\bar{\xi})$ without input matching conditions to the EOM of bulk fluid motion in bubble center frame for fast (left) and slow (right) first-order phase transitions in flat (left) and FLRW (right) background. In flat background, the grey shaded region above $v=\xi$ is forbidden, and the red shaded region is deflagration mode while blue shaded region is detonation mode, which are separated by the shockwave front defined by $\mu(\xi,v)\xi=c_s^2$ and rarefaction front defined by $\mu(\xi,v)=c_s$. The same classification is also shown for FLRW background in comoving coordinate system, except that the solution curves will not be ended at the same improper node point $(\bar{\xi},\bar{v})=(c_s,0)$ as in the flat background any more.}\label{fig:EOM}
\end{figure}

The flat background case \cite{Espinosa:2010hh} can be recovered by noting that $n=0$ for the absence of scale factor. After getting rid of all the bar symbols, the EOM of fluid profile of scalar-fluid system in flat background reads
\begin{align}\label{eq:flatEOM}
2\frac{v}{\xi}=\gamma^2(1-v\xi)\left(\frac{\mu^2}{c_s^2}-1\right)\partial_\xi v,
\end{align}
where $v(\tau)$ and $\xi(\tau)$ can be parameterized in terms of some parameter $\tau$, and above EOM becomes
\begin{align}
\frac{\mathrm{d}v}{\mathrm{d}\tau}&=2vc_s^2(1-v^2)(1-v\xi);\\
\frac{\mathrm{d}\xi}{\mathrm{d}\tau}&=\xi((\xi-v)^2-c_s^2(1-v\xi)^2),
\end{align}
where the first parametrization equation exhibits a fixed point at $(\xi=1,v=1)$, and the second one also exhibits a fixed point at $(\xi=c_s,v=0)$. As you will see, the similarity solution $v(\xi)$ is not a single-valued function, and thus one has to instead solve EOM \eqref{eq:flatEOM} without junction conditions for $\xi(v)$ as presented in the left panel of Fig.\ref{fig:EOM}.

However, the situation becomes difficult for FLRW background in radiation-dominated era with $a(\bar{t})\propto\bar{t}$, where our abbreviation
\begin{align}
n\equiv\bar{t}\frac{\partial_{\bar{t}}a(\bar{t}+\bar{t}_n)}{a(\bar{t}+\bar{t}_n)}=\frac{1}{1+\bar{t}_n/\bar{t}}
\end{align}
is pure time-dependent without similarity, therefore there is simply no way to solve EOM \eqref{eq:FLRWEOM} for $\bar{v}(\bar{\xi})$ as a function of similarity variable $\bar{\xi}$ alone. Nevertheless, there are two limiting cases that we can solve a similarity solution out of EOM \eqref{eq:FLRWEOM}: one is at early-time stage of bubble expansion with $\bar{t}_n\gg\bar{t}$ where $n$ can be approximated as 0; and the other is at the late-time stage of bubble expansion with $\bar{t}_n\ll\bar{t}$ where $n$ can be approximated as 1. Seeking for a general solution for time-dependent $n$ connecting these two similarity solutions with $n=0,1$ is beyond the scope of current paper, which will be pursuit in future.

The $n=0$ case corresponds to the fast first-order phase transition in flat background. In the fast first-order phase transition, bubbles are nucleated with exponential rate so that most of bubbles are nucleated just before the percolation temperature, therefore the elapsed conformal time of bubble expansion is much shorter than the conformal time of bubble nucleations since the beginning of radiation era;
the $n=1$ case one corresponds to the late-time stage of slow first-order phase transition in FLRW background. In the slow first-order phase transition, bubbles are almost simultaneously nucleated around the minimum of bounce action but percolated at later time much longer than the Hubble time when they are nucleated. Take an example from our previous study \cite{Cai:2017tmh}, in the regime of slow first-order phase transition, the nucleation temperature is around $30$ GeV, while the percolation temperature can be as low as $10^{-2}$ GeV, which gives rise to an estimation of
\begin{align}
10^3\simeq\frac{T_\mathrm{nuc}}{T_\mathrm{per}}=\frac{a_\mathrm{per}}{a_\mathrm{nuc}}=\frac{\bar{t}_\mathrm{per}+\bar{t}_n}{\bar{t}_n}\Rightarrow n(\bar{t}_\mathrm{per})=\frac{1}{1+10^{-3}}\simeq1.
\end{align}
Hence in what follows, we will focus on the special case $n=1$, which is equivalent to the late-time stage of slow first-order phase transition in FLRW background and radiation dominated era. The only question remained is that, why is the early-time stage of slow first-order phase transition unimportant ? As we will see in \ref{subsec:analytic}, although the efficiency factor of early-time stage of slow first-order phase transition (equivalent to the fast first-order phase transition in flat background) is larger than that of late-time stage of slow first-order phase transition, the size of bubbles at late-time is much larger than that at early-time. Therefore, the total released vacuum energy at late-time is much larger than that at early-time, consequently, the dissipated energy into bulk fluid motion at late-time is much larger than that at early-time. As a result, the GWs from bulk fluid motion should use the efficiency factor at late-time instead of that at early-time. What we want to do in this paper is to work out the efficiency factor at late-time stage of slow first-order phase transition, namely the $n=1$ case.

The similarity solution of EOM \eqref{eq:FLRWEOM} with $n=1$ is presented in the right panel of Fig.\ref{fig:EOM}, which is significantly different from the case of fast first-order phase transition for flat background. However, as we will see in the next section, the conditions for shockwave front $\bar{\mu}(\bar{\xi},\bar{v})\bar{\xi}=c_s^2$ and rarefaction front $\bar{\mu}(\bar{\xi},\bar{v})=c_s$ are unchanged formally except for the extra bar symbols. As a result, the regions for deflagration (red shaded region) and detonation (blue shaded region) keep the same way as in the flat background.

With the solution of velocity profile $\bar{v}(\bar{\xi})$, one can also obtain the enthalpy profile and temperature profile.
Inserting \eqref{eq:FLRWEOM} into \eqref{eq:FLRWEOM3} and \eqref{eq:FLRWEOM4} gives rise to
\begin{align}
\frac{\partial_{\bar{\xi}}e}{w}&=\frac{\bar{\gamma}^2\bar{\mu}}{c_s^2}\partial_{\bar{\xi}}\bar{v}-\frac{n\bar{v}}{c_s^2(1-\bar{\xi}\bar{v})};\label{eq:FLRWEOM5}\\
\frac{\partial_{\bar{\xi}}p}{w}&=\bar{\gamma}^2\bar{\mu}\partial_{\bar{\xi}}\bar{v}-\frac{n\bar{v}}{1-\bar{\xi}\bar{v}},\label{eq:FLRWEOM6}
\end{align}
which after summing up together becomes
\begin{align}
\partial_{\bar{\xi}}\log w=\bar{\gamma}^2\bar{\mu}\left(\frac{1}{c_s^2}+1\right)\partial_{\bar{\xi}}\bar{v}-\frac{n\bar{v}}{1-\bar{\xi}\bar{v}}\left(\frac{1}{c_s^2}+1\right).
\end{align}
Therefore the enthalpy profile can be directly inferred from the velocity profile via
\begin{align}
w(\bar{\xi})=w(\bar{\xi}_0)\exp\left[\int_{\bar{v}(\bar{\xi}_0)}^{\bar{v}(\bar{\xi})}
\bar{\gamma}^2\bar{\mu}\left(\frac{1}{c_s^2}+1\right)\mathrm{d}\bar{v}(\bar{\xi})
-\int_{\bar{\xi}_0}^{\bar{\xi}}\frac{n\bar{v}}{1-\bar{\xi}\bar{v}}\left(\frac{1}{c_s^2}+1\right)\mathrm{d}\bar{\xi}\right].
\end{align}
As for temperature profile, noting that $w=T\frac{\partial p}{\partial T}$, \eqref{eq:FLRWEOM6} can be rewritten as
\begin{align}
\partial_{\bar{\xi}}\log T=\bar{\gamma}^2\bar{\mu}\partial_{\bar{\xi}}v-\frac{n\bar{v}}{1-\bar{\xi}\bar{v}},
\end{align}
and the temperature profile can be obtained in principle from the velocity profile via
\begin{align}
T(\bar{\xi})=T(\bar{\xi}_0)\exp\left[\int_{\bar{v}(\bar{\xi}_0)}^{\bar{v}(\bar{\xi})}
\bar{\gamma}^2\bar{\mu}\mathrm{d}\bar{v}(\bar{\xi})
-\int_{\bar{\xi}_0}^{\bar{\xi}}\frac{n\bar{v}}{1-\bar{\xi}\bar{v}}\mathrm{d}\bar{\xi}\right]
\end{align}
as long as the junction condition for temperature at the interface is specified. This is beyond the scope of current paper, which will be pursuit further in future works. Therefore, we will only present the velocity profile along with corresponding enthalpy profile in the next section for different expanding modes.

\section{Bubble expansion}\label{sec:modes}

\begin{figure}
\centering
\includegraphics[width=0.49\textwidth]{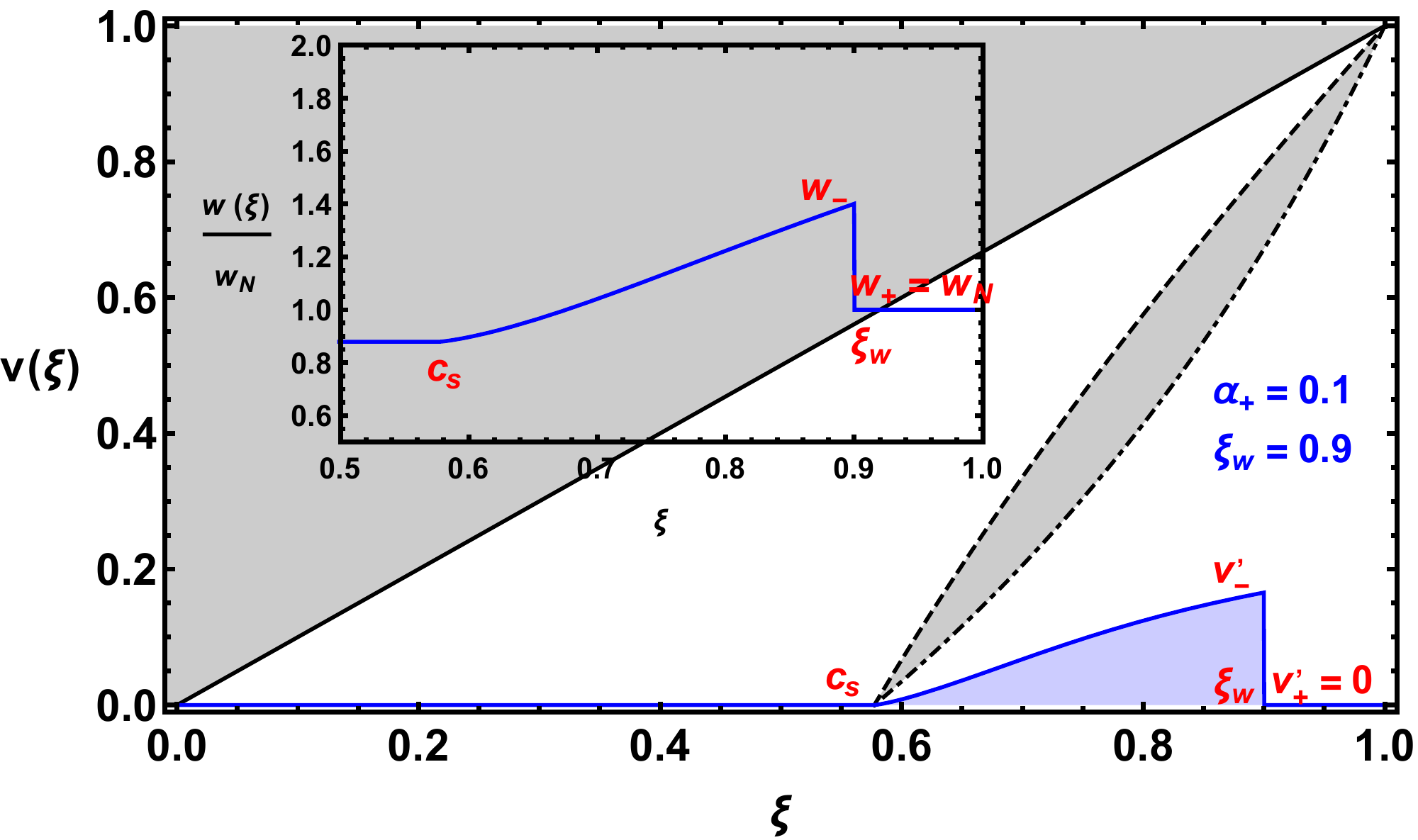}
\includegraphics[width=0.49\textwidth]{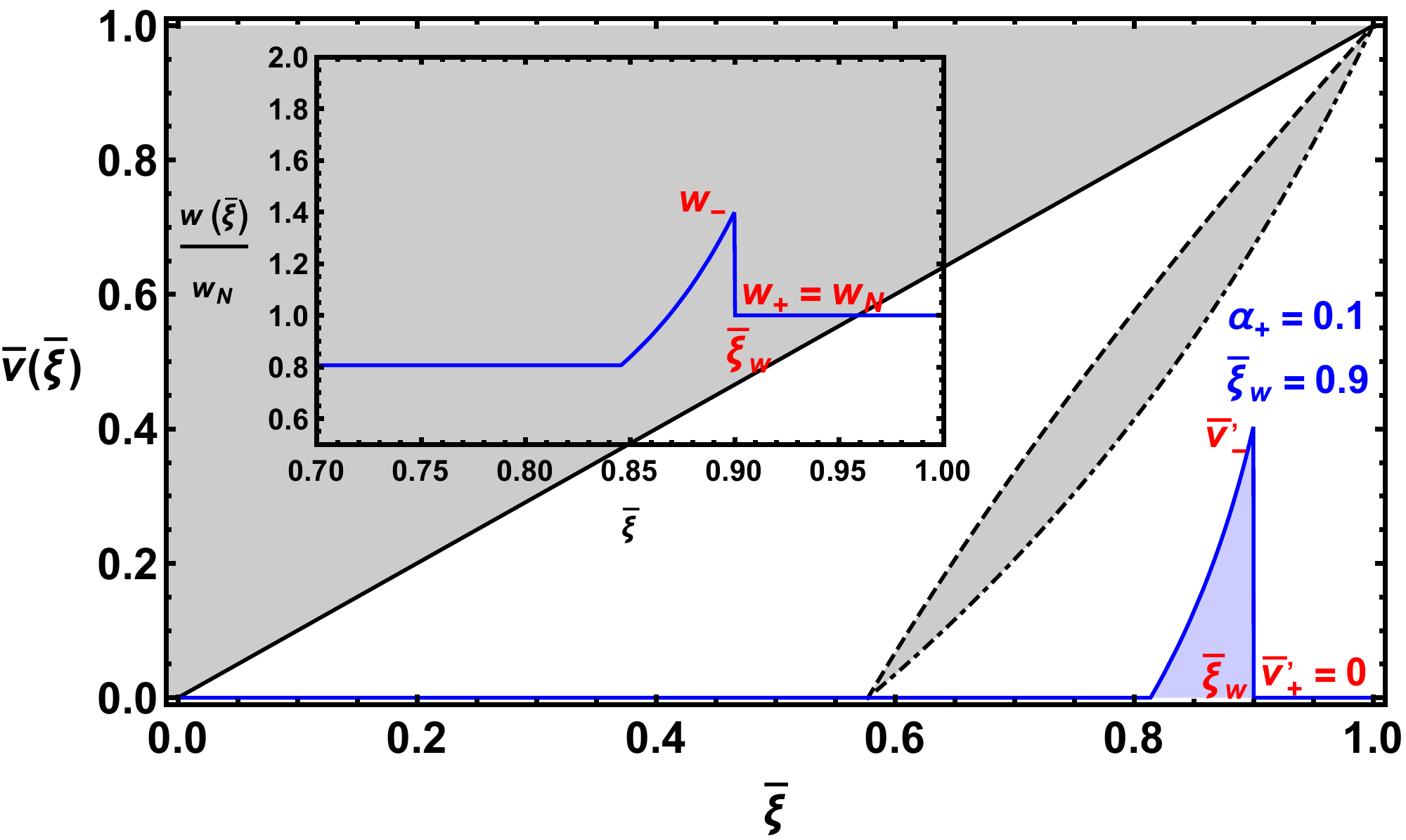}\\
\includegraphics[width=0.49\textwidth]{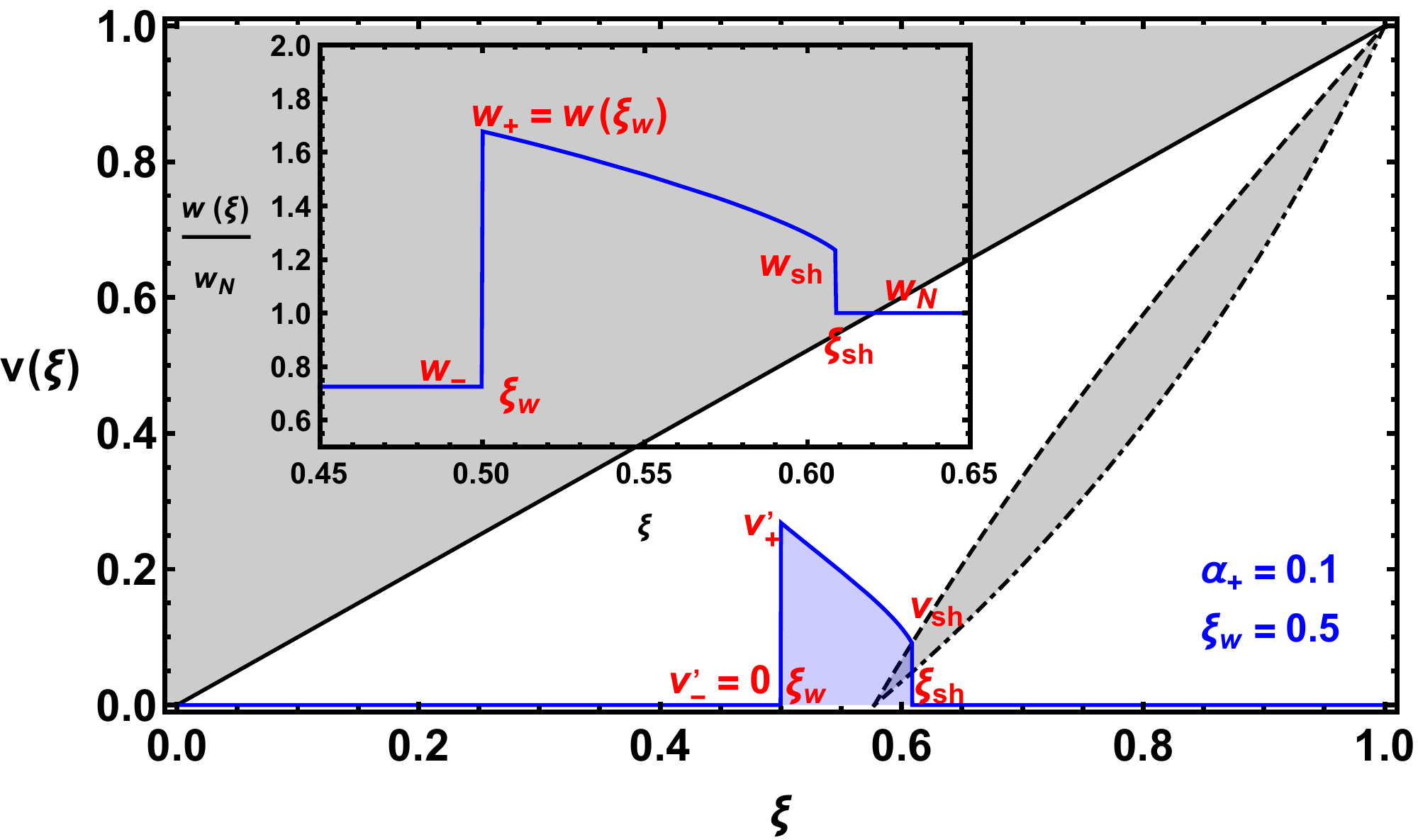}
\includegraphics[width=0.49\textwidth]{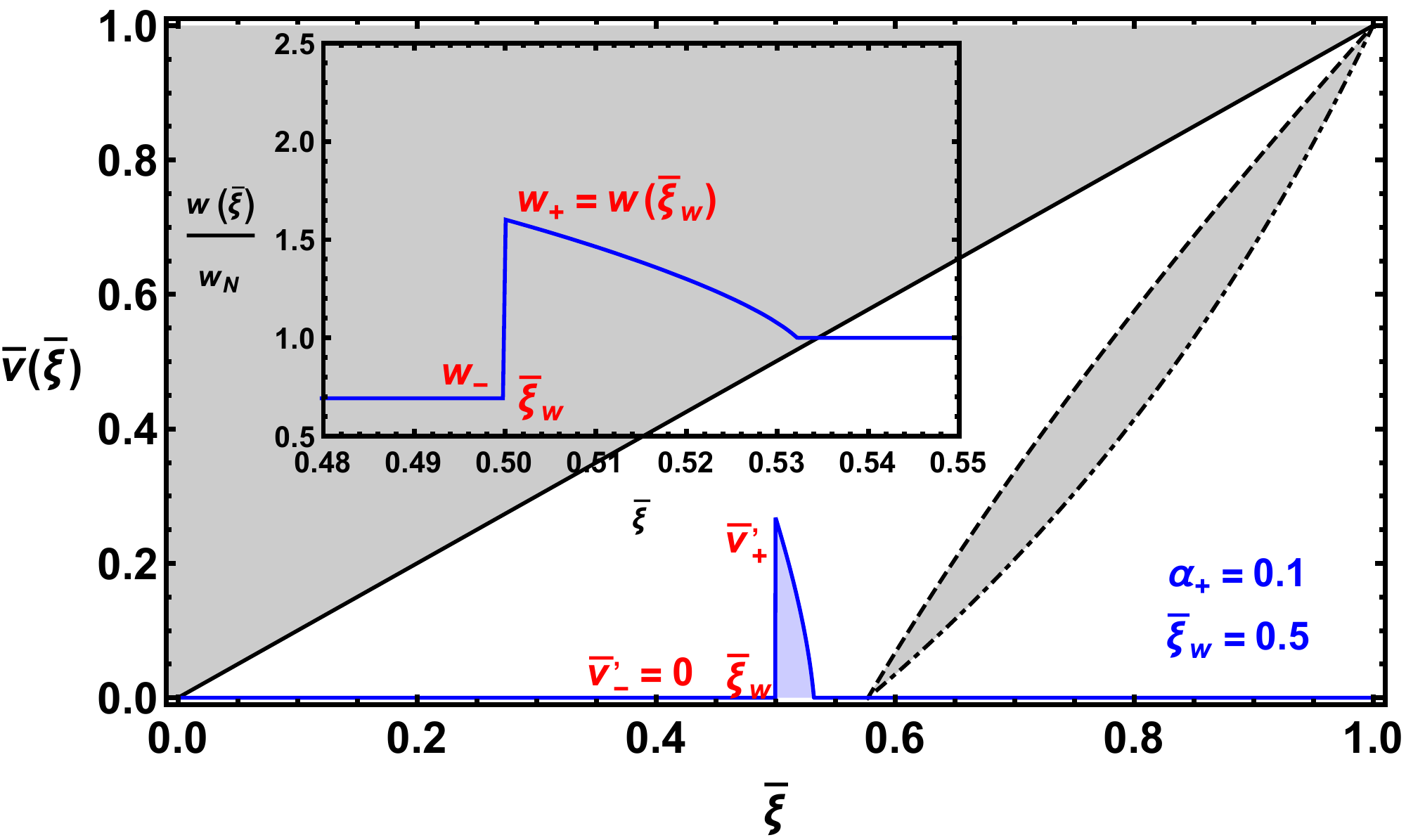}\\
\includegraphics[width=0.49\textwidth]{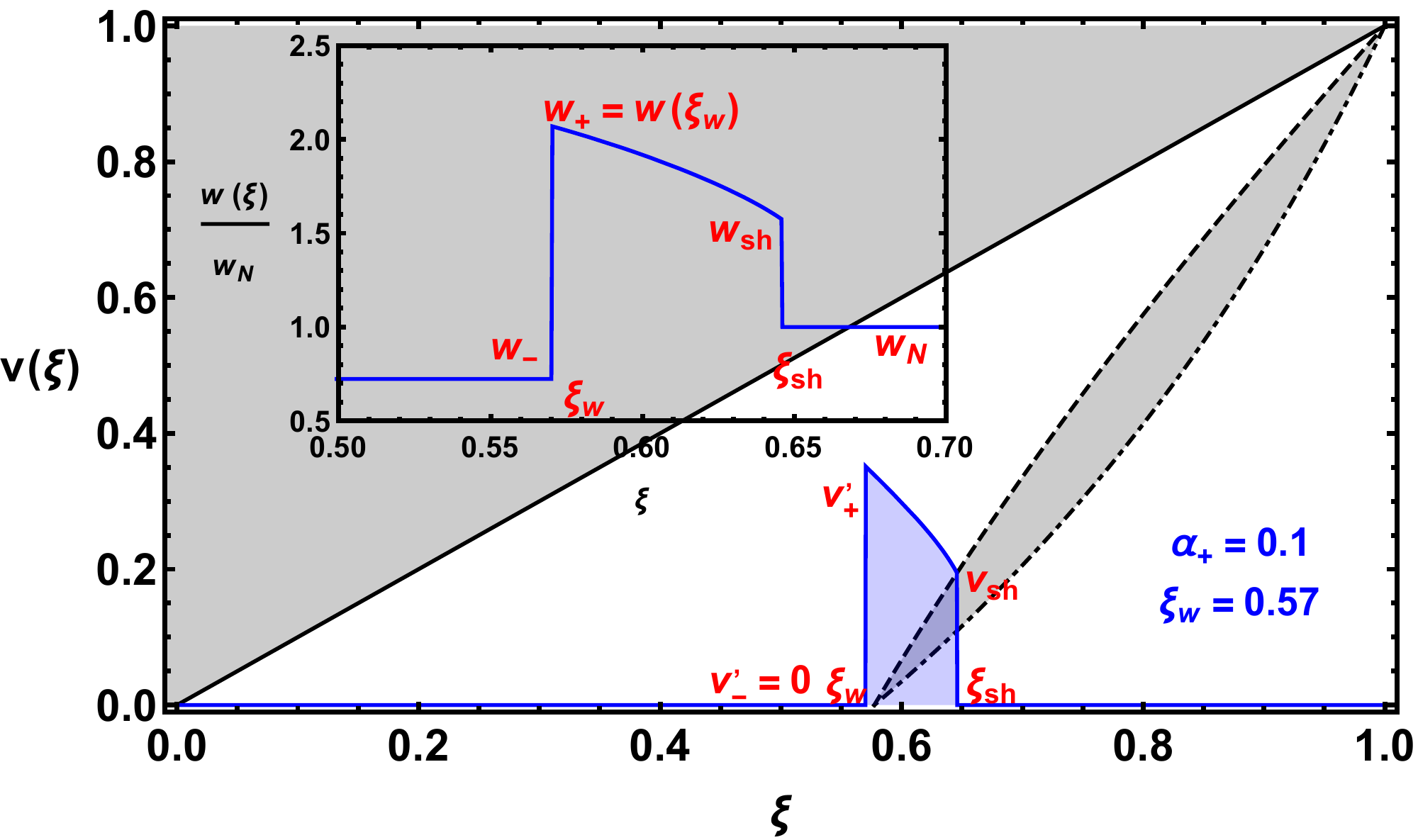}
\includegraphics[width=0.49\textwidth]{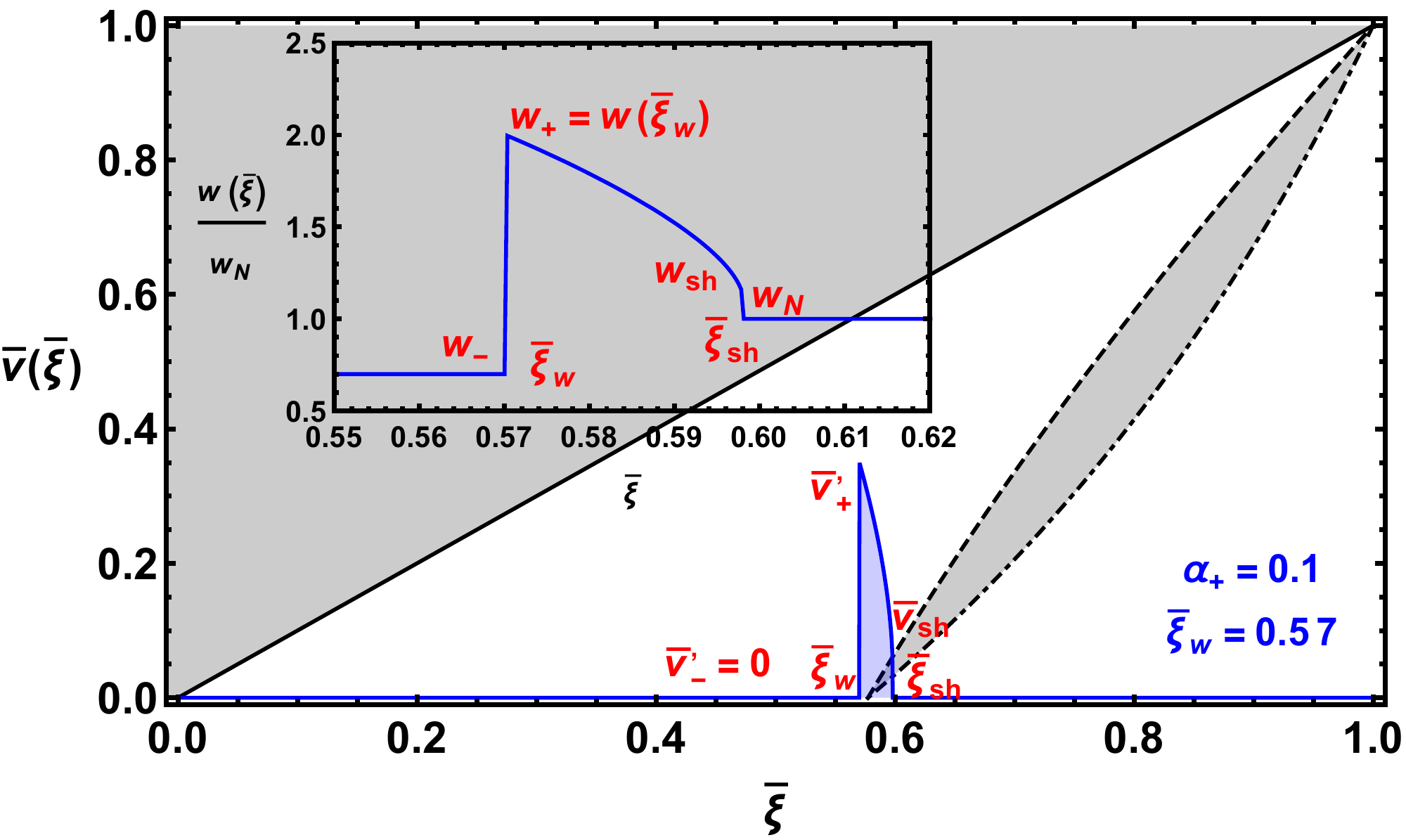}\\
\includegraphics[width=0.49\textwidth]{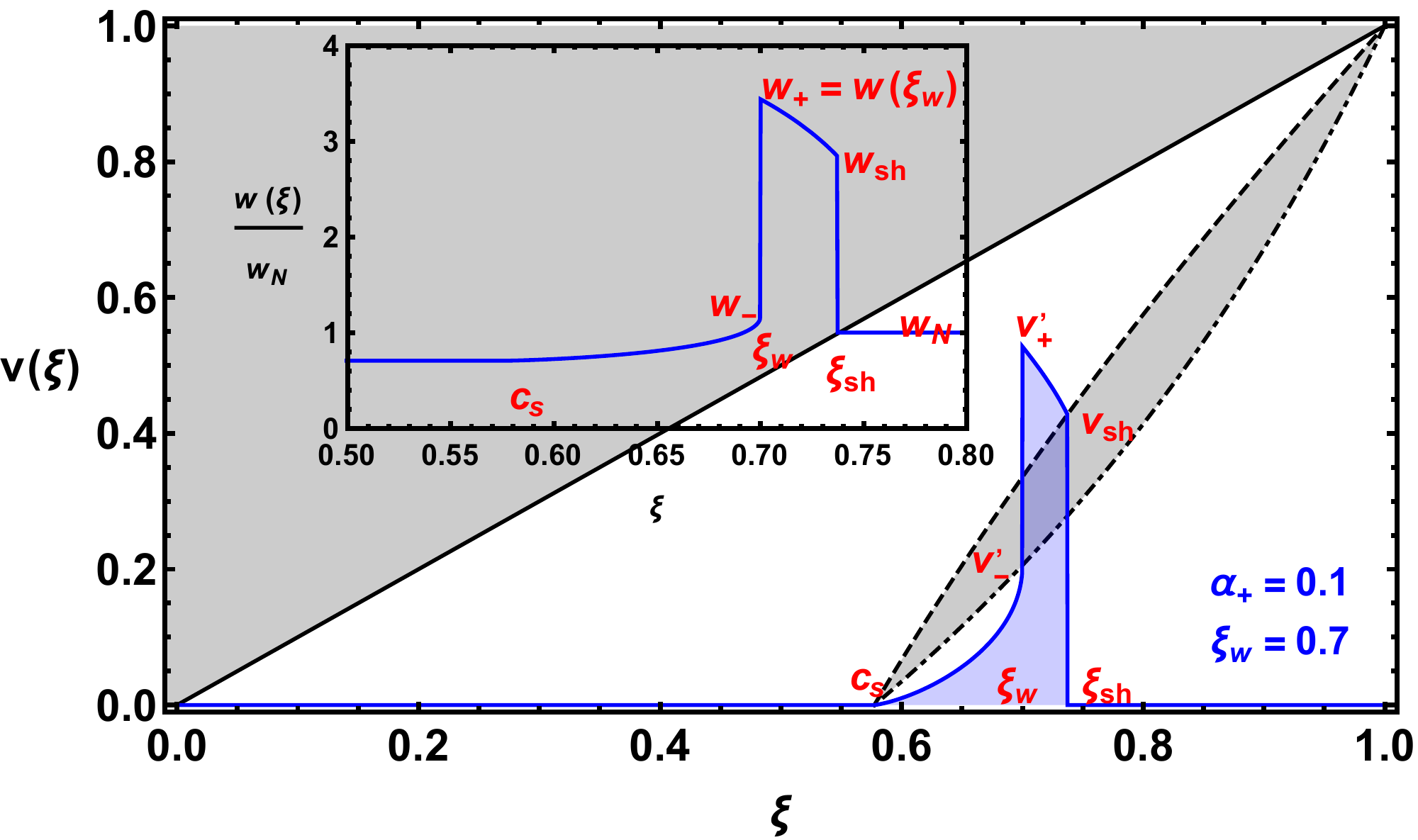}
\includegraphics[width=0.49\textwidth]{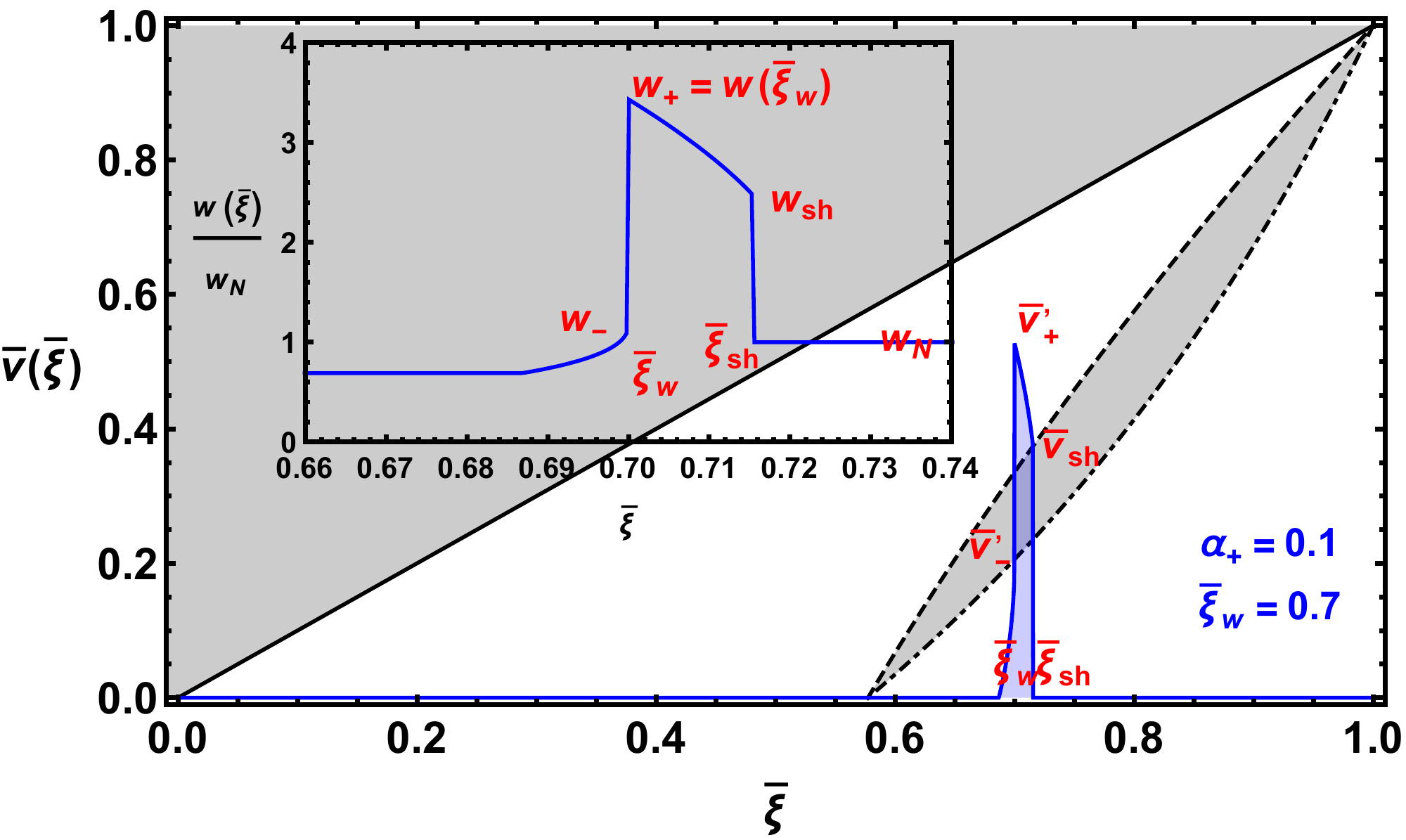}\\
\caption{The velocity profiles $v(\xi)$ (left column) and $\bar{v}(\bar{\xi})$ (right column) of the EOM with input junction conditions from detonation (first line), deflagration (second and third lines) and hybrid (last line) modes in fast (left column) and slow (right column) first-order phase transitions for a given strength factor $\alpha_+$ and a bubble wall velocity $\xi_w$ (left column) and $\bar{\xi}_w$ (right column). The corresponding enthalpy profiles are also presented as small panels. The main difference of velocity profiles is that they are more narrow in the slow first-order phase transition than in the fast first-order phase transition. }\label{fig:profile}
\end{figure}

Solving the EOM \eqref{eq:FLRWEOM} with $n=1$ for slow first-order phase transition under junction conditions \eqref{eq:FLRWjunction} for different modes of bubble expansion, we can acquire the velocity and enthalpy profiles of bulk fluid motion summarized in Fig.\ref{fig:profile}, from which the efficiency factor of energy budget can be obtained in the next section.

\subsection{Detonation}\label{subsec:detona}

Weak detonation wave moves with velocity much larger than the sound speed so that there is no shockwave in the front of the bubble wall, therefore the bulk fluid in the front of this supersonic wall cannot get any warning that there is a wall coming. As a result, the bulk fluid in the front of this supersonic wall is at rest in bubble center frame in comoving coordinate system,
\begin{align}
\bar{v}'_+=\bar{\mu}(\bar{\xi}_w,\bar{v}_+)=0,
\end{align}
which gives us the peculiar velocity of the bubble wall,
\begin{align}
\bar{\xi}_w=\bar{v}_+(\alpha_+,r).
\end{align}
One can also reverse above relation to express $r(\alpha_+,\bar{\xi}_w)$ for given $\alpha_+$ and $\bar{\xi}_w$, thus both $\bar{v}_+$ and $\bar{v}_-$ can be obtained as $\bar{v}_\pm(\alpha_+,r(\alpha_+,\bar{\xi}_w))$ from \eqref{eq:vpvmalphar}. The peculiar velocity of bulk fluid just behind the wall in bubble center frame can be computed by
\begin{align}\label{eq:detonavm}
\bar{v}'_-=\bar{\mu}(\bar{\xi}_w,\bar{v}_-).
\end{align}
To give the velocity profile of rarefaction wave behind the bubble wall, one only needs to find a curve $\bar{v}(\bar{\xi})$ through $(\bar{\xi}_w,\bar{v}(\bar{\xi}_w))$, where $\bar{v}(\bar{\xi}_w))$ is given by $\bar{v}'_-$ from \eqref{eq:detonavm}. The velocity profiles of weak detonation wave are presented in the first line of Fig.\ref{fig:profile} for given strength factor $\alpha_+=0.1$, where the left panel is solved for $v(\xi)$ for fast first-order phase transition in flat background with a bubble wall velocity $\xi_w=0.9$, while the right panel is solved for $\bar{v}(\bar{\xi})$ for slow first-order phase transition in FLRW background with a bubble wall peculiar velocity $\bar{\xi}_w=0.9$.

With velocity profile in hand, one can also obtain the enthalpy profile. Proposing the matching condition
\begin{align}
w_-\bar{v}_-\bar{\gamma}_-^2=w_+\bar{v}_+\bar{\gamma}_+^2
\end{align}
at rarefaction front, namely the bubble wall, with following replacements
\begin{align}
w_-=w(\bar{\xi}_w);\quad \bar{v}_-=\bar{v}_-(\alpha_+,r(\alpha_+,\bar{\xi}_w));\quad w_+=w_N;\quad \bar{v}_+=\bar{\xi}_w,
\end{align}
gives rise to the enthalpy just behind the bubble wall,
\begin{align}
w_-=w_+\frac{\bar{v}_+\bar{\gamma}_+^2}{\bar{v}_-\bar{\gamma}_-^2}=w_N\frac{\bar{\xi}_w}{1-\bar{\xi}_w^2}\frac{1-\bar{v}_-^2}{\bar{v}_-},
\end{align}
where $w_N$ is dubbed as the asymptotic enthalpy far outside the bubble wall. Hence the enthalpy profile can now be obtained by evolving $w(\bar{\xi}_w)$ to $w(\bar{\xi}<\bar{\xi}_w)$ according to
\begin{align}
\frac{w(\bar{\xi})}{w_N}=\frac{\bar{\xi}_w}{1-\bar{\xi}_w^2}\frac{1-\bar{v}_-^2}{\bar{v}_-}
\exp\left[-\int^{\bar{v}(\bar{\xi}_w)}_{\bar{v}(\bar{\xi})}
\bar{\gamma}^2\bar{\mu}\left(\frac{1}{c_s^2}+1\right)\mathrm{d}\bar{v}
+\int^{\bar{\xi}_w}_{\bar{\xi}}\frac{n\bar{v}}{1-\bar{\xi}\bar{v}}\left(\frac{1}{c_s^2}+1\right)\mathrm{d}\bar{\xi}\right].
\end{align}
The enthalpy profiles normalized by the asymptotic enthalpy are also presented as small panels in the first line of Fig.\ref{fig:profile} for the weak detonation wave.

\subsection{Deflagration}\label{subsec:deflag}

Weak deflagration wave moves with velocity smaller than the sound speed so that there forms a compression wave dubbed shockwave in the front of the bubble wall, while in the back of the bubble wall the bulk fluid is at rest in bubble center frame, namely
\begin{align}
\bar{v}'_-=\bar{\mu}(\bar{\xi}_w,\bar{v}_-)\equiv0,
\end{align}
which gives us the peculiar velocity of the bubble wall
\begin{align}
\bar{\xi}_w=\bar{v}_-(\alpha_+,r).
\end{align}
One can also reverse above relation to express $r(\alpha_+,\bar{\xi}_w)$ for given $\alpha_+$ and $\bar{\xi}_w$, thus both $\bar{v}_+$ and $\bar{v}_-$ can be obtained as $\bar{v}_\pm(\alpha_+,r(\alpha_+,\bar{\xi}_w))$ from \eqref{eq:vpvmalphar}. The peculiar velocity of bulk fluid just in the front of the subsonic wall in bubble center frame can be computed by
\begin{align}\label{eq:deflagvp}
\bar{v}'_+=\bar{\mu}(\bar{\xi}_w,\bar{v}_+).
\end{align}
To give the velocity profile of compression wave in the front of the bubble wall, one only needs to find a curve $\bar{v}(\bar{\xi})$ through $(\bar{\xi}_w,\bar{v}(\bar{\xi}_w))$, where $\bar{v}(\bar{\xi}_w)$ is given by $\bar{v}'_+$ from \eqref{eq:deflagvp}. The velocity profile of the compression wave jumps to zero at the shockwave front, in the front of which the bulk fluid peculiar velocity is zero,
\begin{align}
\bar{v}'_+=\bar{\mu}(\bar{\xi}_{sh},\bar{v}_+)\equiv0.
\end{align}
Here the plus and minus signs denote the outside and inside of shockwave front, respectively. The peculiar velocity of shockwave front is thus given by
\begin{align}
\bar{\xi}_{sh}=\bar{v}_+.
\end{align}
At the shockwave front, there is no discontinuity in vacuum energy so that $\alpha_+=0$, which from \eqref{eq:vpandvm} gives rise to $\bar{v}_-\bar{v}_+=1/3$ or equivalently $\bar{v}_-=1/3\bar{\xi}_{sh}$. Therefore, the shockwave front satisfies
\begin{align}
\frac{1}{3\bar{\xi}_{sh}}\equiv\bar{v}_-=\bar{\mu}(\bar{\xi}_{sh},\bar{v}'_-\equiv\bar{v}(\bar{\xi}_{sh})),
\end{align}
namely
\begin{align}\label{eq:shockfront}
\bar{\xi}_{sh}\bar{\mu}(\bar{\xi}_{sh},\bar{v}(\bar{\xi}_{sh}))=c_s^2.
\end{align}
Therefore, the peculiar velocity curve of compression wave through $(\bar{\xi}_w,\bar{v}(\bar{\xi}_w))$ would intersect shockwave front line at a point $(\bar{\xi}_{sh},\bar{v}(\bar{\xi}_{sh}))$ obeying \eqref{eq:shockfront}. As an example, the velocity profiles of deflagration wave are presented in the second and third lines of Fig.\ref{fig:profile} for given strength factor $\alpha_+=0.1$, where the left panels are solved for $v(\xi)$ for fast first-order phase transition in flat background with a bubble wall velocity $\xi_w=0.5$ and $\xi_w=0.57$, while the right panels are solved for $\bar{v}(\bar{\xi})$ for slow first-order phase transition in FLRW background with a bubble wall peculiar velocity $\bar{\xi}_w=0.5$ and $\bar{\xi}_w=0.57$. It is worth noting that for deflagration wave, unlike the non-vanishing bulk fluid velocity at shockwave front for fast first-order phase transition in flat background, the bulk fluid velocity $\bar{v}(\bar{\xi}_{\mathrm{sh}})$ at shockwave front for slow first-order phase transition in FLRW background can be vanished if the bubble wall peculiar velocity is small enough like the right panel in the second line of Fig.\ref{fig:profile}.

The enthalpy profile can also be obtained from the velocity profile. Proposing the matching condition at shockwave front,
\begin{align}
w_-\bar{v}_-\bar{\gamma}_-^2=w_+\bar{v}_+\bar{\gamma}_+^2,
\end{align}
with following replacements
\begin{align}
w_-=w_{sh};\quad \bar{v}_-=\bar{\mu}(\bar{\xi}_{sh},\bar{v}(\bar{\xi}_{sh}));\quad w_+=w_N;\quad \bar{v}_+=\bar{\xi}_{sh},
\end{align}
one has the enthalpy just behind the shockwave front
\begin{align}
w_-=w_+\frac{\bar{v}_+\bar{\gamma}_+^2}{\bar{v}_-\bar{\gamma}_-^2}
=w_N\frac{\bar{\xi}_{sh}}{1-\bar{\xi}_{sh}^2}
\frac{1-\bar{\mu}(\bar{\xi}_{sh},\bar{v}(\bar{\xi}_{sh}))^2}{\bar{\mu}(\bar{\xi}_{sh},\bar{v}(\bar{\xi}_{sh}))}.
\end{align}
Evolving $w_{sh}$ to $w(\bar{\xi}_w<\bar{\xi}<\bar{\xi}_{sh})$ gives rise to the enthalpy profile of form
\begin{align}
\frac{w(\bar{\xi})}{w_N}&=\frac{\bar{\xi}_{sh}}{1-\bar{\xi}_{sh}^2}
\frac{1-\bar{\mu}(\bar{\xi}_{sh},\bar{v}(\bar{\xi}_{sh}))^2}{\bar{\mu}(\bar{\xi}_{sh},\bar{v}(\bar{\xi}_{sh}))}\nonumber\\
&\times\exp\left[-\int_{\bar{v}(\bar{\xi})}^{\bar{v}(\bar{\xi}_{sh})}\left(\frac{1}{c_s^2}+1\right)\bar{\gamma}^2\bar{\mu}\mathrm{d}\bar{v}(\bar{\xi})
+\int_{\bar{\xi}}^{\bar{\xi}_{sh}}\frac{n\bar{v}}{1-\bar{\xi}\bar{v}}\left(\frac{1}{c_s^2}+1\right)\mathrm{d}\bar{\xi}\right].
\end{align}
To see to what value the enthalpy profile jumps at the bubble wall, one propose the matching condition at the bubble wall,
\begin{align}
w_-\bar{v}_-\bar{\gamma}_-^2=w_+\bar{v}_+\bar{\gamma}_+^2,
\end{align}
with following replacements
\begin{align}
\bar{v}_-=\bar{\xi}_w;\quad w_+=w(\bar{\xi}_w);\quad \bar{v}_+=\bar{\mu}(\bar{\xi}_w,\bar{v}_w),
\end{align}
and derives the enthalpy just behind the bubble wall,
\begin{align}
w_-=w_+\frac{\bar{v}_+\bar{\gamma}_+^2}{\bar{v}_-\bar{\gamma}_-^2}
=w(\bar{\xi}_w)\frac{\bar{\mu}(\bar{\xi}_w,\bar{v}_w)}{1-\bar{\mu}(\bar{\xi}_w,\bar{v}_w)^2}\frac{1-\bar{\xi}_w^2}{\bar{\xi}_w},
\end{align}
namely
\begin{align}
\frac{w_-}{w_N}=\frac{w(\bar{\xi}_w)}{w_N}\frac{\bar{\mu}(\bar{\xi}_w,\bar{v}_w)}{1-\bar{\mu}(\bar{\xi}_w,\bar{v}_w)^2}\frac{1-\bar{\xi}_w^2}{\bar{\xi}_w},
\end{align}
where
\begin{align}
\frac{w(\bar{\xi}_w)}{w_N}&=\frac{\bar{\xi}_{sh}}{1-\bar{\xi}_{sh}^2}
\frac{1-\bar{\mu}(\bar{\xi}_{sh},\bar{v}(\bar{\xi}_{sh}))^2}{\bar{\mu}(\bar{\xi}_{sh},\bar{v}(\bar{\xi}_{sh}))}\nonumber\\
&\times\exp\left[-\int_{\bar{v}(\bar{\xi}_w)}^{\bar{v}(\bar{\xi}_{sh})}\left(\frac{1}{c_s^2}+1\right)\bar{\gamma}^2\bar{\mu}\mathrm{d}\bar{v}
+\int_{\bar{\xi}_w}^{\bar{\xi}_{sh}}\frac{n\bar{v}}{1-\bar{\xi}\bar{v}}\left(\frac{1}{c_s^2}+1\right)\mathrm{d}\bar{\xi}\right].
\end{align}
The enthalpy profiles normalized by the asymptotic enthalpy are also presented as small panels in the second and third lines of Fig.\ref{fig:profile} for the weak deflagration wave.

\subsection{Hybrid}\label{subsec:hybrid}

Hybrid wave appears with the presence of both compression shockwave and rarefaction wave in the front and back of the bubble wall when the bubble wall velocity lies between the sound velocity and the so-called Jouguet velocity
\begin{align}
\bar{\xi}_J=\frac{\sqrt{\alpha_+(2+3\alpha_+)}+1}{\sqrt{3}(1+\alpha_+)},
\end{align}
which is determined alone from the Jouguet condition $\bar{v}_-=c_s$ applied to \eqref{eq:vpandvm}. One can also reverse the Jouguet condition $\bar{v}_-(\alpha_+,r)=c_s$ to express $r(\alpha_+)$ and hence the wall frame peculiar velocity just in the front of the bubble wall $\bar{v}_+(\alpha_+,r(\alpha_+))$ for given strength factor $\alpha_+$. As a result, one can derive in bubble center frame the peculiar velocity just in the front of the bubble wall,
\begin{align}
\bar{v}'_+=\bar{\mu}(\bar{\xi}_w,\bar{v}_+),
\end{align}
and the peculiar velocity just behind the bubble wall,
\begin{align}
\bar{v}'_-=\bar{\mu}(\bar{\xi}_w,\bar{v}_-=c_s).
\end{align}
Therefore, the velocity profile of hybrid wave is obtained by finding curves $\bar{v}(\bar{\xi})$ through $(\bar{\xi}_w,\bar{v}'_-)$ and $(\bar{\xi}_w,\bar{v}'_+)$, respectively, where the curve $\bar{v}(\bar{\xi})$ which goes through $(\bar{\xi}_w,\bar{v}'_+)$ intersects with the shockwave front $\bar{\mu}(\bar{\xi},\bar{v})\bar{\xi}=c_s^2$ at $(\bar{\xi}_{sh},\bar{v}(\bar{\xi}_{sh}))$.  As an example, the velocity profiles of hybrid wave are presented in the last line of Fig.\ref{fig:profile} for given strength factor $\alpha_+=0.1$, where the left panel is solved for $v(\xi)$ for fast first-order phase transition in flat background with a bubble wall velocity $\xi_w=0.7$ while the right panel is solved for $\bar{v}(\bar{\xi})$ for slow first-order phase transition in FLRW background with a bubble wall peculiar velocity $\bar{\xi}_w=0.7$.

The corresponding enthalpy profile can also be obtained with the help of velocity profile. Proposing the matching condition at shockwave front,
\begin{align}
w_-\bar{v}_-\bar{\gamma}_-^2=w_+\bar{v}_+\bar{\gamma}_+^2,
\end{align}
with following replacements,
\begin{align}
w_-=w_{sh};\quad \bar{v}_-=\bar{\mu}(\bar{\xi}_{sh},\bar{v}(\bar{\xi}_{sh}));\quad w_+=w_N;\quad \bar{v}_+=\bar{\xi}_{sh},
\end{align}
one obtains the enthalpy just behind the shockwave front,
\begin{align}
w_-=w_+\frac{\bar{v}_+\bar{\gamma}_+^2}{\bar{v}_-\bar{\gamma}_-^2}
=w_N\frac{\bar{\xi}_{sh}}{1-\bar{\xi}_{sh}^2}
\frac{1-\bar{\mu}(\bar{\xi}_{sh},\bar{v}(\bar{\xi}_{sh}))^2}{\bar{\mu}(\bar{\xi}_{sh},\bar{v}(\bar{\xi}_{sh}))}.
\end{align}
Evolving $w_{sh}$ to $w(\bar{\xi}_w<\bar{\xi}<\bar{\xi}_{sh})$ gives rise to the enthalpy profile behind the shockwave front,
\begin{align}
\frac{w(\xi)}{w_N}&=\frac{\bar{\xi}_{sh}}{1-\bar{\xi}_{sh}^2}
\frac{1-\bar{\mu}(\bar{\xi}_{sh},\bar{v}(\bar{\xi}_{sh}))^2}{\bar{\mu}(\bar{\xi}_{sh},\bar{v}(\bar{\xi}_{sh}))}\nonumber\\
&\times\exp\left[-\int_{\bar{v}(\bar{\xi})}^{\bar{v}(\bar{\xi}_{sh})}\left(\frac{1}{c_s^2}+1\right)\bar{\gamma}^2\bar{\mu}\mathrm{d}\bar{v}(\bar{\xi})
+\int_{\bar{\xi}}^{\bar{\xi}_{sh}}\frac{n\bar{v}}{1-\bar{\xi}\bar{v}}\left(\frac{1}{c_s^2}+1\right)\mathrm{d}\bar{\xi}\right].
\end{align}
There is a discontinuity in the enthalpy profile at the bubble wall. Proposing the matching condition at the bubble wall,
\begin{align}
w_-\bar{v}_-\bar{\gamma}_-^2=w_+\bar{v}_+\bar{\gamma}_+^2,
\end{align}
with following replacements,
\begin{align}
\bar{v}_-=c_s;\quad w_+=w(\bar{\xi}_w);\quad \bar{v}_+=\bar{v}_+(\alpha_+,r(\alpha_+,c_s)),
\end{align}
one obtains immediately the enthalpy just behind the bubble wall,
\begin{align}
w_-=w_+\frac{\bar{v}_+\bar{\gamma}_+^2}{\bar{v}_-\bar{\gamma}_-^2}=w(\bar{\xi}_w)\frac{\bar{v}_+}{1-\bar{v}_+^2}\frac{1-c_s^2}{c_s}.
\end{align}
Evolving $w_-$ to $w(\bar{\xi}<\bar{\xi}_w)$ gives rise to the enthalpy profile behind the bubble wall,
\begin{align}
\frac{w(\bar{\xi})}{w_N}&=\frac{w(\bar{\xi}_w)}{w_N}\frac{\bar{v}_+}{1-\bar{v}_+^2}\frac{1-c_s^2}{c_s}\nonumber\\
&\times\exp\left[-\int_{\bar{v}(\bar{\xi})}^{\bar{v}'_-}\left(\frac{1}{c_s^2}+1\right)\bar{\gamma}^2\bar{\mu}\mathrm{d}\bar{v}(\bar{\xi})
+\int_{\bar{\xi}}^{\bar{\xi}_w}\frac{n\bar{v}}{1-\bar{\xi}\bar{v}}\left(\frac{1}{c_s^2}+1\right)\mathrm{d}\bar{\xi}\right]
\end{align}
where
\begin{align}
\frac{w(\bar{\xi}_w)}{w_N}&=\frac{\bar{\xi}_{sh}}{1-\bar{\xi}_{sh}^2}
\frac{1-\bar{\mu}(\bar{\xi}_{sh},\bar{v}(\bar{\xi}_{sh}))^2}{\bar{\mu}(\bar{\xi}_{sh},\bar{v}(\bar{\xi}_{sh}))}\nonumber\\
&\times\exp\left[-\int_{\bar{v}(\bar{\xi}_w)}^{\bar{v}(\bar{\xi}_{sh})}\left(\frac{1}{c_s^2}+1\right)\bar{\gamma}^2\bar{\mu}\mathrm{d}\bar{v}(\bar{\xi})
+\int_{\bar{\xi}_w}^{\bar{\xi}_{sh}}\frac{n\bar{v}}{1-\bar{\xi}\bar{v}}\left(\frac{1}{c_s^2}+1\right)\mathrm{d}\bar{\xi}\right].
\end{align}
The enthalpy profiles normalized by the asymptotic enthalpy are also presented as small panels along with their velocity profiles in the last line of Fig.\ref{fig:profile} for the hybrid wave.

\subsection{Velocity profile}\label{subsec:profile}

\begin{figure}
\centering
\includegraphics[width=0.49\textwidth]{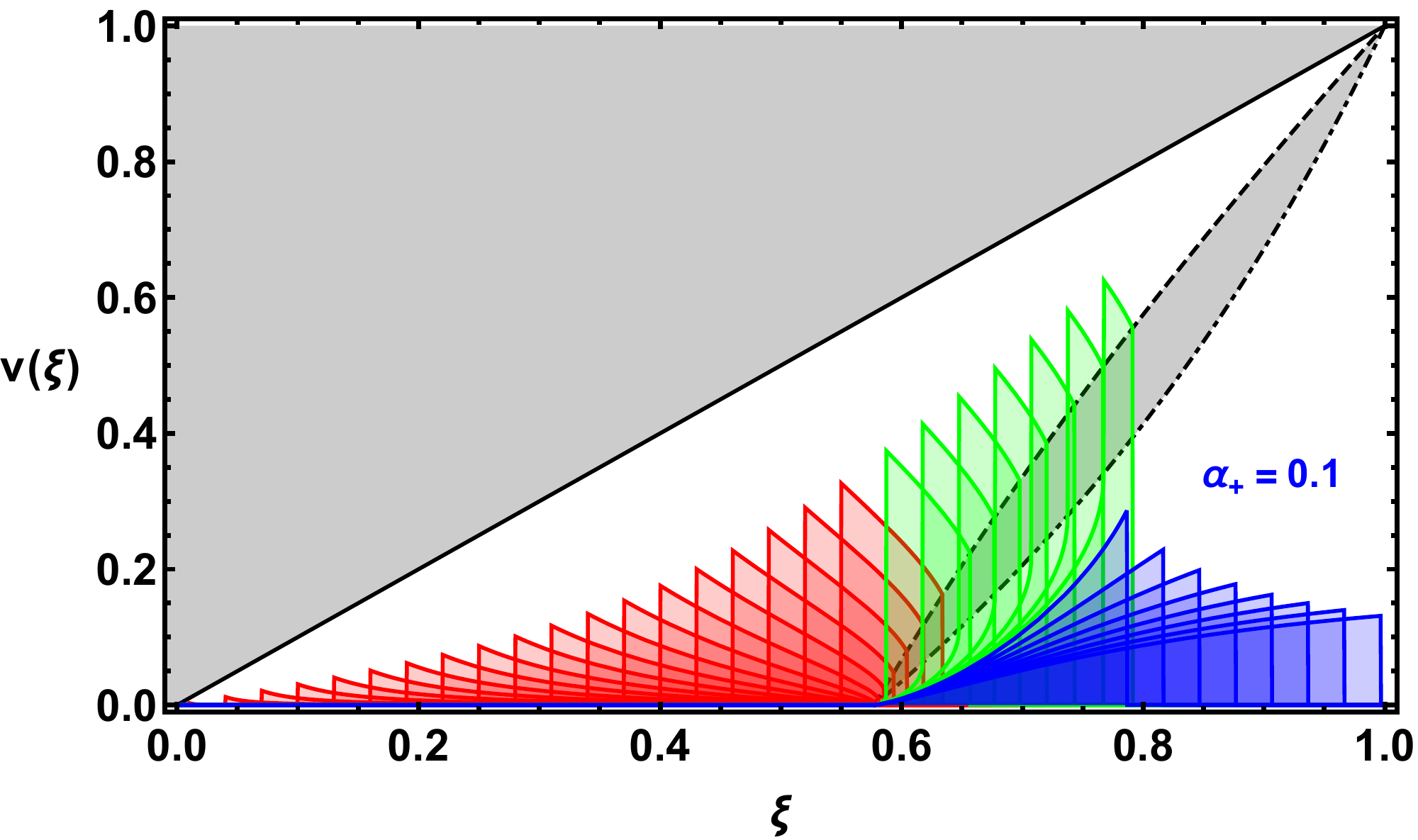}
\includegraphics[width=0.49\textwidth]{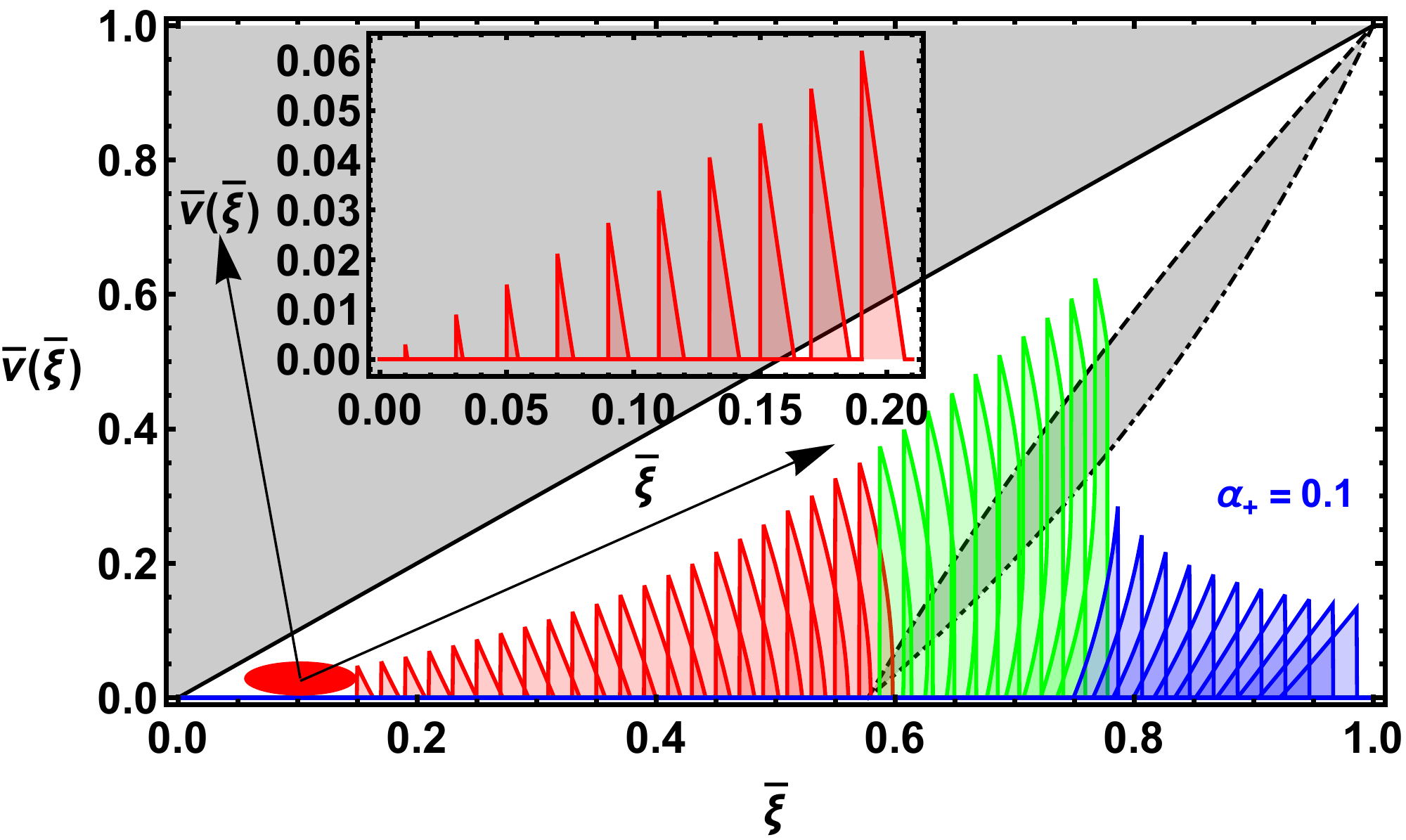}\\
\includegraphics[width=0.49\textwidth]{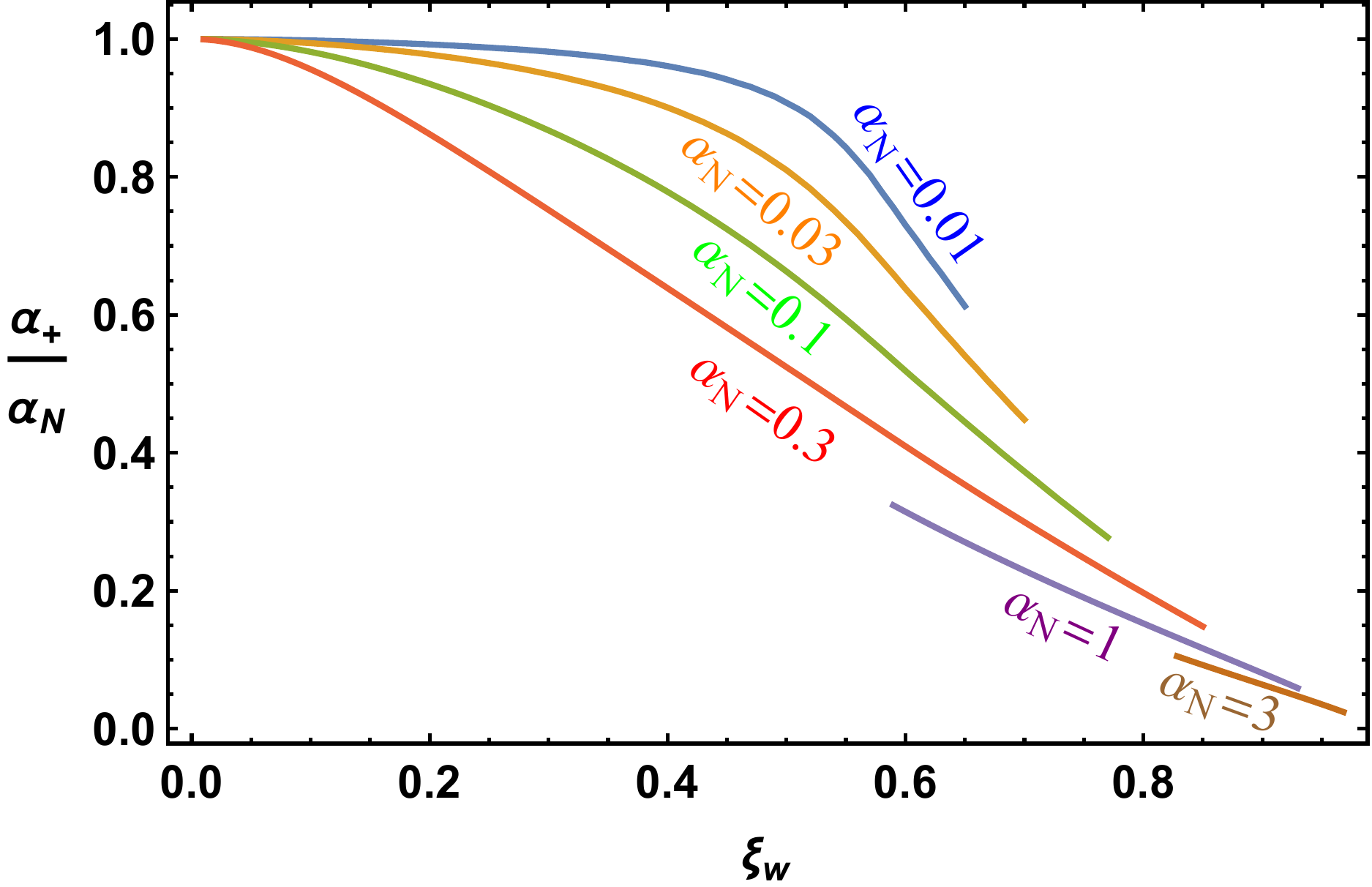}
\includegraphics[width=0.49\textwidth]{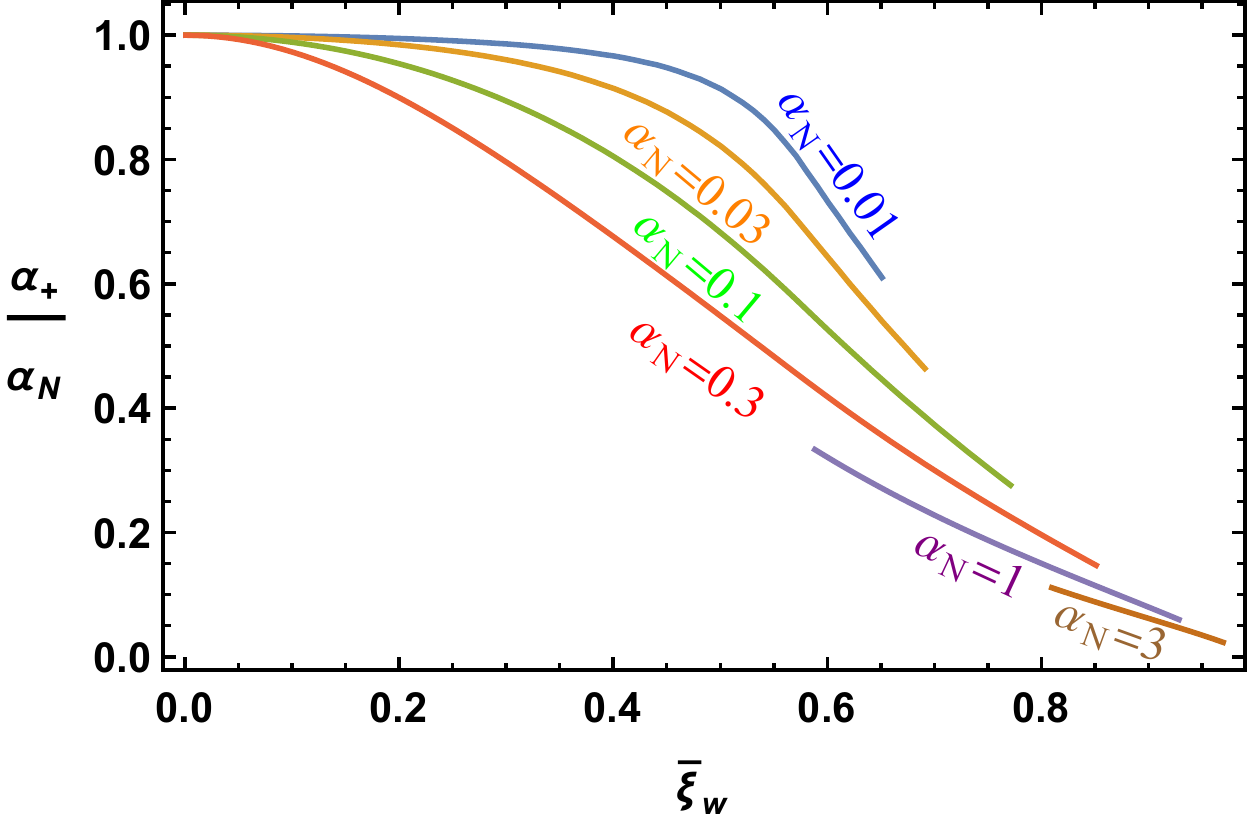}\\
\includegraphics[width=0.49\textwidth]{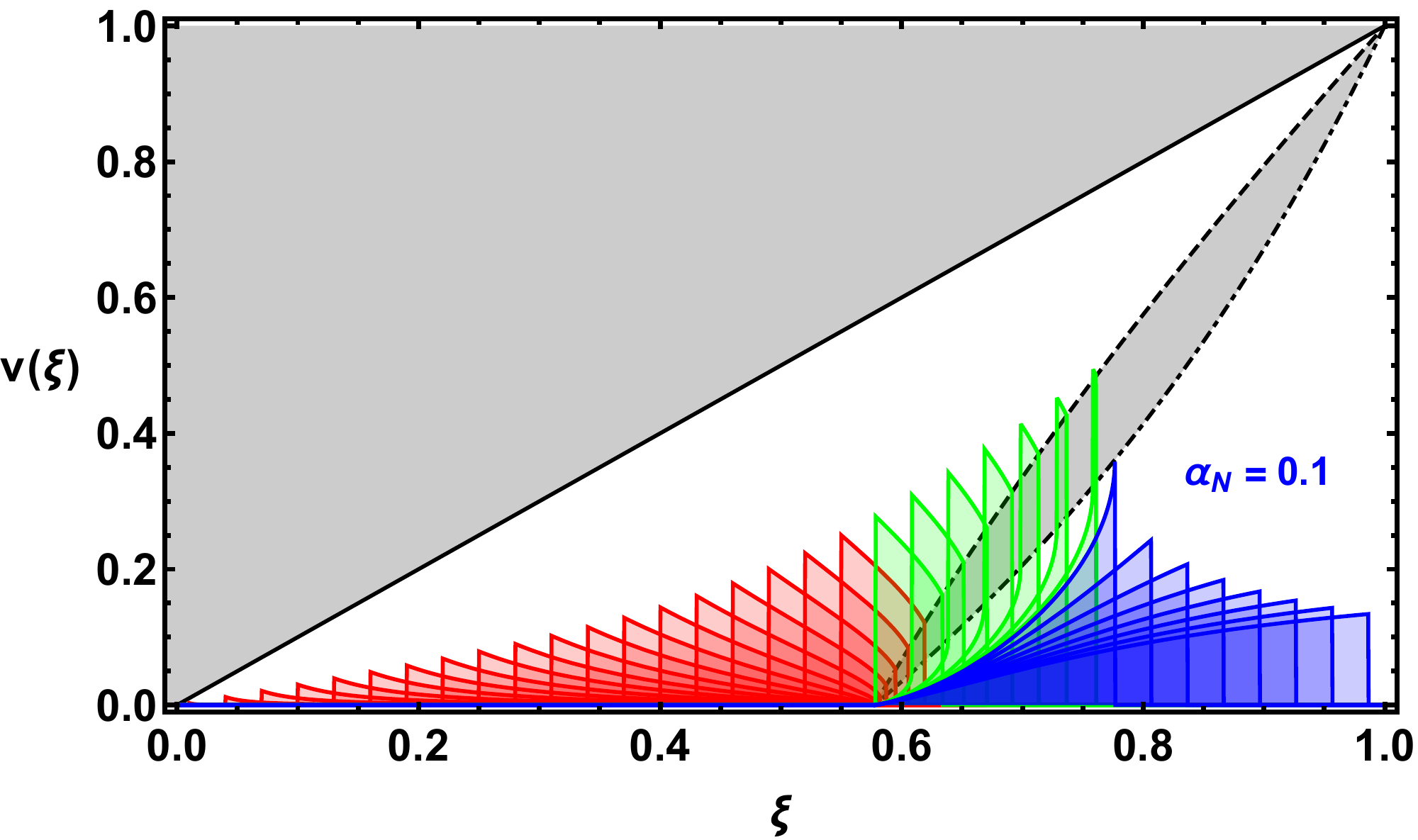}
\includegraphics[width=0.49\textwidth]{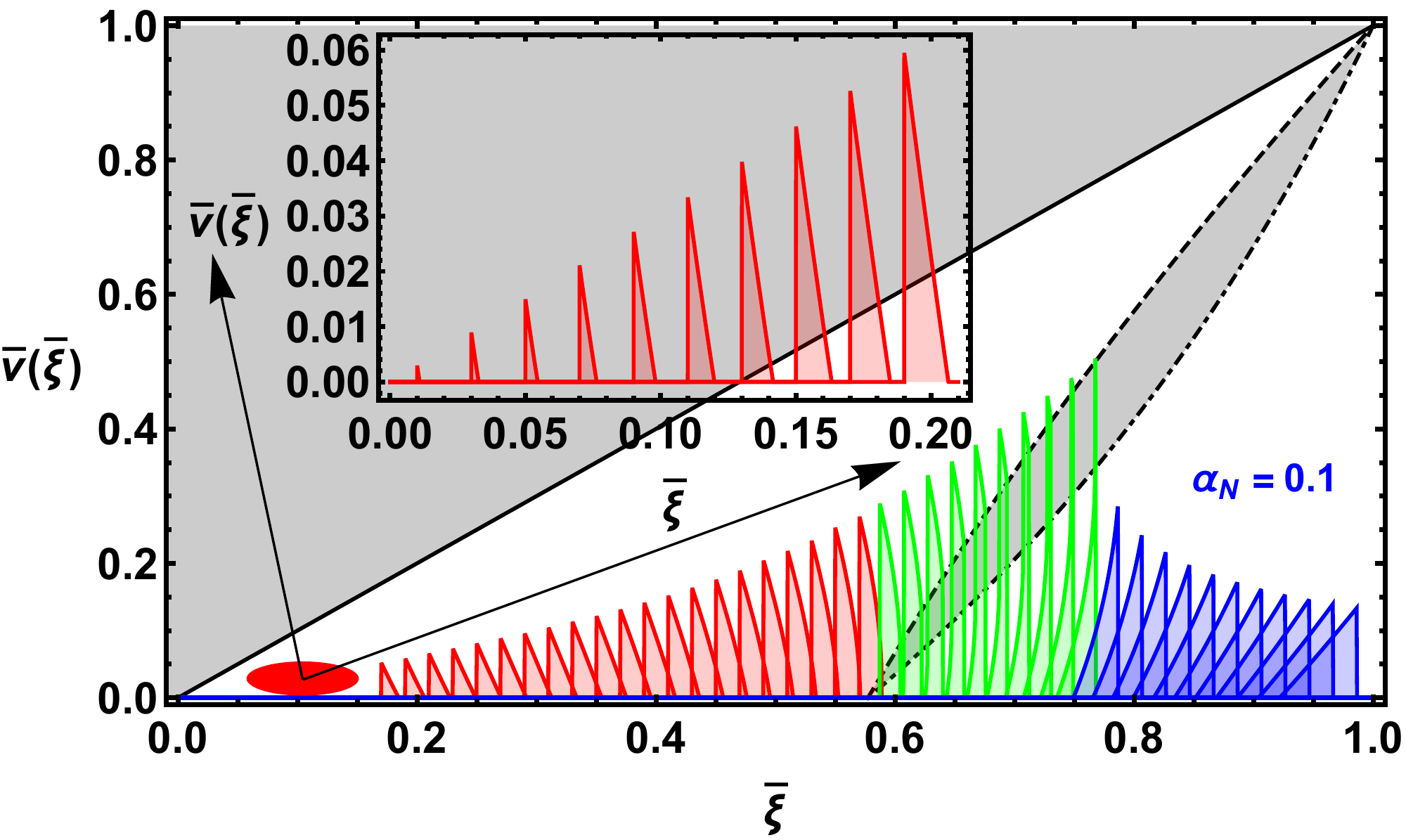}\\
\includegraphics[width=0.49\textwidth]{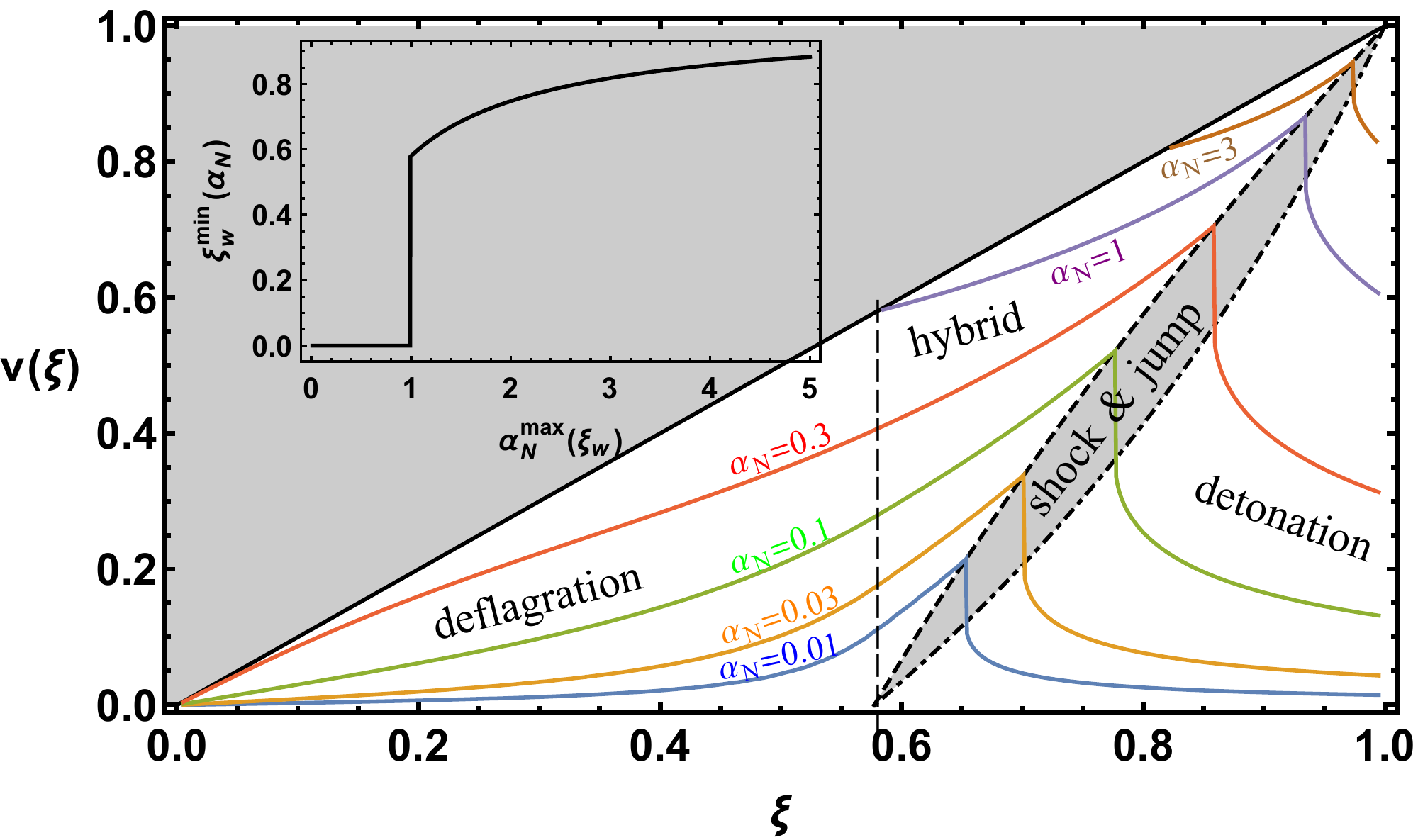}
\includegraphics[width=0.49\textwidth]{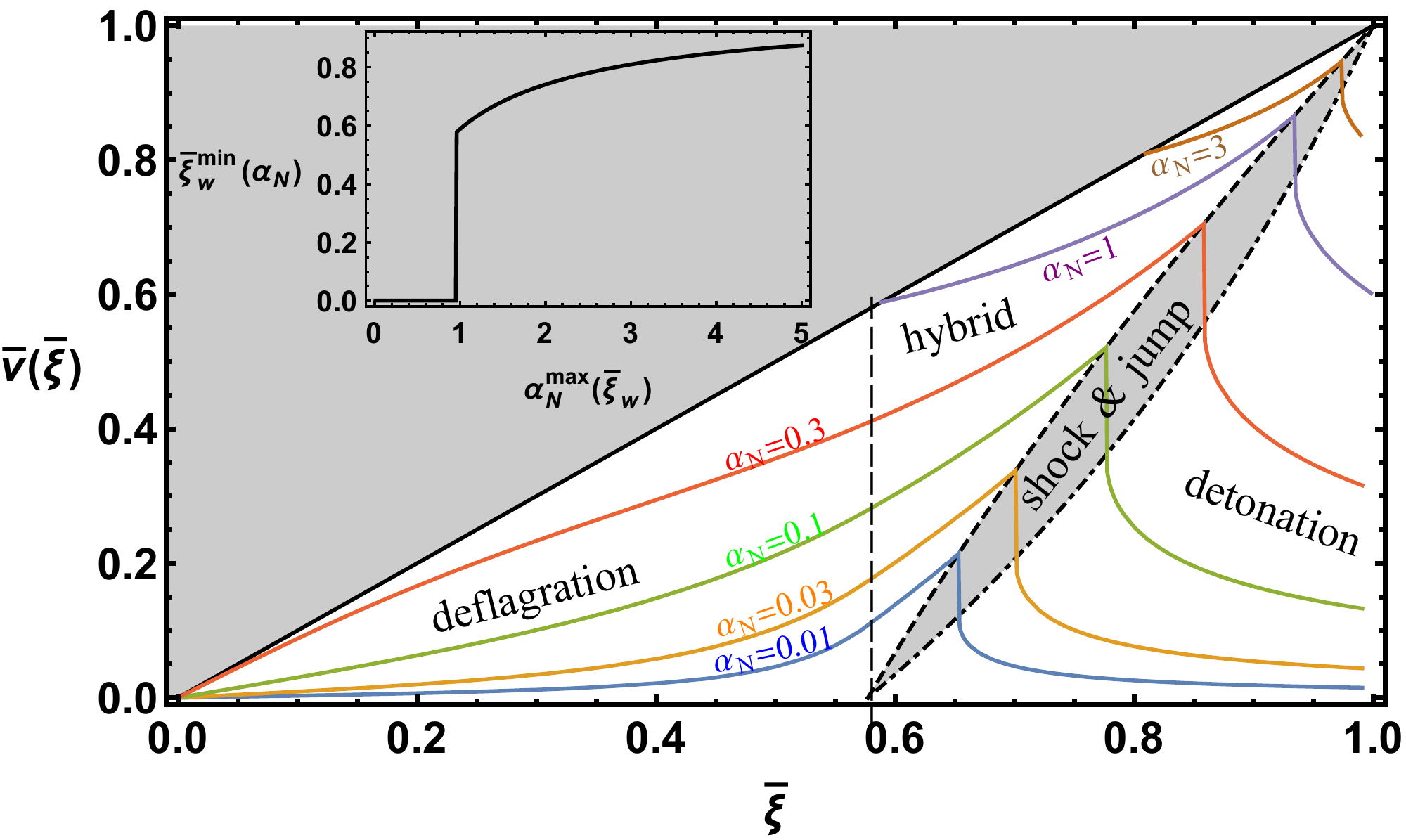}\\
\caption{The velocity profiles are shown in first line for $\alpha_+=0.1$ and different bubble wall velocities. In second line, $\alpha_+$ as a function of the bubble wall velocity are plotted for $\alpha_N=0.01, 0.03, 0.1, 0.3, 1, 3$. The velocity profiles are shown in third line for $\alpha_N=0.1$ and different bubble wall velocities. In last line, the maximal flow velocity in bubble center frame as a function of the bubble wall velocity is shown for $\alpha_N=0.01, 0.03, 0.1, 0.3, 1, 3$. All panels in left and right columns are solved for fast and slow first-order phase transitions in flat and FLRW backgrounds, respectively.}\label{fig:profiles}
\end{figure}

The solutions of velocity profiles in the last section are solved for a given bubble wall peculiar velocity $\bar{\xi}_w$ and the strength factor $\alpha_+$ just in the front of the bubble wall. The velocity profiles for all different modes of bubble expansion with different bubble wall velocities are presented in the first line of Fig.\ref{fig:profiles} with input $\alpha_+=0.1$, where the left and right panels are obtained for fast and slow first-order phase transitions in the flat and FLRW backgrounds, respectively. However, similar with the enthalpy profiles in Fig.\ref{fig:profile}, only in the detonation wave the asymptotic strength factor $\alpha_N$ equals $\alpha_+$. Due to the presence of compression shockwave in the front of the bubble wall, the asymptotic strength factor $\alpha_N$ is in fact unequal with $\alpha_+$ in deflagration and hybrid waves. It would be better to use the asymptotic strength factor $\alpha_N$ instead of the unobservable strength factor $\alpha_+$ hidden inside the compression shockwave.

To express the solutions of velocity profiles in the last section for given $(\bar{\xi}_w,\alpha_N)$ instead of $(\bar{\xi}_w,\alpha_+)$, one notices that there is a simple relation between $\alpha_N$ and $\alpha_+$ from the input profile of enthalpy,
\begin{align}
\frac{\alpha_+}{\alpha_N}=\frac{a_NT_N^4}{a_+T_+^4}=1\left/\frac{w(\bar{\xi}=\bar{\xi}_w^+;\bar{\xi}_w,\alpha_+)}{w_N}\right..
\end{align}
Solving $\alpha_+$ from above equation for given $\alpha_N$, one obtains a function $\alpha_+(\alpha_N)$ for an input bubble wall velocity, which is presented in the second line of Fig.\ref{fig:profiles} with illustrative values of $\alpha_N=0.01, 0.03, 0.1, 0.3, 1, 3$. The results for $\alpha_+/\alpha_N$ as function of asymptotic strength factor and the bubble wall velocity are remained the same for both fast (left) and slow (right) first-order phase transitions in flat (left) and FLRW (right) backgrounds except the overall bar symbols. Hence the velocity profile for given $(\bar{\xi}_w,\alpha_N)$ can be presented as in the third line of Fig.\ref{fig:profiles} with input $\alpha_N=0.1$, where the left and right panels are obtained for fast and slow first-order phase transitions in the flat and FLRW backgrounds, respectively. We also plot the maximal flow velocity in bubble center frame as function of the bubble wall velocity for given asymptotic strength factor $\alpha_N=0.01, 0.03, 0.1, 0.3, 1, 3$. The results for maximal fluid velocity are remained the same for both fast (left) and slow (right) first-order phase transitions in flat (left) and FLRW (right) backgrounds except the overall bar symbols. Last but not the least, As you can see in the last line of Fig.\ref{fig:profiles}, the bubble wall velocity can be zero for $\alpha_N=0.01, 0.03, 0.1, 0.3$, however, for the illustrative values $\alpha_N=1, 3$, there is a minimal value for the bubble wall velocity $\bar{\xi}_w^{\mathrm{min}}$. Equivalently, for a given bubble wall velocity, there is a maximal asymptotic strength factor $\alpha_N^{\mathrm{max}}$ given by
\begin{align}
\alpha_N^{\mathrm{max}}=\alpha_+^{\mathrm{max}}\frac{w(\bar{\xi}=\bar{\xi}^+_w;\bar{\xi}_w,\alpha_+^{\mathrm{max}})}{w_N}
\end{align}
where the maximum value $\alpha_+^{\mathrm{max}}=1/3$ of the strength factor just in the front of the bubble wall for deflagration and hybrid waves can be inferred from the Fig.\ref{fig:WallVelocity}. The maximal asymptotic strength factor $\alpha_N^{\mathrm{max}}$ as a function of the bubble wall velocity $\bar{\xi}_w^{\mathrm{min}}$ is shown as the small panels in the last line of Fig.\ref{fig:profiles}, which are also remained the same for both fast (left) and slow (right) first-order phase transitions in flat (left) and FLRW (right) backgrounds except the overall bar symbols.

\begin{figure}
\centering
\includegraphics[width=0.8\textwidth]{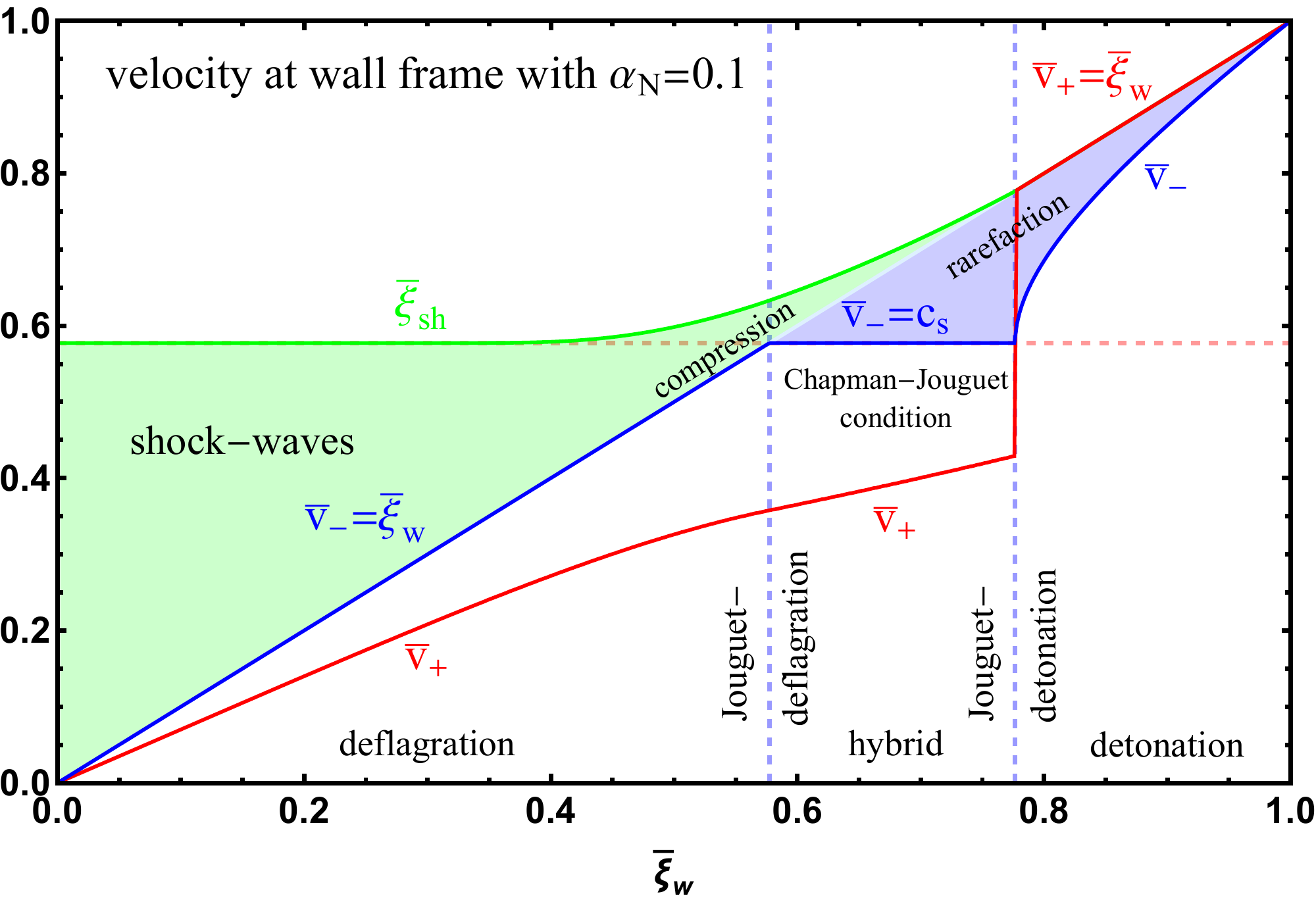}\\
\caption{The fluid peculiar velocities in the bubble wall frame with respect to the bubble wall peculiar velocity for given $\alpha_N=0.1$. The fluid peculiar velocities $\bar{v}_+$ and $\bar{v}_-$ just in the front (red solid line) and back (blue solid line) of the bubble wall are shown along with the rarefaction wave (blue shaded region) and compression shockwave (green shade region) proceeded with shock front peculiar velocity $\bar{\xi}_{\mathrm{sh}}$ (green solid line). }\label{fig:WallVelocities}
\end{figure}

The physical picture of bubble expansion is summarised in Fig.\ref{fig:WallVelocities}, where the fluid peculiar velocities $\bar{v}_+$ and $\bar{v}_-$ just in the front (red solid line) and back (blue solid line) of the bubble wall are shown with respect to the bubble wall peculiar velocity for given $\alpha_N=0.1$. The blue and red shaded region are the rarefaction wave and compression shockwave proceeded with shock front peculiar velocity $\bar{\xi}_{\mathrm{sh}}$ indicated as green solid line. For a subsonic bubble wall peculiar velocity, the bubble expansion proceeds with deflagration wave with compression shockwave in the front of the bubble wall. When the bubble wall peculiar velocity exceeds the sound velocity, there develops rarefaction wave behind the bubble wall and hence forms hybrid wave. With an increasing bubble wall peculiar velocity, the compression shockwave would become narrower and narrower until eventually vanish when the bubble wall peculiar velocity reaches Jouguet velocity. For a bubble wall peculiar velocity larger than the Jouguet velocity, the formed detonation wave would leave only the rarefaction wave behind the bubble wall.

\section{Efficiency factor}\label{sec:efficiency}

In this section, we will calculate the efficiency factor for slow first-order phase transition in FLRW background, which will be compared with that for fast first-order phase transition in flat background. The numerical fitting formulas will also be given for convenient use in future literatures without going to the details of hydrodynamics.

\begin{figure}
\centering
\includegraphics[width=0.70\textwidth]{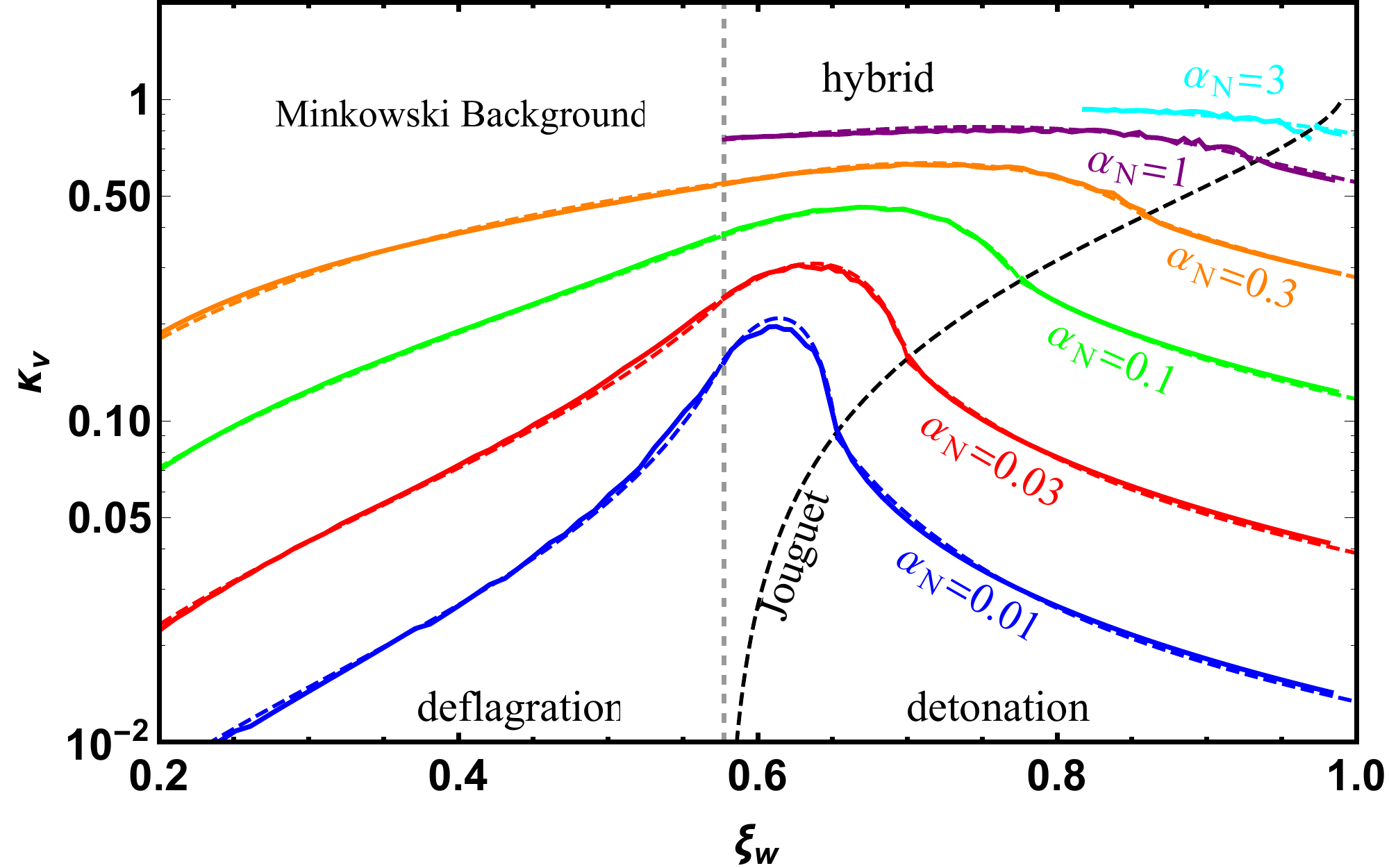}\\
\includegraphics[width=0.70\textwidth]{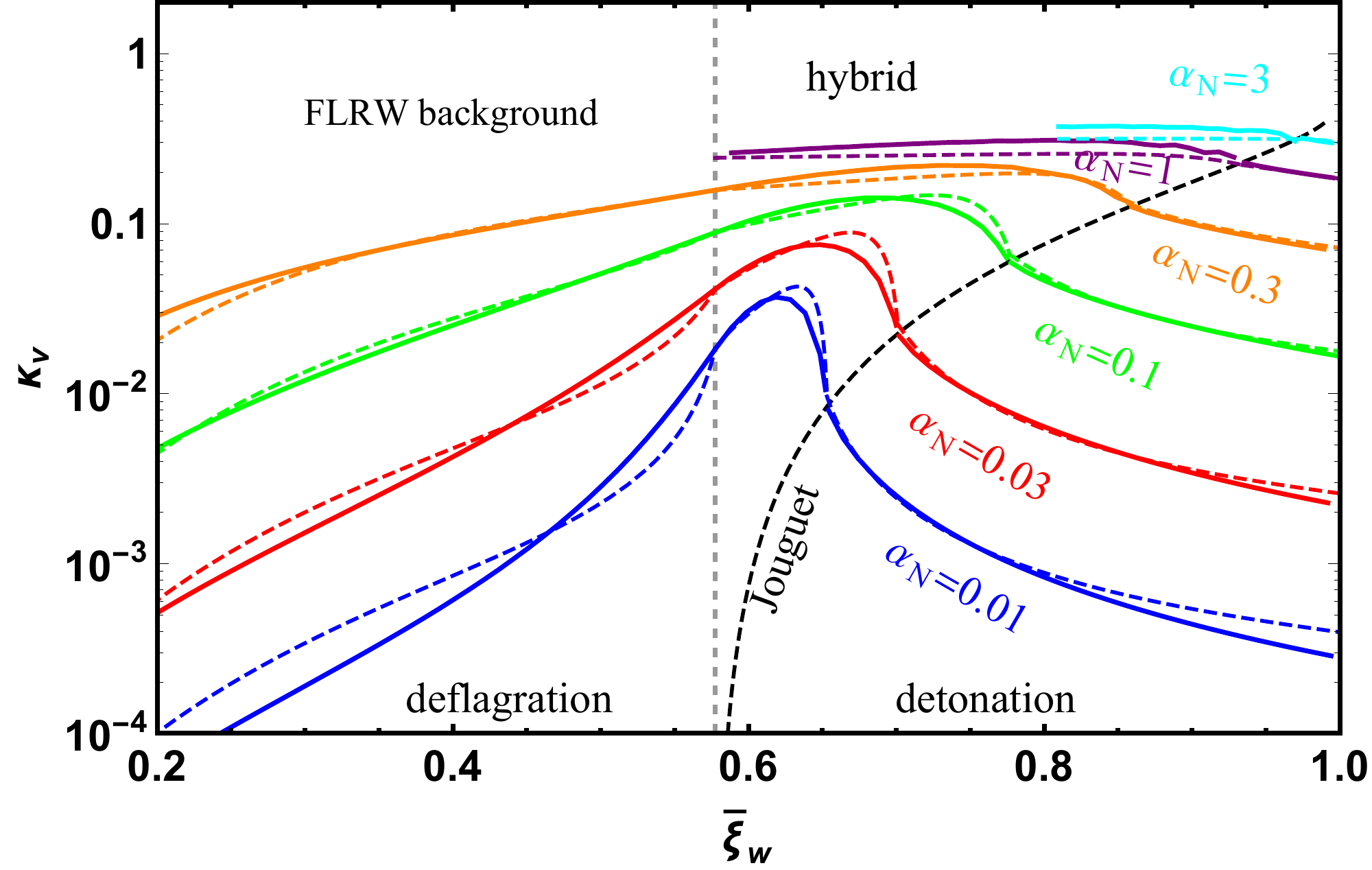}\\
\includegraphics[width=0.70\textwidth]{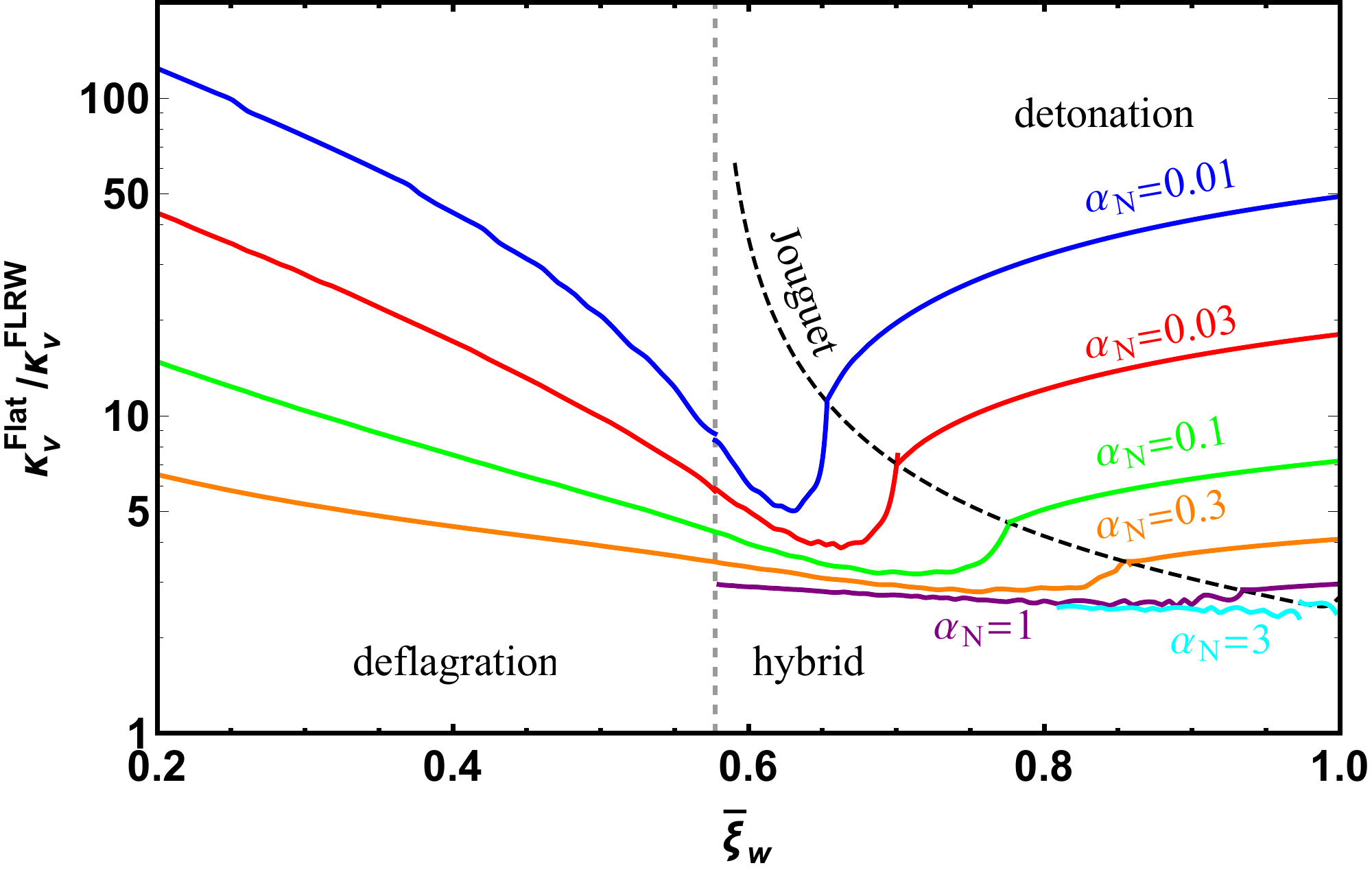}\\
\caption{The efficiency factors with respect to the bubble wall (peculiar) velocity for given asymptotic strength factors for fast (top) and slow (middle) first-order phase transitions and their ratio (bottom).}\label{fig:efficiency}
\end{figure}

\subsection{Analytic results}\label{subsec:analytic}

Before discussing the efficiency factor for fast and slow first-order phase transitions in flat and FLRW backgrounds, one needs to clarify the roles played by the physical velocity and peculiar velocity. The physical coordinate $r$ is related to the comoving coordinate $\bar{r}$ by $r=a\bar{r}$, therefore the physical velocity $v=\mathrm{d}r/\mathrm{d}t$ would receive an extra contribution in addition to the peculiar velocity $\bar{v}=\mathrm{d}\bar{r}/\mathrm{d}\bar{t}$,
\begin{align}
v=a\frac{\mathrm{d}\bar{r}}{\mathrm{d}t}+\bar{r}\frac{\mathrm{d}a}{\mathrm{d}t}
  =\frac{\mathrm{d}\bar{r}}{\mathrm{d}\bar{t}}+\bar{r}a\frac{\mathrm{d}a/\mathrm{d}t}{a},
\end{align}
namely
\begin{align}
v=\bar{v}+\bar{\xi}\,\bar{t}\frac{\mathrm{d}a/\mathrm{d}\bar{t}}{a}=\bar{v}+\bar{\xi}n.
\end{align}
Therefore, the bubble wall could move with superluminal velocity with respect to the bubble center from the view point of physical velocity. However, when two bubbles collide, the extra term $\bar{\xi}n$ would be the same at the colliding point for both bubbles, therefore the relative physical velocity is exactly the relative peculiar velocity. As a result, we will compare the efficiency factor for slow first-order phase transition in FLRW background with respect to that for fast first-order phase transition in flat background for given peculiar velocity.

To define the kinetic energy of bulk fluid, one first uses $\bar{\gamma}^2-1=\bar{v}^2\bar{\gamma}^2$ to split the total energy density into two parts,
\begin{align}
T^{00}(\bar{v})=\frac{1}{a^2}\left((e+p)\bar{\gamma}^2-p\right)
=\frac{1}{a^2}\left(e\bar{\gamma}^2+p\bar{v}^2\bar{\gamma}^2\right)=\frac{1}{a^2}\left(e+w\bar{v}^2\bar{\gamma}^2\right),
\end{align}
where first term in bracket is independent of peculiar velocity, while the second term in bracket is dependent of peculiar velocity.
The bulk fluid kinetic energy due to bubble expansion is therefore defined by
\begin{align}
e_{\bar{v}}=a^2\left[T^{00}(\bar{v})-T^{00}(\bar{v}=0)\right]=w\bar{v}^2\bar{\gamma}^2,
\end{align}
where the scale factor is introduced to eliminate the effect from Hubble expansion, since the fluid element is still comoving with background even without disturbance from bubble expansion. The efficiency factor $\kappa_{\bar{v}}$ is usually defined as the ratio of the integrated kinetic energy of bulk fluid over a sphere region slightly larger than the bubble (so that the shockwave front, if exist, could be included) with respect to the total released vacuum energy over a bubble
\begin{align}
\kappa_{\bar{v}}\frac{4\pi}{3}\bar{r}_w^3a^3\Delta\epsilon=\int w(\bar{r})\bar{v}^2(\bar{r})\bar{\gamma}^2(\bar{v}(\bar{r}))4\pi\bar{r}^2a^3\mathrm{d}\bar{r},
\end{align}
namely
\begin{align}\label{eq:efficiency}
\kappa_{\bar{v}}=\frac{3}{\bar{\xi}_w^3\Delta\epsilon}\int w(\bar{\xi})\bar{v}^2\bar{\gamma}^2\bar{\xi}^2\mathrm{d}\bar{\xi}
=\frac{4}{\bar{\xi}_w^3\alpha_N}\int\frac{w(\bar{\xi})}{w_N}\bar{v}^2(\bar{\xi})\bar{\gamma}^2(\bar{v}(\bar{\xi}))\bar{\xi}^2\mathrm{d}\bar{\xi}.
\end{align}

For a given bubble wall peculiar velocity and asymptotic strength factor, the efficiency factor can be numerically calculated from \eqref{eq:efficiency}, and the results are shown as solid lines in Fig.\ref{fig:efficiency} for some illustrative values of asymptotic strength factor $\alpha_N=0.01, 0.03, 0.1, 0.3, 1, 3$. The top panel is obtained for fast first-order phase transition in flat background, while the middle panel is obtained for slow first-order phase transition in FLRW background in radiation-dominated era. The bottom panel gives the ratio of efficiency factors for fast first-order phase transition in flat background with respect to that for slow first-order phase transition in FLRW background, which manifests a reduction of efficiency factor for slow first-order phase transition in FLRW background compared with that for fast first-order phase transition in flat background. The reduction could be large for smaller asymptotic strength factor and for either non-relativistic or ultra-relativistic peculiar velocity of the bubble wall. For sufficiently large asymptotic strength factor, there seems to exist a minimal reduction between a factor between $2$ and $3$ regardless of the bubble wall peculiar velocity. One way to understand this reduction is that, the velocity profiles for slow first-order phase transition in FLRW background is more narrow than those for fast first-order phase transition in flat background, therefore there is less energy dissipation into the kinetic energy of bulk fluid. The kinetic energy of bulk fluid would then be mostly converted into the GWs energy via sound waves \footnote{It has been argued in \cite{Caprini:2015zlo} that only at most $5-10\%$ of bulk fluid motion is converted into MHD turbulence.}, however, the reduction of efficiency factor indicates that, the contributions from sound waves might not be as large as one previously expected in the literatures \cite{Hindmarsh:2013xza,Hindmarsh:2015qta,Hindmarsh:2017gnf}. Nevertheless, these numerical simulations \cite{Hindmarsh:2013xza,Hindmarsh:2015qta,Hindmarsh:2017gnf} are implemented in flat background. To be accordance with the observations made in the current paper for bubble expansion, we will explore in future the effects from Hubble expansion on bubble percolation in more realistic numerical simulations.

\subsection{Numerical fittings}\label{subsec:fitting}

\begin{figure}
\centering
\includegraphics[width=0.49\textwidth]{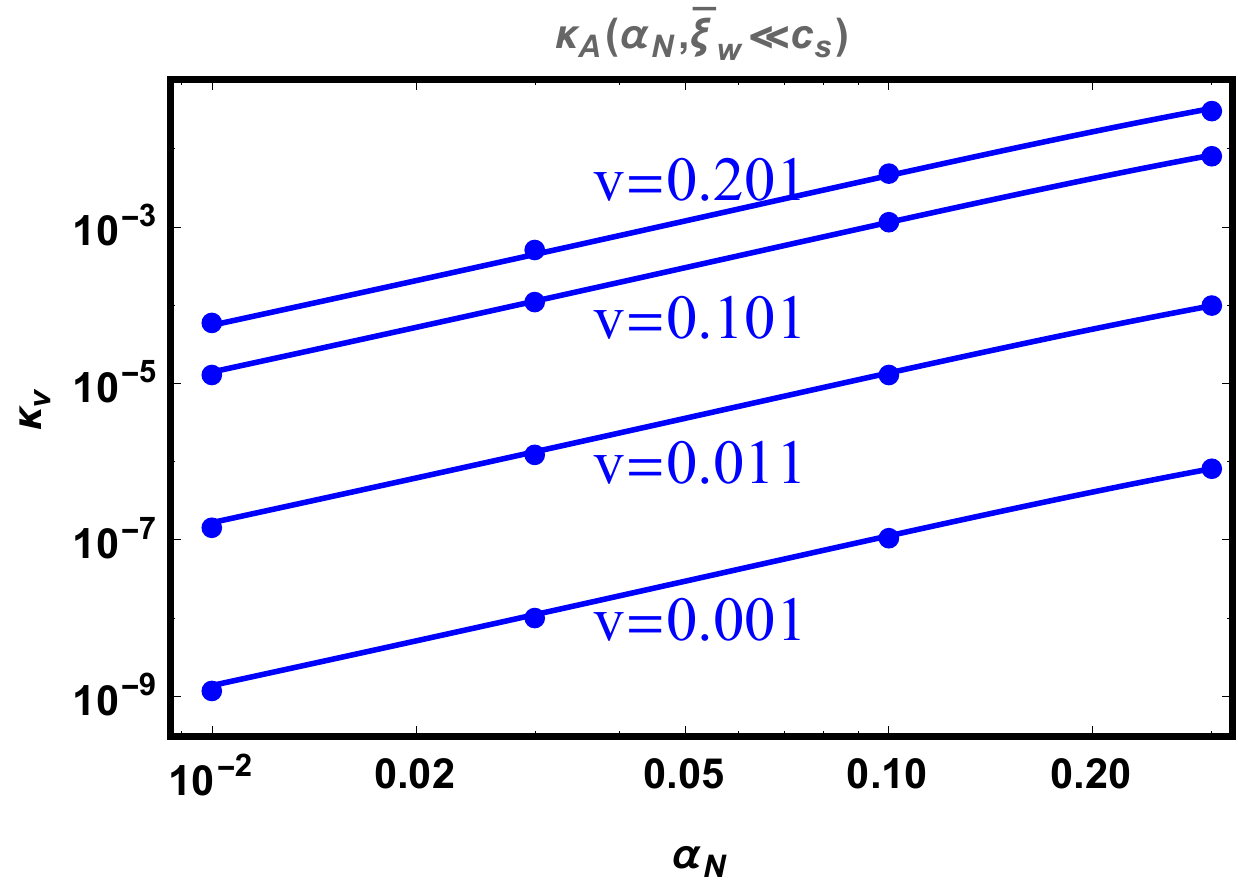}
\includegraphics[width=0.49\textwidth]{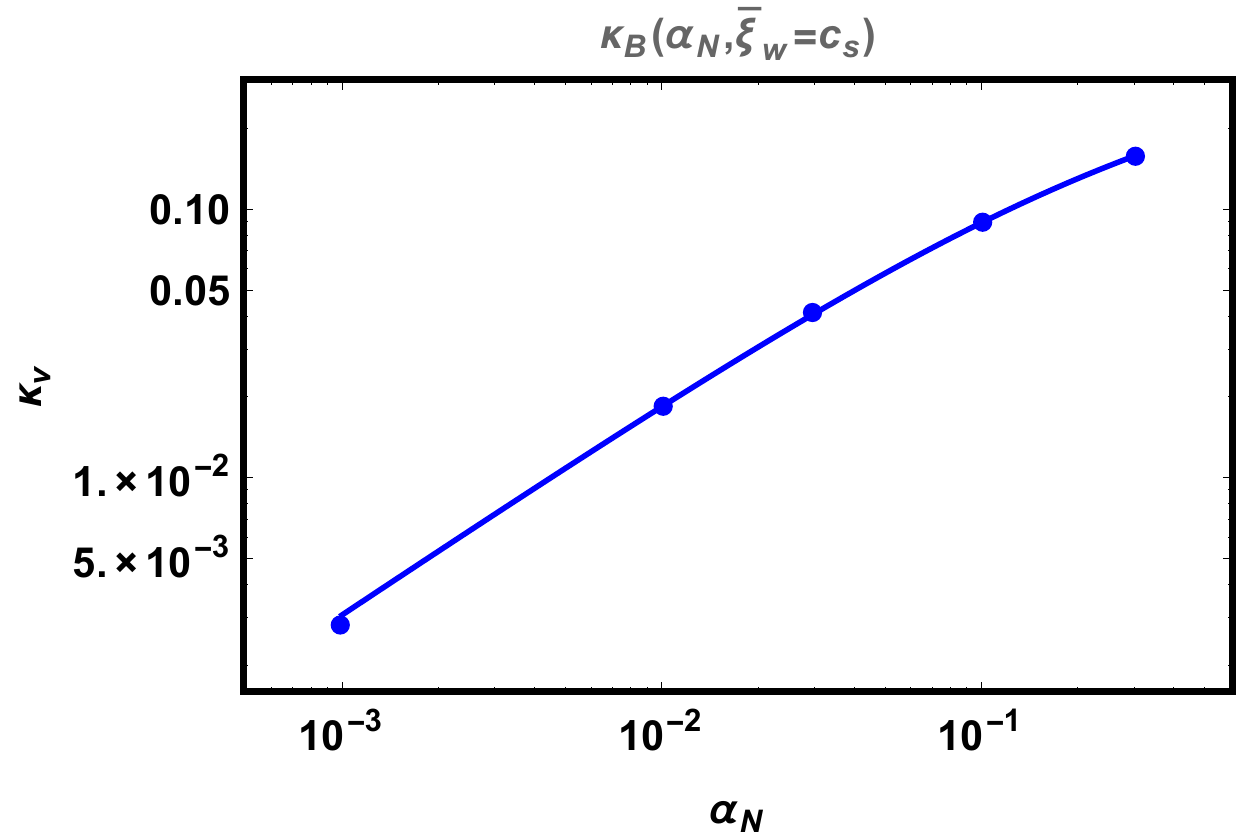}\\
\includegraphics[width=0.49\textwidth]{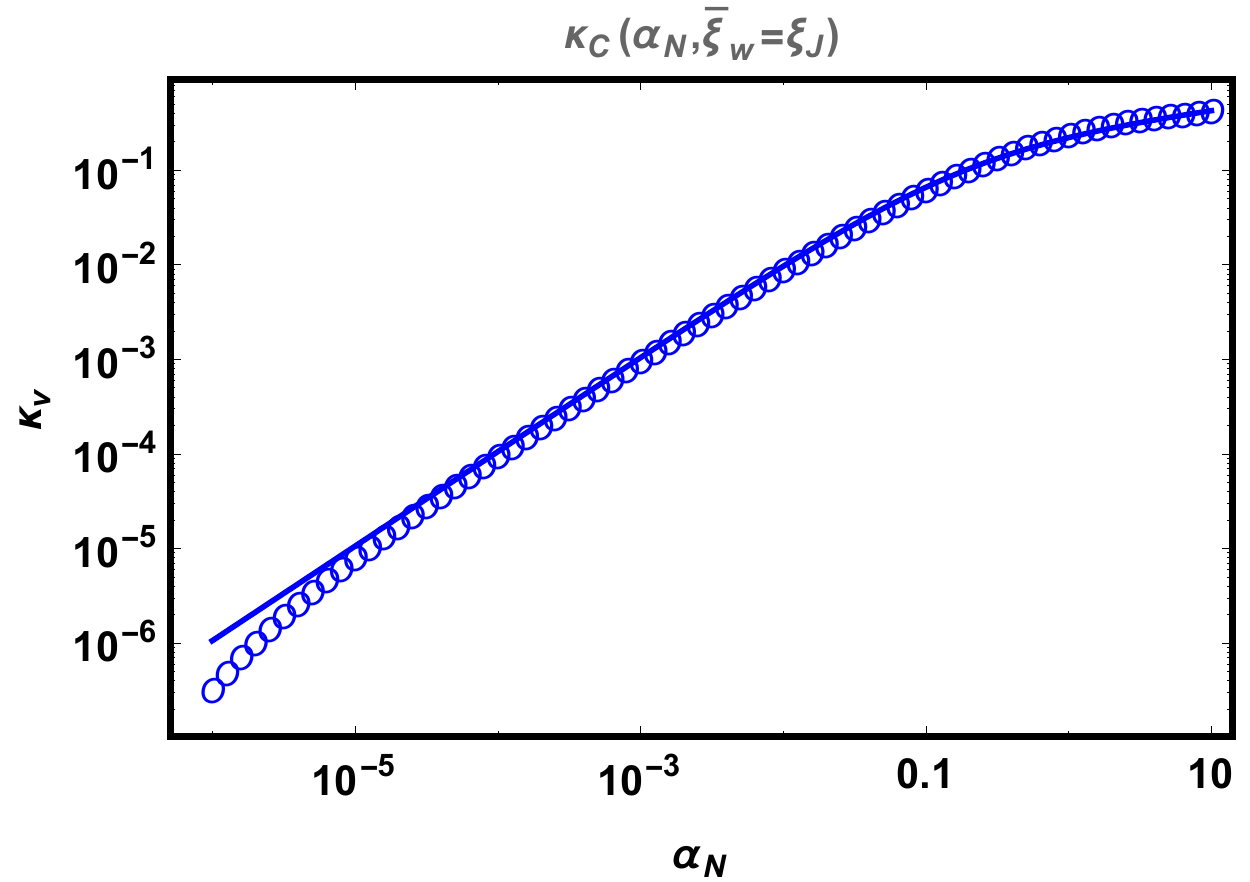}
\includegraphics[width=0.49\textwidth]{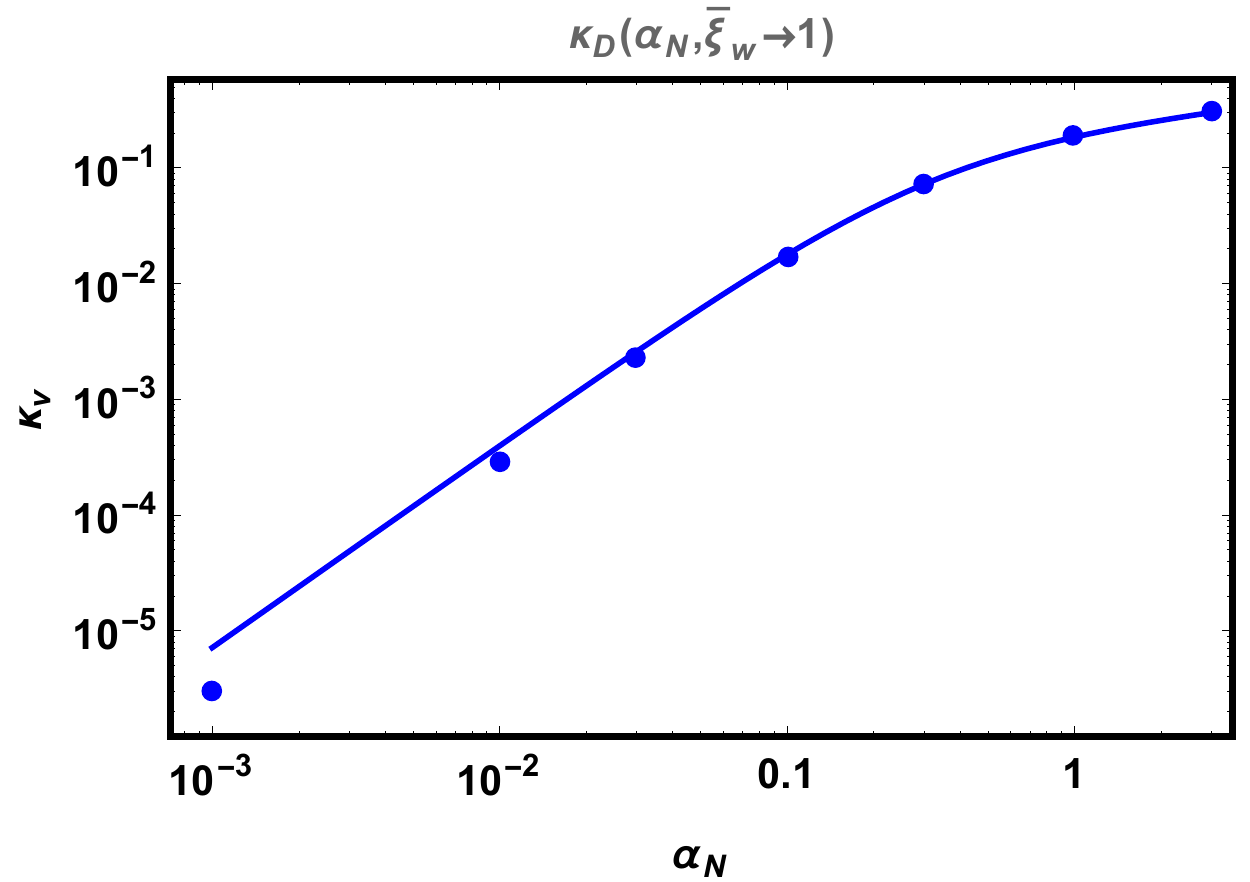}\\
\includegraphics[width=0.49\textwidth]{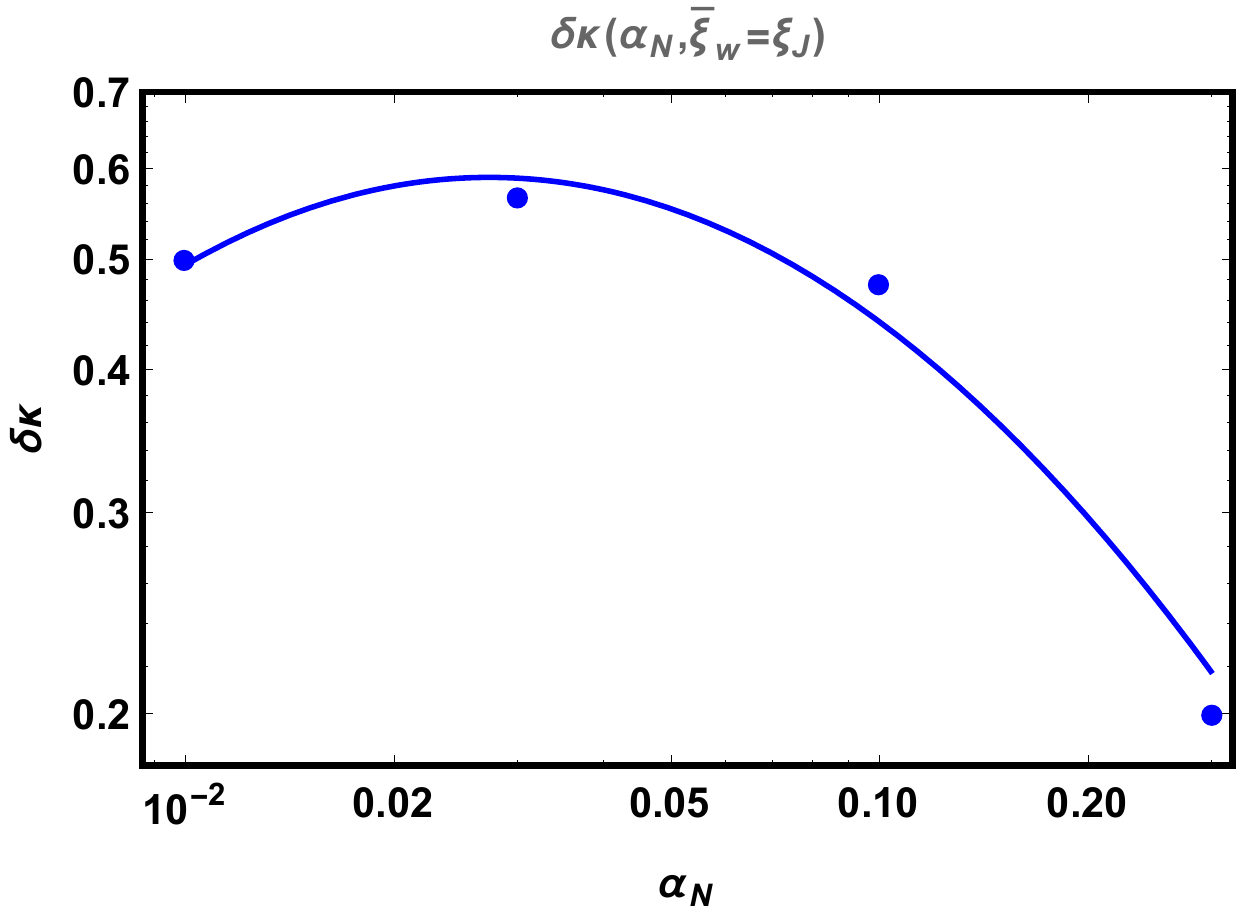}\\
\caption{The numerical fitting of efficiency factors $\kappa_A$, $\kappa_B$, $\kappa_C$ and $\kappa_D$ are shown in the first four panels if the bubble wall velocity is non-relativistic, acoustic, Jouguet and ultra-relativistic, respectively. The slop of efficiency factor at the continuous transition from deflagration region to hybrid region is fitted in the last panel. It is worth noting that the slop of efficiency factor at the transition from hybrid region to detonation region is not continuous. Interpolating these fitting formulas at the boundaries of deflagration, hybrid and detonation regions, one could find the fitting formulas over the whole parameter space of $(\bar{\xi}_w,\alpha_N)$.}\label{fig:fitting}
\end{figure}

The calculations of efficiency factor are rather involved from macroscopic hydrodynamics as we have already seen in above sections. To make life easier for those who just want to know $\kappa_v$ for given $\xi_w$ and $\alpha_N$, some fitting formulas of function $\kappa_v(\xi_w,\alpha_N)$ for fast first-order phase transition in flat background are constructed in \cite{Espinosa:2010hh} for
deflagration wave
\begin{align}
\kappa_v(\xi_w\lesssim c_s)\simeq\frac{c_s^\frac{11}{5}\kappa_A\kappa_B}{(c_s^\frac{11}{5}-\xi_w^\frac{11}{5})\kappa_B+\xi_wc_s^\frac65\kappa_A},
\end{align}
hybrid wave
\begin{align}
\kappa_v(c_s<\xi_w<\xi_J)\simeq\kappa_B+(\xi_w-c_s)\delta\kappa+\frac{(\xi_w-c_s)^3}{(\xi_J-c_s)^3}[\kappa_C-\kappa_B-(\xi_J-c_s)\delta\kappa],
\end{align}
and detonation wave
\begin{align}
\kappa_v(\xi_J\lesssim \xi_w)\simeq\frac{(\xi_J-1)^3\xi_J^\frac52\xi_w^{-\frac52}\kappa_C\kappa_D}{[(\xi_J-1)^3-(\xi_w-1)^3]\xi_J^\frac52\kappa_C+(\xi_w-1)^3\kappa_D},
\end{align}
where $\kappa_A$, $\kappa_B$, $\kappa_C$ and $\kappa_D$ are the fitting formulas for the bubble wall velocity being non-relativistic, acoustic, Jouguet and ultra-relativistic,
\begin{align}
\xi_w\ll c_s&:\quad\kappa_A\simeq \xi_w^\frac65\frac{6.9\alpha_N}{1.36-0.037\sqrt{\alpha_N}+\alpha_N};\\
\xi_w=c_s&:\quad\kappa_B\simeq\frac{\alpha_N^\frac25}{0.017+(0.997+\alpha_N)^\frac25};\\
\xi_w=\xi_J&:\quad\kappa_C\simeq\frac{\sqrt{\alpha_N}}{0.135+\sqrt{0.98+\alpha_N}};\\
\xi_w\rightarrow1&:\quad\kappa_D\simeq\frac{\alpha_N}{0.73+0.083\sqrt{\alpha_N}+\alpha_N},
\end{align}
and $\delta\kappa$ measures the slope of $\kappa_v$ at the continuous transition from deflagration region to hybrid region,
\begin{align}
\delta\kappa\approx-0.9\log\frac{\sqrt{\alpha_N}}{1+\sqrt{\alpha_N}}.
\end{align}

Similar to the case for fast first-order phase transition in flat background, here we also make an attempt to construct some fitting formulas of function $\kappa_{\bar{v}}(\bar{\xi}_w,\alpha_N)$ for slow first-order phase transition in FLRW background for
deflagration region
\begin{align}
\kappa_{\bar{v}}(\bar{\xi}_w\lesssim c_s)\simeq
\frac{c_s^\frac{33}{5}\kappa_A^\frac32\kappa_B}{(c_s^\frac{33}{5}-\bar{\xi}_w^\frac{33}{5})\kappa_B+\bar{\xi}_w^{-\frac15}c_s^\frac{34}{5}\kappa_A^\frac32},
\end{align}
Hybrid region
\begin{align}
\kappa_{\bar{v}}(c_s<\bar{\xi}_w<\bar{\xi}_J)\simeq
\kappa_B+(\bar{\xi}_w-c_s)\delta\kappa+\frac{(\bar{\xi}_w-c_s)^9}{(\bar{\xi}_J-c_s)^9}[\kappa_C-\kappa_B-(\bar{\xi}_J-c_s)\delta\kappa],
\end{align}
Detonation region
\begin{align}
\kappa_{\bar{v}}(\bar{\xi}_J\lesssim \bar{\xi}_w)\simeq
\frac{(\bar{\xi}_J-1)^3\bar{\xi}_J^\frac{14}{5}\bar{\xi}_w^{-\frac{14}{5}}\kappa_C\kappa_D}{[(\bar{\xi}_J-1)^3-(\bar{\xi}_w-1)^3]\bar{\xi}_J^\frac{14}{5}\kappa_C+(\bar{\xi}_w-1)^3\kappa_D},
\end{align}
where
\begin{align}
\bar{\xi}_w\ll c_s&:\quad\kappa_A=\bar{\xi}_w^2\frac{2.305\alpha_N}{0.500-0.863\sqrt{\alpha_N}+\alpha_N};\\
\bar{\xi}_w=c_s&:\quad\kappa_B=\frac{\alpha_N^\frac45}{1.283+2.825\alpha_N^\frac34};\\
\bar{\xi}_w=\bar{\xi}_J&:\quad\kappa_C=\frac{\alpha_N}{0.952+3.579\alpha_N^\frac45};\\
\bar{\xi}_w\rightarrow1&:\quad\kappa_D=\frac{\alpha_N^\frac74}{0.789+3.663\alpha_N^\frac32+\alpha_N};\\
\bar{\xi}_w=c_s&:\quad\delta\kappa=0.063\alpha_N^{-0.175\log\alpha_N-1.25}.
\end{align}
The outcomes of our fitting formulas for slow first-order phase transition in FLRW background are presented in Fig.\ref{fig:efficiency} as dashed lines. However, compared with the solid lines, the precision we achieved is not as good as those obtained for fast first-order phase transition in flat background, which is better than $15\%$ in the region $10^{-3}<\alpha_N<10$.

\section{Conclusions}\label{sec:conclusion}

The cosmological first-order phase transition could contribute to the stochastic GWs backgrounds from colliding bubble walls, sound waves and MHD turbulences, which might be detected in future space-borne GW detectors. In the early literatures, only the contributions from colliding bubble walls and MHD turbulences are appreciated, however, the recent numerical simulations \cite{Hindmarsh:2013xza,Hindmarsh:2015qta,Hindmarsh:2017gnf} indicate that, the sound waves from bulk fluid motion could be the main source of GWs. The efficiency factor is thus defined to characterize the amount of energy liberated into bulk fluid motion compared with the total released vacuum energy. The previous calculation \cite{Espinosa:2010hh} of this efficiency factor from macroscopic hydrodynamics was implemented in flat background, so were those numerical simulations, which can be applied to the case of fast first-order phase transition. In this paper, we take a closer look at the effect of Hubble expansion on bubble expansion applied to the case of slow first-order phase transition. It is found that, for given peculiar velocity, the efficiency factor is significantly reduced in slow first-order phase transition than that in fast first-order phase transition, which will result in less GWs contributions from the sound waves.

\appendix

\section{Bubble wall velocity}

In the previous study of macroscopic hydrodynamics, we have assumed that the bubble wall expansion has reached the stationary state in bubble center frame with a presupposed bubble wall peculiar velocity. In this appendix, we outline the usually-adopted model-independent approach for estimating the bubble wall velocity by taking into account two new features: one is that, the bubble wall may never runaway in bubble center frame according to the recent claim in \cite{Bodeker:2017cim} due to extra friction from transition radiation; the other one is that, the background spacetime has experienced the Hubble expansion, therefore it is necessary to check the form of the Boltzmann equation and the EOM of the scalar-fluid system, which turns out to be unchanged with appropriate redefinition in bubble center frame.

\subsection{Boltzmann equation}

In the bubble center frame with comoving coordinates $\mathrm{d}s^2=a(\bar{t}+\bar{t}_n)^2(-\mathrm{d}\bar{t}^2+\delta_{ij}\mathrm{d}\bar{x}^i\mathrm{d}\bar{x}^j)$, the corresponding 4-momentum is defined by $p^\mu=\mathrm{d}\bar{x}^\mu/\mathrm{d}\lambda$. For particle of mass $m$, the wordline parameter $\lambda$ is chose as $\tau/m$, where the proper time $\tau$ is defined by $-\mathrm{d}\tau^2=\mathrm{d}s^2$. Then the components of 4-momentum $p^\mu=m\bar{\gamma}(1,\bar{v}^i)/a$ is written by the Lorentz factor $\bar{\gamma}=1/\sqrt{1-\bar{v}^2}$ of the norm $\bar{v}^2=\delta_{ij}\bar{v}^i\bar{v}^j$ of peculiar 3-velocity $\bar{v}^i=\mathrm{d}\bar{x}^i/\mathrm{d}\bar{t}$. Then the norm of the 3-momentum $\mathbf{p}^2=g_{ij}p^ip^j=m^2\bar{\gamma}^2\bar{v}^2$ is equal to the norm $\mathbf{\bar{p}}^2=\delta_{ij}\bar{p}^i\bar{p}^j$ of barred 3-momentum defined by $\bar{p}_i=p_i/a=ap^i=\bar{p}^i=m\bar{\gamma}\bar{v}^i$. The on-shell relation is then $p^2=g_{\mu\nu}p^\mu p^\nu=-E^2+\mathbf{p}^2=-m^2$ with $E^2=-g_{00}(p^0)^2$ and $\mathbf{p}^2=g_{ij}p^ip^j=\delta_{ij}\bar{p}^i\bar{p}^j=\mathbf{\bar{p}}^2$. Conventionally, the unbarred 3-momentum is referred as the physical peculiar momentum, and the barred 3-momentum is referred as the comoving peculiar momentum, which can be raised up and down as if they are in the Euclidean space. The physical meaning of the comoving peculiar velocity is that they are actually the relative velocity with respect to the comoving frame (a frame comoving with Hubble expansion, not bubble expansion).

The distribution function $f(\bar{x}^\mu,p_\mu)$ in phase space evolves by the Boltzmann equation
\begin{align}
\frac{\mathrm{D}}{\mathrm{d}\lambda}f
\equiv\left(\frac{\mathrm{D}\bar{x}^\mu}{\mathrm{d}\lambda}\frac{\partial}{\partial \bar{x}^\mu}+\frac{\mathrm{D}p_\mu}{\mathrm{d}\lambda}\frac{\partial}{\partial p_\mu}\right)f
=C[f],
\end{align}
where $C[f]$ is the usual collision term, and the directional covariant derivatives are of form
\begin{align}
\frac{\mathrm{D}\bar{x}^\mu}{\mathrm{d}\lambda}&=\frac{\mathrm{d}\bar{x}^\mu}{\mathrm{d}\lambda}=p^\mu;\\
\frac{\mathrm{D}p_\mu}{\mathrm{d}\lambda}&=\frac{\mathrm{d}p_\mu}{\mathrm{d}\lambda}-\Gamma^\sigma_{\mu\nu}p_\sigma p^\nu\equiv mF_\mu,
\end{align}
The external force can be defined by the geodesic equation
\begin{align}
\frac{\mathrm{d}p^\mu}{\mathrm{d}\lambda}+\Gamma^\mu_{\nu\sigma}p^\nu p^\sigma\equiv mF^\mu,
\end{align}
which has a special form $F_\mu=-\partial_\mu m$ resulted from the directional covariant derivative of on-shell relation for spatial-dependent effective mass term. Therefore, the form of Boltzmann equation is unchanged as in the flat background, namely
\begin{align}\label{eq:Boltzmann}
\left(p^\mu\frac{\partial}{\partial x^\mu}+mF_\mu\frac{\partial}{\partial p_\mu}\right)f=C[f].
\end{align}

\subsection{Equation-of-motion}

In the curved spacetime, the energy-momentum tensor for the thermal fluid is of form
\begin{align}
T^{\mu\nu}_f=\sum\limits_{i=\mathrm{B},\mathrm{F}}g_i\int\frac{\mathrm{d}p_1\mathrm{d}p_2\mathrm{d}p_3}{(2\pi)^3\sqrt{-g}p^0}p^\mu p^\nu f_i
=\sum\limits_{i=\mathrm{B},\mathrm{F}}g_i\int\frac{\mathrm{d}\bar{p}_1\mathrm{d}\bar{p}_2\mathrm{d}\bar{p}_3}{(2\pi)^3E_i}p^\mu p^\nu f_i,
\end{align}
where $\sqrt{-g}=\sqrt{-g_{00}}a^3$, $\sqrt{-g_{00}}p^0=E_i=\sqrt{\mathbf{\bar{p}}^2+m_i^2}$, and $p_i/a=\bar{p}_i$ have been used. The above form is covariant by noting that
\begin{align}\label{eq:d4p}
\int\frac{\mathrm{d}^3\mathbf{p}}{\sqrt{-g}2p^0}=\int\frac{\mathrm{d}^3\mathbf{\bar{p}}}{2E}
=\int\mathrm{d}^3\mathbf{\bar{p}}\int_{-\infty}^\infty\mathrm{d}E\delta(E^2-\mathbf{\bar{p}}^2-m^2)\theta(E)
\end{align}
Multiplying both side of Eq.\eqref{eq:Boltzmann} by Eq.\eqref{eq:d4p} with extra multiplier $p_\nu$, and then summing over the particle of species $i$ gives rise to
\begin{align}
\sum\limits_ig_i\int\frac{\mathrm{d}^3\mathbf{\bar{p}}}{(2\pi)^3E_i}&p^\mu p_\nu\partial_\mu f_i=\nabla_\mu T_{f\nu}^\mu
=-\nabla_\mu T_{\phi\nu}^\mu=-\nabla_\nu\phi(\nabla_\mu\nabla^\mu\phi-\frac{\partial V_0}{\partial\phi});\\
\sum\limits_ig_i\int\frac{\mathrm{d}^3\mathbf{\bar{p}}}{(2\pi)^3E_i}&mF_\mu p_\nu\frac{\partial}{\partial p_\mu}f_i
=\nabla_\nu\phi\sum\limits_ig_i\frac{\mathrm{d}m_i^2}{\mathrm{d}\phi}\int\frac{\mathrm{d}^3\mathbf{\bar{p}}}{(2\pi)^32E_i}f_i,
\end{align}
where total conservation law $\nabla_\mu(T_f^{\mu\nu}+T_\phi^{\mu\nu})=0$ is used in the first line, and integration by part with $F_\mu=-\partial_\mu m$ is used in the second line. The collision term simply vanishes upon above manipulations if collisions of particles happen at some points connected with geodesic equation. Therefore, the EOM of scalar-fluid system is obtained as
\begin{align}\label{eq:EOM1}
-\nabla_\mu\nabla^\mu\phi+\frac{\partial V_0}{\partial\phi}+\sum\limits_i g_i\frac{\mathrm{d}m_i^2}{\mathrm{d}\phi}\int\frac{\mathrm{d}^3\mathbf{\bar{p}}}{(2\pi)^32E_i}f_i=0.
\end{align}

Splitting the distribution function into equilibrium and non-equilibrium parts $f_i=f_i^\mathrm{eq}+\delta f_i$ with $f_i^\mathrm{eq}=\exp(-E_i/T)/(1\mp\exp(-E_i/T))$, one found that the derivative of the finite temperature part of effective potential,
\begin{align}
\frac{\partial V_T}{\partial\phi}&=\frac{\partial}{\partial\phi}\sum\limits_i\pm g_iT\int\frac{\mathrm{d}^3\mathbf{\bar{p}}}{(2\pi)^3}(1\mp\exp(-E_i/T));\\
&=\sum\limits_i\pm g_iT\int\frac{\mathrm{d}^3\mathbf{\bar{p}}}{(2\pi)^3}\frac{\mp f_i^\mathrm{eq}}{T}\left(-\frac{\mathrm{d}E_i}{\mathrm{d}\phi}\right);\\
&=\sum\limits_ig_i\frac{\mathrm{d}m_i^2}{\mathrm{d}\phi}\int\frac{\mathrm{d}^3\mathbf{\bar{p}}}{(2\pi)^32E_i}f_i^\mathrm{eq},
\end{align}
is exactly the equilibrium part of the third term in the left hand side of Eq.\eqref{eq:EOM1}, therefore the EOM of scalar-fluid system reads
\begin{align}\label{eq:EOM2}
-\nabla_\mu\nabla^\mu\phi+\frac{\partial\mathcal{F}}{\partial\phi}-\mathcal{K}(\phi)=0,
\end{align}
where the driving term comes from the free energy density, and the friction term comes from the departure from equilibrium,
\begin{align}
\mathcal{K}(\phi)=-\sum\limits_i g_i\frac{\mathrm{d}m_i^2}{\mathrm{d}\phi}\int\frac{\mathrm{d}^3\mathbf{\bar{p}}}{(2\pi)^32E_i}\delta f_i.
\end{align}
Due to a recent finding in \cite{Bodeker:2017cim}, the friction term should contain a Lorentz factor that grows without bound for an accelerating bubble wall, thus leading to an eventual balance between driving force and friction force. Therefore the bubble wall cannot runaway in this case. Such phenomenological parametrization of friction term
\begin{align}
\mathcal{K}(\phi)=T_N\widetilde{\eta}u^\mu\partial_\mu\phi
\end{align}
has already been proposed in \cite{Ignatius:1993qn,Megevand:2009gh} before a modified parametrization of friction term \cite{Espinosa:2010hh,Megevand:2013hwa} for runaway behavior \cite{Bodeker:2009qy}.

\subsection{Expansion equation}

To write down the explicit form of the EOM \eqref{eq:EOM2}, it is usually conventional and convenient to work under planar limit where the bubble wall moves along $z$ direction. With use of the comoving coordinate system, the bubble center frame (comoving with Hubble expansion) is presented by $(\bar{t},\bar{z})$, and the bubble wall frame (comoving with bubble expansion) is presented by $(\bar{t}',\bar{z}')$. The comoving position of the bubble wall is thus presented by $\bar{z}_w(\bar{t})$ with comoving peculiar velocity $\bar{v}_w(\bar{t})$ \footnote{It should not be confused with the fluid velocity $\bar{v}(\bar{\xi}_w)$ at the bubble wall in the previous sections.} and its corresponding Lorentz factor $\bar{\gamma}_w(\bar{t})$. In the bubble wall frame, the scalar profile depends only on $\bar{z}'$ through $\phi(\bar{z}')$ with suitable boundary conditions $\phi(\bar{z}'=-\infty)=\phi_-$, $\phi(\bar{z}'=0)=\phi_-/2$, $\phi(\bar{z}'=+\infty)=\phi_+$. When written in the bubble center frame, the scalar profile depends both on the $\bar{t}$ and $\bar{z}$ by $\phi(\bar{t},\bar{z})=\phi(\bar{\gamma}_w(\bar{t})[\bar{z}-\bar{z}_w(\bar{t})])\equiv\phi(\bar{z}')$ through a local Lorentz transformation
\begin{align}
\bar{t}'&=\bar{\gamma}_w(\bar{t})[\bar{t}-\bar{v}_w(\bar{t})\bar{z}];\\
\bar{z}'&=\bar{\gamma}_w(\bar{t})[\bar{z}-\bar{z}_w(\bar{t})].
\end{align}
The time derivatives of scalar profile in bubble center frame are computed directly as
\begin{align}
\frac{\partial}{\partial\bar{t}}\phi(\bar{t},\bar{z})&=\phi'(\bar{z}')[\dot{\bar{\gamma}}_w(\bar{z}-\bar{z}_w)-\bar{\gamma}_w\dot{\bar{z}}_w];\\
\frac{\partial^2}{\partial\bar{t}^2}\phi(\bar{t},\bar{z})
&=\phi''(\bar{z}')[\dot{\bar{\gamma}}_w(\bar{z}-\bar{z}_w)-\bar{\gamma}_w\dot{\bar{z}}_w]^2
+\phi'(\bar{z}')[\ddot{\bar{\gamma}}_w(\bar{z}-\bar{z}_w)-2\dot{\bar{\gamma}}_w\dot{\bar{z}}_w-\bar{\gamma}_w\ddot{\bar{z}}_w].
\end{align}
Therefore, the first term in EOM \eqref{eq:EOM2} can be worked out in bubble center frame as
\begin{align}
\nabla_\mu\nabla^\mu\phi&=\partial_\mu\partial^\mu\phi+\Gamma^\mu_{\mu\nu}\partial^\nu\phi
=\frac{1}{a(\bar{t}+\bar{t}_n)^2}\left(\frac{\partial^2}{\partial\bar{z}^2}-\frac{\partial^2}{\partial\bar{t}^2}\right)\phi(\bar{t},\bar{z})\\
&=\frac{1}{a^2}\left(\bar{\gamma}_w^2\phi''(\bar{z}')
-\phi''(\bar{z}')[\dot{\bar{\gamma}}_w(\bar{z}-\bar{z}_w)-\bar{\gamma}_w\dot{\bar{z}}_w]^2
-\phi'(\bar{z}')[\ddot{\bar{\gamma}}_w(\bar{z}-\bar{z}_w)-2\dot{\bar{\gamma}}_w\dot{\bar{z}}_w-\bar{\gamma}_w\ddot{\bar{z}}_w]\right).
\end{align}
To evaluate above expression, one can introduce the surface tension as
\begin{align}
\sigma\equiv\int_{-\infty}^\infty\mathrm{d}\bar{z}'\phi'(\bar{z}')^2,
\end{align}
then the mean value of some quantity $F(\bar{z}')$ cross the bubble wall can be defined by
\begin{align}
\langle F\rangle\equiv\frac{1}{\sigma}\int_{-\infty}^\infty\mathrm{d}\bar{z}'\phi'(\bar{z}')^2F(\bar{z}').
\end{align}
If $F$ is an odd function across the bubble wall, then its mean value $\langle F\rangle$ should be zero. As an example, the bubble wall position can be defined in this way by
\begin{align}
\int\mathrm{d}\bar{z}(\partial_{\bar{z}}\phi)^2(\bar{z}-\bar{z}_w)
=\int\mathrm{d}\bar{z}'\frac{\mathrm{d}\bar{z}}{\mathrm{d}\bar{z}'}\phi'(\bar{z}')^2\bar{\gamma}_w^2(\bar{z}-\bar{z}_w)
=\int_{-\infty}^\infty\mathrm{d}\bar{z}'\phi'(\bar{z}')^2\bar{z}'
=\langle\bar{z}'\rangle\sigma=0.
\end{align}
Note that the boundary conditions imply
\begin{align}
\int_{-\infty}^\infty\mathrm{d}\bar{z}'\phi'(\bar{z}')\phi''(\bar{z}')=\int_{-\infty}^\infty\mathrm{d}\phi\,\phi''=\phi'|_{-\infty}^\infty=0.
\end{align}
Therefore, one can evaluate the first term in EOM \eqref{eq:EOM2} multiplied by $\phi'(\bar{z}')$ and integrated across the bubble wall as
\begin{align}
\int_{-\infty}^\infty\mathrm{d}\bar{z}'\phi'(\bar{z}')\nabla_\mu\nabla^\mu\phi
=\frac{1}{a(\bar{t}+\bar{t}_n)^2}\int_{-\infty}^\infty\mathrm{d}\bar{z}'\phi'(\bar{z}')^2[2\dot{\bar{\gamma}}_w\dot{\bar{z}}_w+\bar{\gamma}_w\ddot{\bar{z}}_w]
\equiv\frac{\sigma}{a^2}\bar{\gamma}^3_w(1+\dot{\bar{z}}^2_w)\ddot{\bar{z}}_w.
\end{align}
For stationary expansion of the bubble wall in the bubble center frame, this term is simply zero.

Next, the second term in EOM \eqref{eq:EOM2}, when multiplied by $\phi'(\bar{z}')$ and integrated across the bubble wall, gives rise to the driving force,
\begin{align}
F_\mathrm{dr}
&\equiv\int_{-\infty}^\infty\mathrm{d}\bar{z}'\phi'(\bar{z}')\frac{\partial\mathcal{F}}{\partial\phi}(\phi(\bar{z}'),T(\bar{z}'))\\
&=\int_{-\infty}^\infty\mathrm{d}\bar{z}'\left(\frac{\mathrm{d}\mathcal{F}}{\mathrm{d}\bar{z}'}-\frac{\partial\mathcal{F}}{\partial T}T'(\bar{z}')\right);\\
&=\mathcal{F}|_-^+-\int_{T_-}^{T_+}\mathrm{d}T^2\frac{\partial\mathcal{F}}{\partial T^2};\\
&\simeq\epsilon|_-^+-\left\langle\frac{\partial\mathcal{F}}{\partial T^2}\right\rangle(T_+^2-T_-^2);\\
&=a_+T_+^4\alpha_+-\frac13(a_+-a_-)T_+^2T_-^2;\\
&=a_+T_+^4\left[\alpha_+-\frac13\left(1-\frac{a_-}{a_+}\right)\frac{T_-^2}{T_+^2}\right],
\end{align}
where in the forth line the integral is approximated by its average value across the wall,
\begin{align}
\left\langle\frac{\partial\mathcal{F}}{\partial T^2}\right\rangle
\equiv\frac12\left(\frac{\partial\mathcal{F}_+}{\partial T_+^2}+\frac{\partial\mathcal{F}_-}{\partial T_-^2}\right),
\end{align}
and in the last line the ratio of temperatures across the wall can be inferred from \eqref{eq:FLRWjunction},
\begin{align}
\frac{w_-}{w_+}=\frac{a_-T_-^4}{a_+T_+^4}=\frac{\bar{v}_+\bar{\gamma}_+^2}{\bar{v}_-\bar{\gamma}_-}\Rightarrow\frac{T_-^2}{T_+^2}
=\sqrt{\frac{a_+}{a_-}\frac{\bar{v}_+}{\bar{v}_-}\frac{\bar{\gamma}_+^2}{\bar{\gamma}_-^2}}.
\end{align}

The last term in EOM \eqref{eq:EOM2}, when multiplied by $\phi'(\bar{z}')$ and integrated across the bubble wall, gives rise to the friction force,
\begin{align}
F_\mathrm{fr}
&\equiv\int_{-\infty}^\infty\mathrm{d}\bar{z}'\phi'(\bar{z}')T_N\widetilde{\eta}u^\mu\partial_\mu\phi;\\
&=\int_{-\infty}^\infty\mathrm{d}\bar{z}'\phi'(\bar{z}')T_N\widetilde{\eta}\frac{\bar{\gamma}(\bar{v})}{a(\bar{t}+\bar{t}_n)}
\left(\phi'(\bar{z}')[\dot{\bar{\gamma}}_w(\bar{z}-\bar{z}_w)-\bar{\gamma}_w\dot{\bar{z}}_w]+\bar{v}\phi'(\bar{z}')\bar{\gamma}_w\right);\\
&=\frac{T_N\widetilde{\eta}}{a}\int_{-\infty}^\infty\mathrm{d}\bar{z}'\phi'(\bar{z}')^2
(\bar{\gamma}\bar{v}\bar{\gamma}_w-\bar{\gamma}\bar{\gamma}_w\bar{v}_w);\\
&=\frac{\sigma}{a}T_N\widetilde{\eta}\left(\bar{\gamma}_w\langle\bar{\gamma}\bar{v}\rangle-\bar{\gamma}_w\bar{v}_w\langle\bar{\gamma}\rangle\right)
\end{align}
If one introduces $\eta$ to simply parameterize the friction term as $\eta a_NT_N^4\bar{\gamma}_w\langle\bar{\gamma}\bar{v}\rangle$,
then the peculiar wall velocity of a stationary bubble expansion can be obtained from $F_\mathrm{dr}=F_\mathrm{fr}$, namely
\begin{align}
\alpha_+-\frac13\left(1-\frac{a_-}{a_+}\right)\frac{T_-^2}{T_+^2}=\eta\frac{\alpha_+}{\alpha_N}\bar{\gamma}_w\langle\bar{\gamma}\bar{v}\rangle,
\end{align}
which can be readily solved for given $\alpha_N(\alpha_+)$ and $\eta$. In practice, $\alpha_N(\alpha_+)$ and $\bar{v}_w$ are input into above equation to see if the outcome of $\eta$ could match the estimation from the microphysics of specific model.

\acknowledgments
SJW would like to thank the invitation, support and warm hospitality from Mark Hindmarsh, Kari Rummukainen and David J. Weir during his visit at Helsinki institute of physics, Helsinki university, Finland. SJW wants to thank Thomas Konstandin for his valuable observation so that the application range of our conclusion in the second version of this manuscript has shrunk down to the slow first-order phase transition. SJW also wants to thank Huai-Ke Guo, Run-Qiu Yang for helpful discussions. RGC is supported in part by the National Natural Science Foundation of China Grants No.11690022, No.11435006, No.11447601 and No.11647601, and by the Strategic Priority Research Program of CAS Grant No.XDB23030100, and by the Peng Huanwu Innovation Research Center for Theoretical Physics Grant No.11747601, and by the Key Research Program of Frontier Sciences of CAS. We acknowledge the use of HPC Cluster of ITP-CAS.

\bibliographystyle{JHEP}
\bibliography{ref}

\providecommand{\href}[2]{#2}\begingroup\raggedright\begin{thebibliography}{10}

\bibitem{Cohen:1990py}
A.~G. Cohen, D.~B. Kaplan and A.~E. Nelson, \emph{{WEAK SCALE BARYOGENESIS}},
  \href{http://dx.doi.org/10.1016/0370-2693(90)90690-8}{\emph{Phys. Lett.} {\bf
  B245} (1990) 561--564}.

\bibitem{Cohen:1990it}
A.~G. Cohen, D.~B. Kaplan and A.~E. Nelson, \emph{{Baryogenesis at the weak
  phase transition}},
  \href{http://dx.doi.org/10.1016/0550-3213(91)90395-E}{\emph{Nucl. Phys.} {\bf
  B349} (1991) 727--742}.

\bibitem{Nelson:1991ab}
A.~E. Nelson, D.~B. Kaplan and A.~G. Cohen, \emph{{Why there is something
  rather than nothing: Matter from weak interactions}},
  \href{http://dx.doi.org/10.1016/0550-3213(92)90440-M}{\emph{Nucl. Phys.} {\bf
  B373} (1992) 453--478}.

\bibitem{Cohen:1994ss}
A.~G. Cohen, D.~B. Kaplan and A.~E. Nelson, \emph{{Diffusion enhances
  spontaneous electroweak baryogenesis}},
  \href{http://dx.doi.org/10.1016/0370-2693(94)00935-X}{\emph{Phys. Lett.} {\bf
  B336} (1994) 41--47}, [\href{https://arxiv.org/abs/hep-ph/9406345}{{\tt
  hep-ph/9406345}}].

\bibitem{Cohen:1993nk}
A.~G. Cohen, D.~B. Kaplan and A.~E. Nelson, \emph{{Progress in electroweak
  baryogenesis}},
  \href{http://dx.doi.org/10.1146/annurev.ns.43.120193.000331}{\emph{Ann. Rev.
  Nucl. Part. Sci.} {\bf 43} (1993) 27--70},
  [\href{https://arxiv.org/abs/hep-ph/9302210}{{\tt hep-ph/9302210}}].

\bibitem{Witten:1984rs}
E.~Witten, \emph{{Cosmic Separation of Phases}},
  \href{http://dx.doi.org/10.1103/PhysRevD.30.272}{\emph{Phys. Rev.} {\bf D30}
  (1984) 272--285}.

\bibitem{Hogan:1986qda}
C.~J. Hogan, \emph{{Gravitational radiation from cosmological phase
  transitions}}, {\emph{Mon. Not. Roy. Astron. Soc.} {\bf 218} (1986)
  629--636}.

\bibitem{Kosowsky:1991ua}
A.~Kosowsky, M.~S. Turner and R.~Watkins, \emph{{Gravitational radiation from
  colliding vacuum bubbles}},
  \href{http://dx.doi.org/10.1103/PhysRevD.45.4514}{\emph{Phys. Rev.} {\bf D45}
  (1992) 4514--4535}.

\bibitem{Kosowsky:1992rz}
A.~Kosowsky, M.~S. Turner and R.~Watkins, \emph{{Gravitational waves from first
  order cosmological phase transitions}},
  \href{http://dx.doi.org/10.1103/PhysRevLett.69.2026}{\emph{Phys. Rev. Lett.}
  {\bf 69} (1992) 2026--2029}.

\bibitem{Kosowsky:1992vn}
A.~Kosowsky and M.~S. Turner, \emph{{Gravitational radiation from colliding
  vacuum bubbles: envelope approximation to many bubble collisions}},
  \href{http://dx.doi.org/10.1103/PhysRevD.47.4372}{\emph{Phys. Rev.} {\bf D47}
  (1993) 4372--4391}, [\href{https://arxiv.org/abs/astro-ph/9211004}{{\tt
  astro-ph/9211004}}].

\bibitem{Kamionkowski:1993fg}
M.~Kamionkowski, A.~Kosowsky and M.~S. Turner, \emph{{Gravitational radiation
  from first order phase transitions}},
  \href{http://dx.doi.org/10.1103/PhysRevD.49.2837}{\emph{Phys. Rev.} {\bf D49}
  (1994) 2837--2851}, [\href{https://arxiv.org/abs/astro-ph/9310044}{{\tt
  astro-ph/9310044}}].

\bibitem{Hogan:1983zz}
C.~J. Hogan, \emph{{Magnetohydrodynamic Effects of a First-Order Cosmological
  Phase Transition}},
  \href{http://dx.doi.org/10.1103/PhysRevLett.51.1488}{\emph{Phys. Rev. Lett.}
  {\bf 51} (1983) 1488--1491}.

\bibitem{Quashnock:1988vs}
J.~M. Quashnock, A.~Loeb and D.~N. Spergel, \emph{{Magnetic Field Generation
  During the Cosmological QCD Phase Transition}},
  \href{http://dx.doi.org/10.1086/185528}{\emph{Astrophys. J.} {\bf 344} (1989)
  L49--L51}.

\bibitem{Vachaspati:1991nm}
T.~Vachaspati, \emph{{Magnetic fields from cosmological phase transitions}},
  \href{http://dx.doi.org/10.1016/0370-2693(91)90051-Q}{\emph{Phys. Lett.} {\bf
  B265} (1991) 258--261}.

\bibitem{Cheng:1994yr}
B.-l. Cheng and A.~V. Olinto, \emph{{Primordial magnetic fields generated in
  the quark - hadron transition}},
  \href{http://dx.doi.org/10.1103/PhysRevD.50.2421}{\emph{Phys. Rev.} {\bf D50}
  (1994) 2421--2424}.

\bibitem{Baym:1995fk}
G.~Baym, D.~Bodeker and L.~D. McLerran, \emph{{Magnetic fields produced by
  phase transition bubbles in the electroweak phase transition}},
  \href{http://dx.doi.org/10.1103/PhysRevD.53.662}{\emph{Phys. Rev.} {\bf D53}
  (1996) 662--667}, [\href{https://arxiv.org/abs/hep-ph/9507429}{{\tt
  hep-ph/9507429}}].

\bibitem{Hawking:1982ga}
S.~W. Hawking, I.~G. Moss and J.~M. Stewart, \emph{{Bubble Collisions in the
  Very Early Universe}},
  \href{http://dx.doi.org/10.1103/PhysRevD.26.2681}{\emph{Phys. Rev.} {\bf D26}
  (1982) 2681}.

\bibitem{Kodama:1982sf}
H.~Kodama, M.~Sasaki and K.~Sato, \emph{{Abundance of Primordial Holes Produced
  by Cosmological First Order Phase Transition}},
  \href{http://dx.doi.org/10.1143/PTP.68.1979}{\emph{Prog. Theor. Phys.} {\bf
  68} (1982) 1979}.

\bibitem{Moss:1994iq}
I.~G. Moss, \emph{{Singularity formation from colliding bubbles}},
  \href{http://dx.doi.org/10.1103/PhysRevD.50.676}{\emph{Phys. Rev.} {\bf D50}
  (1994) 676--681}.

\bibitem{Binetruy:2012ze}
P.~Binetruy, A.~Bohe, C.~Caprini and J.-F. Dufaux, \emph{{Cosmological
  Backgrounds of Gravitational Waves and eLISA/NGO: Phase Transitions, Cosmic
  Strings and Other Sources}},
  \href{http://dx.doi.org/10.1088/1475-7516/2012/06/027}{\emph{JCAP} {\bf 1206}
  (2012) 027}, [\href{https://arxiv.org/abs/1201.0983}{{\tt 1201.0983}}].

\bibitem{Caprini:2015zlo}
C.~Caprini et~al., \emph{{Science with the space-based interferometer eLISA.
  II: Gravitational waves from cosmological phase transitions}},
  \href{http://dx.doi.org/10.1088/1475-7516/2016/04/001}{\emph{JCAP} {\bf 1604}
  (2016) 001}, [\href{https://arxiv.org/abs/1512.06239}{{\tt 1512.06239}}].

\bibitem{Cai:2017cbj}
R.-G. Cai, Z.~Cao, Z.-K. Guo, S.-J. Wang and T.~Yang, \emph{{The
  Gravitational-Wave Physics}},
  \href{http://dx.doi.org/10.1093/nsr/nwx029}{\emph{National Science Review}
  {\bf 4} (2017) 687--706}, [\href{https://arxiv.org/abs/1703.00187}{{\tt
  1703.00187}}].

\bibitem{Weir:2017wfa}
D.~J. Weir, \emph{{Gravitational waves from a first order electroweak phase
  transition: a brief review}},
  \href{http://dx.doi.org/10.1098/rsta.2017.0126}{\emph{Phil. Trans. Roy. Soc.
  Lond.} {\bf 376} (2018) 20170126},
  [\href{https://arxiv.org/abs/1705.01783}{{\tt 1705.01783}}].

\bibitem{Kobakhidze:2017mru}
A.~Kobakhidze, C.~Lagger, A.~Manning and J.~Yue, \emph{{Gravitational waves
  from a supercooled electroweak phase transition and their detection with
  pulsar timing arrays}},
  \href{http://dx.doi.org/10.1140/epjc/s10052-017-5132-y}{\emph{Eur. Phys. J.}
  {\bf C77} (2017) 570}, [\href{https://arxiv.org/abs/1703.06552}{{\tt
  1703.06552}}].

\bibitem{Cai:2017tmh}
R.-G. Cai, M.~Sasaki and S.-J. Wang, \emph{{The gravitational waves from the
  first-order phase transition with a dimension-six operator}},
  \href{http://dx.doi.org/10.1088/1475-7516/2017/08/004}{\emph{JCAP} {\bf 1708}
  (2017) 004}, [\href{https://arxiv.org/abs/1707.03001}{{\tt 1707.03001}}].

\bibitem{Megevand:2016lpr}
A.~Megevand and S.~Ramirez, \emph{{Bubble nucleation and growth in very strong
  cosmological phase transitions}},
  \href{http://dx.doi.org/10.1016/j.nuclphysb.2017.03.009}{\emph{Nucl. Phys.}
  {\bf B919} (2017) 74--109}, [\href{https://arxiv.org/abs/1611.05853}{{\tt
  1611.05853}}].

\bibitem{Jinno:2017ixd}
R.~Jinno, S.~Lee, H.~Seong and M.~Takimoto, \emph{{Gravitational waves from
  first-order phase transitions: Towards model separation by bubble nucleation
  rate}}, \href{http://dx.doi.org/10.1088/1475-7516/2017/11/050}{\emph{JCAP}
  {\bf 1711} (2017) 050}, [\href{https://arxiv.org/abs/1708.01253}{{\tt
  1708.01253}}].

\bibitem{Megevand:2017vtb}
A.~M\'{e}gevand and S.~Ram\'{\i}rez, \emph{{Bubble nucleation and growth in
  slow cosmological phase transitions}},
  \href{https://arxiv.org/abs/1710.06279}{{\tt 1710.06279}}.

\bibitem{Coleman:1977py}
S.~R. Coleman, \emph{{The Fate of the False Vacuum. 1. Semiclassical Theory}},
  \href{http://dx.doi.org/10.1103/PhysRevD.15.2929,
  10.1103/PhysRevD.16.1248}{\emph{Phys. Rev.} {\bf D15} (1977) 2929--2936}.

\bibitem{Callan:1977pt}
C.~G. Callan, Jr. and S.~R. Coleman, \emph{{The Fate of the False Vacuum. 2.
  First Quantum Corrections}},
  \href{http://dx.doi.org/10.1103/PhysRevD.16.1762}{\emph{Phys. Rev.} {\bf D16}
  (1977) 1762--1768}.

\bibitem{Linde:1980tt}
A.~D. Linde, \emph{{Fate of the False Vacuum at Finite Temperature: Theory and
  Applications}},
  \href{http://dx.doi.org/10.1016/0370-2693(81)90281-1}{\emph{Phys. Lett.} {\bf
  B100} (1981) 37--40}.

\bibitem{Linde:1981zj}
A.~D. Linde, \emph{{Decay of the False Vacuum at Finite Temperature}},
  \href{http://dx.doi.org/10.1016/0550-3213(83)90293-6,
  10.1016/0550-3213(83)90072-X}{\emph{Nucl. Phys.} {\bf B216} (1983) 421}.

\bibitem{Steinhardt:1981ct}
P.~J. Steinhardt, \emph{{Relativistic Detonation Waves and Bubble Growth in
  False Vacuum Decay}},
  \href{http://dx.doi.org/10.1103/PhysRevD.25.2074}{\emph{Phys. Rev.} {\bf D25}
  (1982) 2074}.

\bibitem{Laine:1993ey}
M.~Laine, \emph{{Bubble growth as a detonation}},
  \href{http://dx.doi.org/10.1103/PhysRevD.49.3847}{\emph{Phys. Rev.} {\bf D49}
  (1994) 3847--3853}, [\href{https://arxiv.org/abs/hep-ph/9309242}{{\tt
  hep-ph/9309242}}].

\bibitem{Moore:1995ua}
G.~D. Moore and T.~Prokopec, \emph{{Bubble wall velocity in a first order
  electroweak phase transition}},
  \href{http://dx.doi.org/10.1103/PhysRevLett.75.777}{\emph{Phys. Rev. Lett.}
  {\bf 75} (1995) 777--780}, [\href{https://arxiv.org/abs/hep-ph/9503296}{{\tt
  hep-ph/9503296}}].

\bibitem{Moore:1995si}
G.~D. Moore and T.~Prokopec, \emph{{How fast can the wall move? A Study of the
  electroweak phase transition dynamics}},
  \href{http://dx.doi.org/10.1103/PhysRevD.52.7182}{\emph{Phys. Rev.} {\bf D52}
  (1995) 7182--7204}, [\href{https://arxiv.org/abs/hep-ph/9506475}{{\tt
  hep-ph/9506475}}].

\bibitem{John:2000zq}
P.~John and M.~G. Schmidt, \emph{{Do stops slow down electroweak bubble
  walls?}}, \href{http://dx.doi.org/10.1016/S0550-3213(00)00768-9,
  10.1016/S0550-3213(02)01014-3}{\emph{Nucl. Phys.} {\bf B598} (2001)
  291--305}, [\href{https://arxiv.org/abs/hep-ph/0002050}{{\tt
  hep-ph/0002050}}].

\bibitem{Cline:2000nw}
J.~M. Cline, M.~Joyce and K.~Kainulainen, \emph{{Supersymmetric electroweak
  baryogenesis}},
  \href{http://dx.doi.org/10.1088/1126-6708/2000/07/018}{\emph{JHEP} {\bf 07}
  (2000) 018}, [\href{https://arxiv.org/abs/hep-ph/0006119}{{\tt
  hep-ph/0006119}}].

\bibitem{Carena:2000id}
M.~Carena, J.~M. Moreno, M.~Quiros, M.~Seco and C.~E.~M. Wagner,
  \emph{{Supersymmetric CP violating currents and electroweak baryogenesis}},
  \href{http://dx.doi.org/10.1016/S0550-3213(01)00032-3}{\emph{Nucl. Phys.}
  {\bf B599} (2001) 158--184},
  [\href{https://arxiv.org/abs/hep-ph/0011055}{{\tt hep-ph/0011055}}].

\bibitem{Carena:2002ss}
M.~Carena, M.~Quiros, M.~Seco and C.~E.~M. Wagner, \emph{{Improved results in
  supersymmetric electroweak baryogenesis}},
  \href{http://dx.doi.org/10.1016/S0550-3213(02)01065-9}{\emph{Nucl. Phys.}
  {\bf B650} (2003) 24--42}, [\href{https://arxiv.org/abs/hep-ph/0208043}{{\tt
  hep-ph/0208043}}].

\bibitem{Konstandin:2005cd}
T.~Konstandin, T.~Prokopec, M.~G. Schmidt and M.~Seco, \emph{{MSSM electroweak
  baryogenesis and flavor mixing in transport equations}},
  \href{http://dx.doi.org/10.1016/j.nuclphysb.2005.11.028}{\emph{Nucl. Phys.}
  {\bf B738} (2006) 1--22}, [\href{https://arxiv.org/abs/hep-ph/0505103}{{\tt
  hep-ph/0505103}}].

\bibitem{Cirigliano:2006dg}
V.~Cirigliano, S.~Profumo and M.~J. Ramsey-Musolf, \emph{{Baryogenesis,
  Electric Dipole Moments and Dark Matter in the MSSM}},
  \href{http://dx.doi.org/10.1088/1126-6708/2006/07/002}{\emph{JHEP} {\bf 07}
  (2006) 002}, [\href{https://arxiv.org/abs/hep-ph/0603246}{{\tt
  hep-ph/0603246}}].

\bibitem{Kozaczuk:2015owa}
J.~Kozaczuk, \emph{{Bubble Expansion and the Viability of Singlet-Driven
  Electroweak Baryogenesis}},
  \href{http://dx.doi.org/10.1007/JHEP10(2015)135}{\emph{JHEP} {\bf 10} (2015)
  135}, [\href{https://arxiv.org/abs/1506.04741}{{\tt 1506.04741}}].

\bibitem{Huber:2013kj}
S.~J. Huber and M.~Sopena, \emph{{An efficient approach to electroweak bubble
  velocities}},  \href{https://arxiv.org/abs/1302.1044}{{\tt 1302.1044}}.

\bibitem{Konstandin:2014zta}
T.~Konstandin, G.~Nardini and I.~Rues, \emph{{From Boltzmann equations to
  steady wall velocities}},
  \href{http://dx.doi.org/10.1088/1475-7516/2014/09/028}{\emph{JCAP} {\bf 1409}
  (2014) 028}, [\href{https://arxiv.org/abs/1407.3132}{{\tt 1407.3132}}].

\bibitem{KurkiSuonio:1984ba}
H.~Kurki-Suonio, \emph{{Deflagration Bubbles in the Quark - Hadron Phase
  Transition}},
  \href{http://dx.doi.org/10.1016/0550-3213(85)90135-X}{\emph{Nucl. Phys.} {\bf
  B255} (1985) 231--252}.

\bibitem{Ignatius:1993qn}
J.~Ignatius, K.~Kajantie, H.~Kurki-Suonio and M.~Laine, \emph{{The growth of
  bubbles in cosmological phase transitions}},
  \href{http://dx.doi.org/10.1103/PhysRevD.49.3854}{\emph{Phys. Rev.} {\bf D49}
  (1994) 3854--3868}, [\href{https://arxiv.org/abs/astro-ph/9309059}{{\tt
  astro-ph/9309059}}].

\bibitem{Megevand:2009ut}
A.~Megevand and A.~D. Sanchez, \emph{{Detonations and deflagrations in
  cosmological phase transitions}},
  \href{http://dx.doi.org/10.1016/j.nuclphysb.2009.05.007}{\emph{Nucl. Phys.}
  {\bf B820} (2009) 47--74}, [\href{https://arxiv.org/abs/0904.1753}{{\tt
  0904.1753}}].

\bibitem{Megevand:2009gh}
A.~Megevand and A.~D. Sanchez, \emph{{Velocity of electroweak bubble walls}},
  \href{http://dx.doi.org/10.1016/j.nuclphysb.2009.09.019}{\emph{Nucl. Phys.}
  {\bf B825} (2010) 151--176}, [\href{https://arxiv.org/abs/0908.3663}{{\tt
  0908.3663}}].

\bibitem{Espinosa:2010hh}
J.~R. Espinosa, T.~Konstandin, J.~M. No and G.~Servant, \emph{{Energy Budget of
  Cosmological First-order Phase Transitions}},
  \href{http://dx.doi.org/10.1088/1475-7516/2010/06/028}{\emph{JCAP} {\bf 1006}
  (2010) 028}, [\href{https://arxiv.org/abs/1004.4187}{{\tt 1004.4187}}].

\bibitem{Megevand:2013hwa}
A.~M\'{e}gevand, \emph{{Friction forces on phase transition fronts}},
  \href{http://dx.doi.org/10.1088/1475-7516/2013/07/045}{\emph{JCAP} {\bf 1307}
  (2013) 045}, [\href{https://arxiv.org/abs/1303.4233}{{\tt 1303.4233}}].

\bibitem{Bodeker:2009qy}
D.~Bodeker and G.~D. Moore, \emph{{Can electroweak bubble walls run away?}},
  \href{http://dx.doi.org/10.1088/1475-7516/2009/05/009}{\emph{JCAP} {\bf 0905}
  (2009) 009}, [\href{https://arxiv.org/abs/0903.4099}{{\tt 0903.4099}}].

\bibitem{Bodeker:2017cim}
D.~Bodeker and G.~D. Moore, \emph{{Electroweak Bubble Wall Speed Limit}},
  \href{http://dx.doi.org/10.1088/1475-7516/2017/05/025}{\emph{JCAP} {\bf 1705}
  (2017) 025}, [\href{https://arxiv.org/abs/1703.08215}{{\tt 1703.08215}}].

\bibitem{Leitao:2010yw}
L.~Leitao and A.~Megevand, \emph{{Spherical and non-spherical bubbles in
  cosmological phase transitions}},
  \href{http://dx.doi.org/10.1016/j.nuclphysb.2010.11.012}{\emph{Nucl. Phys.}
  {\bf B844} (2011) 450--470}, [\href{https://arxiv.org/abs/1010.2134}{{\tt
  1010.2134}}].

\bibitem{Leitao:2014pda}
L.~Leitao and A.~Megevand, \emph{{Hydrodynamics of phase transition fronts and
  the speed of sound in the plasma}},
  \href{http://dx.doi.org/10.1016/j.nuclphysb.2014.12.008}{\emph{Nucl. Phys.}
  {\bf B891} (2015) 159--199}, [\href{https://arxiv.org/abs/1410.3875}{{\tt
  1410.3875}}].

\bibitem{Megevand:2013yua}
A.~Megevand and F.~A. Membiela, \emph{{Stability of cosmological deflagration
  fronts}}, \href{http://dx.doi.org/10.1103/PhysRevD.89.103507}{\emph{Phys.
  Rev.} {\bf D89} (2014) 103507}, [\href{https://arxiv.org/abs/1311.2453}{{\tt
  1311.2453}}].

\bibitem{Megevand:2014yua}
A.~Megevand and F.~A. Membiela, \emph{{Stability of cosmological detonation
  fronts}}, \href{http://dx.doi.org/10.1103/PhysRevD.89.103503}{\emph{Phys.
  Rev.} {\bf D89} (2014) 103503}, [\href{https://arxiv.org/abs/1402.5791}{{\tt
  1402.5791}}].

\bibitem{Megevand:2014dua}
A.~Megevand, F.~A. Membiela and A.~D. Sanchez, \emph{{Lower bound on the
  electroweak wall velocity from hydrodynamic instability}},
  \href{http://dx.doi.org/10.1088/1475-7516/2015/03/051}{\emph{JCAP} {\bf 1503}
  (2015) 051}, [\href{https://arxiv.org/abs/1412.8064}{{\tt 1412.8064}}].

\bibitem{Jackson:2018maa}
G.~Jackson and M.~Laine, \emph{{Hydrodynamic fluctuations from a weakly coupled
  scalar field}},  \href{https://arxiv.org/abs/1803.01871}{{\tt 1803.01871}}.

\bibitem{Huber:2008hg}
S.~J. Huber and T.~Konstandin, \emph{{Gravitational Wave Production by
  Collisions: More Bubbles}},
  \href{http://dx.doi.org/10.1088/1475-7516/2008/09/022}{\emph{JCAP} {\bf 0809}
  (2008) 022}, [\href{https://arxiv.org/abs/0806.1828}{{\tt 0806.1828}}].

\bibitem{Weir:2016tov}
D.~J. Weir, \emph{{Revisiting the envelope approximation: gravitational waves
  from bubble collisions}},
  \href{http://dx.doi.org/10.1103/PhysRevD.93.124037}{\emph{Phys. Rev.} {\bf
  D93} (2016) 124037}, [\href{https://arxiv.org/abs/1604.08429}{{\tt
  1604.08429}}].

\bibitem{Caprini:2007xq}
C.~Caprini, R.~Durrer and G.~Servant, \emph{{Gravitational wave generation from
  bubble collisions in first-order phase transitions: An analytic approach}},
  \href{http://dx.doi.org/10.1103/PhysRevD.77.124015}{\emph{Phys. Rev.} {\bf
  D77} (2008) 124015}, [\href{https://arxiv.org/abs/0711.2593}{{\tt
  0711.2593}}].

\bibitem{Caprini:2009fx}
C.~Caprini, R.~Durrer, T.~Konstandin and G.~Servant, \emph{{General Properties
  of the Gravitational Wave Spectrum from Phase Transitions}},
  \href{http://dx.doi.org/10.1103/PhysRevD.79.083519}{\emph{Phys. Rev.} {\bf
  D79} (2009) 083519}, [\href{https://arxiv.org/abs/0901.1661}{{\tt
  0901.1661}}].

\bibitem{Jinno:2016vai}
R.~Jinno and M.~Takimoto, \emph{{Gravitational waves from bubble collisions:
  analytic derivation}},
  \href{http://dx.doi.org/10.1103/PhysRevD.95.024009}{\emph{Phys. Rev.} {\bf
  D95} (2017) 024009}, [\href{https://arxiv.org/abs/1605.01403}{{\tt
  1605.01403}}].

\bibitem{Jinno:2017fby}
R.~Jinno and M.~Takimoto, \emph{{Gravitational waves from bubble dynamics:
  Beyond the Envelope}},  \href{https://arxiv.org/abs/1707.03111}{{\tt
  1707.03111}}.

\bibitem{Hindmarsh:2013xza}
M.~Hindmarsh, S.~J. Huber, K.~Rummukainen and D.~J. Weir, \emph{{Gravitational
  waves from the sound of a first order phase transition}},
  \href{http://dx.doi.org/10.1103/PhysRevLett.112.041301}{\emph{Phys. Rev.
  Lett.} {\bf 112} (2014) 041301}, [\href{https://arxiv.org/abs/1304.2433}{{\tt
  1304.2433}}].

\bibitem{Hindmarsh:2015qta}
M.~Hindmarsh, S.~J. Huber, K.~Rummukainen and D.~J. Weir, \emph{{Numerical
  simulations of acoustically generated gravitational waves at a first order
  phase transition}},
  \href{http://dx.doi.org/10.1103/PhysRevD.92.123009}{\emph{Phys. Rev.} {\bf
  D92} (2015) 123009}, [\href{https://arxiv.org/abs/1504.03291}{{\tt
  1504.03291}}].

\bibitem{Hindmarsh:2017gnf}
M.~Hindmarsh, S.~J. Huber, K.~Rummukainen and D.~J. Weir, \emph{{Shape of the
  acoustic gravitational wave power spectrum from a first order phase
  transition}}, \href{http://dx.doi.org/10.1103/PhysRevD.96.103520}{\emph{Phys.
  Rev.} {\bf D96} (2017) 103520}, [\href{https://arxiv.org/abs/1704.05871}{{\tt
  1704.05871}}].

\bibitem{Cutting:2018tjt}
D.~Cutting, M.~Hindmarsh and D.~J. Weir, \emph{{Gravitational waves from vacuum
  first-order phase transitions: from the envelope to the lattice}},
  \href{https://arxiv.org/abs/1802.05712}{{\tt 1802.05712}}.

\bibitem{Kosowsky:2001xp}
A.~Kosowsky, A.~Mack and T.~Kahniashvili, \emph{{Gravitational radiation from
  cosmological turbulence}},
  \href{http://dx.doi.org/10.1103/PhysRevD.66.024030}{\emph{Phys. Rev.} {\bf
  D66} (2002) 024030}, [\href{https://arxiv.org/abs/astro-ph/0111483}{{\tt
  astro-ph/0111483}}].

\bibitem{Dolgov:2002ra}
A.~D. Dolgov, D.~Grasso and A.~Nicolis, \emph{{Relic backgrounds of
  gravitational waves from cosmic turbulence}},
  \href{http://dx.doi.org/10.1103/PhysRevD.66.103505}{\emph{Phys. Rev.} {\bf
  D66} (2002) 103505}, [\href{https://arxiv.org/abs/astro-ph/0206461}{{\tt
  astro-ph/0206461}}].

\bibitem{Nicolis:2003tg}
A.~Nicolis, \emph{{Relic gravitational waves from colliding bubbles and cosmic
  turbulence}},
  \href{http://dx.doi.org/10.1088/0264-9381/21/4/L05}{\emph{Class. Quant.
  Grav.} {\bf 21} (2004) L27}, [\href{https://arxiv.org/abs/gr-qc/0303084}{{\tt
  gr-qc/0303084}}].

\bibitem{Caprini:2006jb}
C.~Caprini and R.~Durrer, \emph{{Gravitational waves from stochastic
  relativistic sources: Primordial turbulence and magnetic fields}},
  \href{http://dx.doi.org/10.1103/PhysRevD.74.063521}{\emph{Phys. Rev.} {\bf
  D74} (2006) 063521}, [\href{https://arxiv.org/abs/astro-ph/0603476}{{\tt
  astro-ph/0603476}}].

\bibitem{Gogoberidze:2007an}
G.~Gogoberidze, T.~Kahniashvili and A.~Kosowsky, \emph{{The Spectrum of
  Gravitational Radiation from Primordial Turbulence}},
  \href{http://dx.doi.org/10.1103/PhysRevD.76.083002}{\emph{Phys. Rev.} {\bf
  D76} (2007) 083002}, [\href{https://arxiv.org/abs/0705.1733}{{\tt
  0705.1733}}].

\bibitem{Caprini:2009yp}
C.~Caprini, R.~Durrer and G.~Servant, \emph{{The stochastic gravitational wave
  background from turbulence and magnetic fields generated by a first-order
  phase transition}},
  \href{http://dx.doi.org/10.1088/1475-7516/2009/12/024}{\emph{JCAP} {\bf 0912}
  (2009) 024}, [\href{https://arxiv.org/abs/0909.0622}{{\tt 0909.0622}}].

\bibitem{Niksa:2018ofa}
P.~Niksa, M.~Schlederer and G.~Sigl, \emph{{Gravitational Waves produced by
  Compressible MHD Turbulence from Cosmological Phase Transitions}},
  \href{https://arxiv.org/abs/1803.02271}{{\tt 1803.02271}}.

\bibitem{KurkiSuonio:1996rk}
H.~Kurki-Suonio and M.~Laine, \emph{{Real time history of the cosmological
  electroweak phase transition}},
  \href{http://dx.doi.org/10.1103/PhysRevLett.77.3951}{\emph{Phys. Rev. Lett.}
  {\bf 77} (1996) 3951--3954},
  [\href{https://arxiv.org/abs/hep-ph/9607382}{{\tt hep-ph/9607382}}].

\bibitem{Chodos:1974je}
A.~Chodos, R.~L. Jaffe, K.~Johnson, C.~B. Thorn and V.~F. Weisskopf, \emph{{A
  New Extended Model of Hadrons}},
  \href{http://dx.doi.org/10.1103/PhysRevD.9.3471}{\emph{Phys. Rev.} {\bf D9}
  (1974) 3471--3495}.

\end{thebibliography}\endgroup

\end{document}